\RequirePackage{rotating}

\documentclass[12pt,a4paper,fleqn]{article}
\usepackage[utf8]{inputenc}
\usepackage{authblk}
\usepackage{geometry}
\usepackage{xcolor}
\usepackage[round]{natbib}
\usepackage{graphicx}
\usepackage{pdflscape}  
\usepackage{appendix}
\usepackage{threeparttable}
\usepackage{booktabs}
\usepackage{longtable}
\usepackage{amssymb}
\usepackage{amsbsy}
\usepackage{bm}
\usepackage{amsmath}
\usepackage{amsthm}
\usepackage{graphicx}
\usepackage{mathtools}
\usepackage{setspace}
\theoremstyle{definition}

\DeclareMathOperator*{\argmax}{arg\,max}
\DeclareMathOperator*{\argmin}{arg\,min}

\usepackage{color, colortbl}
\definecolor{ashgrey}{rgb}{0.7, 0.75, 0.71}
\definecolor{columbiablue}{rgb}{0.61, 0.87, 1.0}
\definecolor{coral}{rgb}{1.0, 0.5, 0.31}

\definecolor{plotred}{rgb}{1,0,0}
\definecolor{plotblue}{rgb}{0,0,0.55}
\definecolor{plotgreen}{rgb}{0,0.39,0}
\definecolor{plotgrey}{rgb}{.5,.5,.5}

\usepackage{algorithm}
\usepackage{algpseudocode}

\newlength\myindent
\usepackage{enumitem}
\newlist{steps}{enumerate}{1}
\setlist[steps,1]{label = Step \arabic*:}

\usepackage{adjustbox}
\usepackage{dcolumn}
\newcolumntype{d}[1]{D..{#1}} % for alignment of numbers on decimal marker

\usepackage{xcolor,colortbl}

\definecolor{Gray}{gray}{0.85}
\definecolor{LightCyan}{rgb}{0.88,1,1}

\def\approxprop{%
  \def\p{%
    \setbox0=\vbox{\hbox{$\propto$}}%
    \ht0=0.6ex \box0 }%
  \def\s{%
    \vbox{\hbox{$\sim$}}%
  }%
  \mathrel{\raisebox{0.7ex}{%
      \mbox{$\underset{\s}{\p}$}%
    }}%
}

\newcolumntype{a}{>{\columncolor{Gray}}c}
\newcolumntype{b}{>{\columncolor{white}}c}

\usepackage{multirow}

\usepackage{caption}
\usepackage{subcaption}
\definecolor{nblue}{HTML}{000660}
\usepackage[colorlinks=true,urlcolor=nblue,linkcolor=nblue,citecolor=nblue]{hyperref}
\newcommand*{\myeqref}[2][Equation~]{%
  \hyperref[{#2}]{#1(\ref*{#2})}%
}
\def\equationautorefname#1#2\null{%
  Eq.#1(#2\null)%
}
\def\sectionautorefname#1#2\null{%
  Section#1#2%\null)%
}
\def\subsectionautorefname#1#2\null{%
  Subsection#1#2%\null)%
}
\def\algorithmautorefname#1#2\null{%
  Algorithm#1#2%\null)%
}

\begin{document}
%\title{\textbf{Correcting Bayesian Vector Autoregressions for Incorrect Specification}

%\title{\textbf{Coarsening Incorrectly Specified Bayesian Vector Autoregressions}

\title{\textbf{ Coarsened Bayesian VARs \\
\small Correcting BVARs for Incorrect Specification}
\thanks{
\textit{Corresponding author}: Florian Huber, Department of Economics, University of Salzburg. \textit{Address}: Mönchsberg 2A, 5020-Salzburg, Austria. \textit{Email}: \href{mailto:florian.huber@plus.ac.at}{florian.huber@plus.ac.at}. 
 Huber gratefully acknowledges financial support from the Austrian Science Fund (FWF, grant no. ZK 35). Scheckel gratefully acknowledges funding from the Jubiläumsfond of the Oesterreichische Nationalbank (OeNB, grant no. JF-18763). This study was funded by the European Union (NextGenerationEU, Mission 4, Component 2), in the framework of the GRINS (Growing Resilient, INclusive and Sustainable) project (GRINS PE00000018, CUP B43C22000760006). The views and opinions expressed are solely those of the authors and do not necessarily reflect those of the European Union, nor can the European Union be held responsible for them. We would like to thank Atsushi Inoue, the handling editor, an anonymous associate editor, two referees, Luca Barbaglia, Tony Chernis, Todd Clark, Niko Hauzenberger, Gary Koop, Dimitris Korobilis, Fabian Kr\"{u}ger, Christian Matthes, Stuart McIntyre, Elmar Mertens, Luca Onorante, Anthoulla Phella, Francesco Ravazollo, Cathy Yi-Hsuan Chen and participants at research seminars at the University of Strathclyde and University of Glasgow for helpful comments and suggestions. The authors acknowledge the computational resources and services provided by Salzburg Collaborative Computing (SCC), funded by the Federal Ministry of Education, Science and Research (BMBWF) and the State of Salzburg.}
 }
%\author{}
\author[a]{Florian \textsc{Huber}}
\author[b,c]{Massimiliano \textsc{Marcellino}}
\author[a,b]{Tobias \textsc{Scheckel}}
\affil[a]{\textit{University of Salzburg}}
\affil[b]{\textit{Bocconi University}}
\affil[c]{CEPR, IGIER, BAFFI, BIDSA}
%\affil[d]{WIFO}
%\affil[c]{\textit{International Institute for Applied Systems Analysis}}

\date{\today}

\maketitle\thispagestyle{empty}\normalsize\vspace*{-2em}\small\linespread{1.5}
\begin{center}
\begin{minipage}{0.8\textwidth}
\begin{center}
    \small \textbf{Abstract}\\
\end{center}\vspace{-0.3cm}
 \noindent Model misspecification in multivariate econometric models can strongly influence estimates of quantities of interest such as structural parameters, forecast distributions or responses to structural shocks, even more so if higher-order forecasts or responses are considered, due to parameter convolution.  We propose a simple method for addressing these  specification issues in the context of Bayesian VARs. Our method, called coarsened Bayesian VARs (cBVARs), replaces the exact likelihood with a coarsened likelihood that takes into account that the model might be misspecified along important but unknown dimensions. Since endogenous variables in a VAR can feature different degrees of misspecification, our model allows for this and automatically detects the degree of misspecification. The resulting cBVARs  perform well in simulations for several types of misspecification. Applied to US data, cBVARs improve point and density forecasts compared to standard BVARs. 

\end{minipage}
\end{center}

\normalsize\newpage
\doublespacing

%\clearpage
\section{Introduction}
When working with multivariate econometric models, challenges like measurement errors, structural breaks, outliers, or non-Gaussian characteristics, including asymmetric and fat-tailed errors, frequently arise. Overlooking these in the model formulation results in misspecification. In a Bayesian framework, this suggests that the posterior distribution concentrates on the parameter value that reduces the Kullback-Leibler divergence between the true data-generating process and the chosen model. Consequently, the estimate found is essentially a pseudo-true parameter, which can be deceptive, especially if misspecification is significant \citep[see, e.g.,][]{muller2013risk}. Under such circumstances, posterior uncertainty assessments, such as credible intervals, might become excessively narrow around incorrect values, creating a false sense of certainty.

Especially in recessionary episodes or when large shocks hit the economy (e.g., the global financial crisis or the Covid-19 pandemic), violations of the standard model assumptions become prevalent. Moreover, selecting appropriate variables is necessary to strike a balance between a large model, which might include many irrelevant predictors, and a small tractable model, which could omit important information. These issues can be partly addressed by making the model larger, more flexible, and non-Gaussian, but this typically leads to substantial increases in computational complexity and risks overfitting the data. For example, \cite{cogley2005drifts, primiceri2005time, kalli2018bayesian, huber2020nowcasting, chkmp2021tail, goulet2022machine, korobilis2021,karlsson2023vector, huber2024fast} propose models that allow for time-variation and nonlinearities in the conditional mean, the conditional variance, or both.  These methods often outperform their linear counterparts in terms of reliability of estimation and predictive accuracy.  However, they are difficult to tune and the computational burden increases markedly with the size of the dataset. Similarly, to address the omitted variable problem, it is common to rely on factor models \citep[e.g.,][]{stock2002macroeconomic, bernanke2005measuring, kaufmann2019bayesian} or large Bayesian VARs \citep[see, e.g.,][]{banbura2010large, koop2013forecasting, chan2022asymmetric, gefang2023forecasting}. However, factor models require the selection of the number of factors and call for filtering techniques to estimate the latent factors. Large VARs, on the other hand, are often specified to be linear and homoskedastic. Both assumptions are necessary to retain conjugacy and thus enable fast estimation, but could be unrealistic. And one could introduce a separate stochastic process to capture measurement errors \citep[see, e.g.,][]{cogley2015measuring}, but doing so requires knowledge of the nature of the measurement error and is thus not easy to apply in general.   

In summary, while there exist methods to handle specific types of misspecification of standard, simple, econometric models, these methods are not commonly used in practice, due to their complexity and computational costs, in addition to the fact that the type of misspecification is often unknown. Small, linear, Gaussian econometric models, such as vector autoregressions (VARs), remain the workhorse of applied macroeconomists. 

There is also a literature on robust estimation of VARs and Bayesian VARs (BVARs). For instance, \cite{schorfheide2005var} analyzes multi-step-ahead forecasting with VARs under a dynamically misspecified data generating process (DGP).  More recently,  \cite{gonzalez2025misspecification} propose setting the hyperparameters of a Bayesian VAR using robust loss functions that can be tailored to the specific application at hand (i.e. whether the focus is on using the VAR for forecasting or structural inference).% [\textbf{Note for Tobi: perhaps we can add one or two additional misspecified VARs paper here. In principle, we can quote the Bayesian LP paper and by doing so directly quote the Mueller paper.}]

Our objective in this paper is to modify BVARs to make them more robust to general and unknown forms of misspecification, without changing the simple model specification and retaining computational simplicity and efficiency. We build on a recent paper, \cite{miller2018robust},  and propose a robust version of a conjugate VAR. We call this model coarsened Bayesian VAR (cBVAR). The cBVAR replaces the exact likelihood with a coarsened likelihood that takes general and unknown forms of model misspecification into account. The key idea is that instead of conditioning on the observed data $\bm Y$, one conditions on the event that the difference in the sampling distribution of the observed data $P_{\bm Y}$ and of the idealized data $P_{\bm Y^*}$ (with $\bm {Y^*}$ denoting the idealized data) is smaller than a constant number $c$, where the idealized data are such that all the model assumptions are valid. The conditioning event effectively requires a distance metric and we use relative entropy. \cite{miller2018robust} provide a simple approximation of the coarsened likelihood under relative entropy distance that reduces to raising the standard likelihood to a fraction, called the learning rate.\footnote{\cite{baumeister2019structural} downweight earlier observations in their sample by raising the corresponding likelihood contributions to a fraction smaller than one.} This approximation links the coarsening approach to the literature on power (or Gibbs) posteriors \citep{holmes2017assigning, gruenwald_van-ommen_2017_ba, bhattacharya2019bayesian}. The learning rate has a particularly simple interpretation. The smaller the extent of the misspecification, the closer the fraction is to one. Vice versa, for large misspecification the fraction is close to zero, putting less weight on the data.

What sets us apart from the literature on power posteriors is that these papers consider univariate regression models whereas our goal is to develop robust multivariate time series models. 
Using a single coarsening parameter for the full system induces the same degree of misspecification for all endogenous variables. We avoid this issue by introducing a separate coarsening parameter per equation. This implies that different equations can feature different amounts of misspecification. However, the introduction of separate coarsening parameters breaks up the convenient conjugacy of the multivariate Gaussian likelihood with the Normal-inverse Wishart prior commonly used in BVARs. As a solution, we use the asymmetric conjugate prior proposed in \cite{chan2022asymmetric} which gives more flexibility with respect to the prior and likelihood specification. 

However, this raises a serious issue. As noted in \cite{miller2018robust}, setting the learning rate appropriately is challenging. This issue is exacerbated in our framework since we have to decide on various learning rates, calling for a solution that requires little input from the researcher. We achieve this using the SafeBayes algorithm \citep{gruenwald_van-ommen_2017_ba} to set the learning rate on an equation-by-equation basis. This works since the asymmetric conjugate prior of \cite{chan2022asymmetric} avoids full system estimation by augmenting each equation with the contemporaneous values of the endogenous variables of the preceding equations while maintaining conjugacy plus order-invariance. This enables equation-by-equation estimation of all model parameters, including the hyperparameters of the prior and the equation-specific learning rates.

To assess the empirical performance of the cBVAR, we first use simulated data, generated from a large variety of data generating processes (DGPs), characterized by different types of misspecification. For each of the DGPs, we compare density forecasts obtained from the cBVAR and from a standard BVAR. It turns out that in the presence of misspecification using coarsening improves predictive accuracy appreciably, particularly so for longer forecast horizons. When comparing differences across DGPs we find that coarsening helps particularly in the case of omitted (exogenous) variables or not considered MA errors.

Next, we consider a forecasting application to show that coarsening can often lessen the effects of misspecification for actual US data. Specifically, we focus on point and density forecasting monthly US unemployment, inflation and short-term interest rates, using three model sizes (small, medium, and large), Our findings indicate that gains from coarsening the likelihood are largest if models are small (i.e. the risk of omitted variable bias is particularly pronounced), with  relative gains decreasing if the model becomes larger, but still present and systematic, in particular for longer forecast horizons. As macroeconomists tend to prefer small models for empirical analysis, this finding can be of substantial practical use. Moreover, the gains from coarsening are generally larger when the Covid period is included in the evaluation sample, in line with the increased misspecification of the standard BVAR during uncommon times.  

The remainder of the paper is structured as follows. The next section provides an intuitive introduction to the coarsening idea of \cite{miller2018robust} and links it to the literature on generalized posteriors. This section sets the stage for our model developments that follow in \autoref{sec:cBVAR} in which we derive the coarsened likelihood for the BVAR and then back out the corresponding posterior distributions. In addition, this section also includes how we set the learning rates and hyperparameters of the prior. \autoref{sec:sim_study} provides evidence from synthetic data, while \autoref{sec:forecasting} includes our real data forecast exercise. The final section offers a brief summary and conclusions. %\textbf{MAYBE IT WOULD BE CLEANER HERE TO ALSO REFERENCE THE FINAL SECTION WITH AUTOREF? (SCHECKELTON)}

\section{Coarsened Posterior Distributions}\label{sec: coarsening}
This section motivates our use of the coarsened likelihood function to derive the coarsened posterior distribution in the univariate case. To set the stage, the data we observe is defined as $\bm y = (y_1, \dots, y_T)'$ which is a $T \times 1$ matrix with empirical sampling distribution $P_{\bm y} = \frac{1}{T}\sum_{t=1}^T \delta_ {y_t}$. However, the unobserved idealized dataset is given by $\bm {y^*}$ with associated sampling distribution $P_{\bm y^*}$ defined analogously to $P_{\bm y}$. Suppose that $\bm y$ arises from $\bm y^*$ through a stochastic process which is, unfortunately, unknown and let $d(P_{\bm y}, P_{\bm y^*})$ denote a distance function with $d(\cdot, \cdot) \ge 0$. We assume that $d(P_{\bm y}, P_{\bm y^*}) < c$ for some threshold parameter $c \ge 0$. Particular examples of such stochastic processes are simple measurement error models that assume that $y_t = y_t^* + \bm \varpi$ with $\bm \varpi$ denoting a random measurement error with a particular distribution such as a Gaussian or multivariate student t or nonlinear models that assume that $ y_t = g( y_t^*)$ for some nonlinear function $g$.

Standard Bayesian practice would specify a likelihood $p(\bm y | \bm \vartheta)$, with $\bm \vartheta$ denoting a vector of parameters, and a prior on $\bm \vartheta \sim p(\bm \vartheta)$, both of which are then used to back out the posterior distribution:
\begin{equation*}
    p(\bm \vartheta | \bm y) \propto p(\bm y | \bm \vartheta) ~ p(\bm \vartheta).
\end{equation*}
This procedure, however, neglects the fact that $\bm y$ is a corrupted version of $\bm y^*$.  An alternative would be to set up an auxiliary model $p(\bm y| \bm y^*)$. However, this is not feasible since, in applications with actual data, the process giving rise to $\bm y$ from $\bm y^*$ is not known (or might render the resulting model computationally involved). \cite{miller2018robust} propose a simple alternative. Instead of conditioning on $\bm y$ when forming the posterior, one could condition on  $d(P_{\bm y}, P_{\bm y^*}) < C$ instead.  Since $C$ is typically unknown to the researcher one could specify a prior on it, i.e. $C \sim \pi$.  Doing so leads to the coarsened posterior:
\begin{equation*}
    p(\bm \vartheta | \hat{d}(P_{\bm y}, P_{\bm y^*}) < C) \propto \mathbb{P}(\hat{d}(P_{\bm y}, P_{\bm y^*}) < C|\bm \vartheta) ~ p(\bm \vartheta),
\end{equation*}
where the probability $\mathbb{P}(\hat{d}(P_{\bm y}, P_{\bm y^*}) < C|\bm \vartheta)$ can be interpreted as a likelihood function. Notice that this coarsened likelihood is generally not a probability distribution of $\bm y$ given $\bm \vartheta$.

Different alternatives for the distance function $d$ can be used. The choice of the discrepancy function can be based on the expected (or rather feared) type of misspecification. For example, as noted by \cite{miller2018robust}, robustness to outliers requires a discrepancy function that is little sensitive to movements of small amounts of probability mass to the outlying region (e.g., the first Wasserstein distance). Among the different distance functions $d$, one particularly attractive variant stands out, which considers differences in the entire likelihood function: the relative entropy. Moreover, using the relative entropy between $P_{\bm y}$ and $P_{\bm y^*}$ and an exponential prior on $C \sim Exp(\alpha)$ leads to a particularly simple and accurate approximation of the coarsened posterior. In this case, it can be approximated as follows \citep[see][for a proof]{miller2018robust}:
\begin{equation}
    p(\bm \vartheta | \hat{d}(P_{\bm y}, P_{\bm y^*}) < C) \approxprop p(\bm \vartheta)~\prod_{t=1}^T p(\bm y_t|\bm \vartheta, \bm y_{1}, \dots, \bm y_{t-1})^{\phi}, \label{eq: cpost}
\end{equation}
with $\approxprop$ denoting  approximately proportional to. \myeqref{eq: cpost} implies that the coarsened posterior is simply equal to the prior times a tempered likelihood.\footnote{This particular form of the coarsened posterior resembles the  intermediate approximating distribution of sequential Monte Carlo (SMC) methods. The main difference is that the learning rate in SMC grows from $0$ (i.e. the likelihood plays no role) to $1$ (i.e. one obtains the uncoarsened posterior). If one would fix the learning rate in SMC to $\zeta_T$, the result would be a sequential coarsened posterior.} This is a power posterior which has been shown to be robust to misspecification of a general form \citep[see, e.g.][]{holmes2017assigning, gruenwald_van-ommen_2017_ba, bhattacharya2019bayesian} and hence using this approximation directly links the coarsening approach to this literature.

The standard likelihood is raised by a learning rate $\phi$. To understand the role of this learning rate, it is useful to note that the resulting posterior corresponds to shrinking the sample size from $T$ to $\phi T$. Hence, when $\phi$ is small and coarsening is relevant, the sample size is substantially reduced and the posterior is much less concentrated. This is sensible, as substantial coarsening should be associated with a larger extent of misspecification, and hence uncertainty should indeed be larger. Yet, if $\phi$ is set to a small number when the model is instead approximately correct, posterior credible sets will be too large and model complexity likely under-estimated. Therefore, the choice of the parameter $\phi$ is very important for coarsening to be helpful and we will discuss a method to select it in a fully automatic manner later in the paper.

It is worth stressing that this form of the coarsened posterior does not require the computation of the relative entropy term  since it is absorbed in the constant of proportionality (and thus independent of $\bm \vartheta$). %In this regard,  \cite{miller2018robust} suggest that as a simple starting point one could set $\alpha = T$, implying that you need to observe at least $T$ additional observations to be robust to a presumed perturbation. Alternatively, $\alpha$ could be set as a decreasing function in $T$, effectively capturing the notion that longer series tend to feature more outliers, structural breaks and other non-linear features.

An interesting question is whether the choice of a specific prior, in particular a robust one, could make coarsening less relevant. As discussed by  \cite{miller2018robust}, this is not the case since a robust prior makes the results less sensitive to the choice of the prior but the importance of misspecification of the likelihood remains the same, and the likelihood dominates the prior when the sample size grows (even though the coarsened posterior does not concentrate as $T$ diverges just because of coarsening). A related issue is whether the choice of the prior matters more in the coarsened than in the standard case, since the "weight" of the likelihood decreases. Intuitively, this makes sense. Conditional on the prior, if we believe that the information contained in the likelihood is severely corrupted (either through wrongly specifying the likelihood or measurement errors, both of which fall into our definition of misspecification) a good choice would be to downweight this piece of information.

It is also worth discussing the relationship of coarsening with robust control theory, to clarify they they are quite different. In robust control theory \citep[see, e.g.,][]{Hansen06, Hansen08} the decision maker has one reference model but she evaluates a decision rule under a set of alternative models that are perturbed versions of the reference model. \cite{Hansen06} measure the difference between the reference and perturbed models using relative entropy, taking the maximum value of the difference as a parameter that measures the set of perturbations against which the decision maker seeks robustness (and restricts the extent of model misspeciﬁcation). They also provide conditions that permit to consider the perturbed models as the multiple priors that appear in the max–min expected utility theory of \cite{Gilboa89}. Instead, in our context, the prior is unique, we condition on the fact that the reference model can be at a certain (entropic) distance from the true model, and we  contaminate the likelihood to take that into account when forming the posterior distribution of the model parameters. 

%Entropic tilting (see, e.g., \cite{Robertson05} instead minimizes the Kullback-Leibler distance between the predictive density under restrictions and the unconditional one. Coarsening does not aim to reduce the distance between the model in use and the true one, rather it takes it as given and assumes that it is smaller than a certain value, and tries to robustify estimation and inference for the model in use.

\section{Coarsened Bayesian VAR}\label{sec:cBVAR}

In this section we develop the cBVAR model for the multivariate case. Inference is complicated by the fact that the degree of misspecification may vary across equations. This warrants the introduction of a distinct learning rate for each endogenous variable in the VAR. We start by discussing the coarsened likelihood of the VAR in \autoref{subsec:likelihood}. In \autoref{subsec:posteriors} we sketch our prior setup and provide details on the full conditional posterior distributions. Finally we move on to discussing how to set the learning rates and other hyperparameters in \autoref{subsec:hyperparms}.

\subsection{Coarsened Likelihood}\label{subsec:likelihood}

Let $\bm y_t = (y_{t,1}, \dots, y_{t,M})'$ be the vector of $M$ economic variables at time $t$. We consider a linear and homoskedastic VAR model of order $P$, with  reduced form given by:
\begin{equation} \label{eq:var_reduced}
    \bm y_t = \bm a+ \bm A_1 \bm y_{t-1} + \dots + \bm A_P \bm y_{t-P} + \bm u_t,\quad \bm u_t \sim \mathcal{N}(\bm 0, \bm \Sigma),
\end{equation}
with $\bm \alpha$ denoting the intercept vector, $\bm A_p ~(p=1, \dots, P)$ being $M \times M$ matrices of autoregressive coefficients, and $\bm u_t$ denotes a Gaussian vector white noise process with zero mean and $M \times M$-dimensional variance-covariance matrix $\bm \Sigma$.  The model in \autoref{eq:var_reduced} contains $k=(PM+1) M$ regression coefficients and $v = M (M+1)/2$ free elements in the error variance-covariance matrix. %If $M$ and $P$ are moderately large, the number of coefficients quickly exceeds the available number of observations $T$, making the VAR prone to overfitting.

Using the Cholesky decomposition of $\bm \Sigma =  \bm B_0^{-1} \tilde{\bm \Sigma} \bm B_0^{-1'}$, where $\bm B_0$ is the lower uni-triangular and $\tilde{\bm \Sigma} = \text{diag}(\sigma_1^2, \dots, \sigma_M^2)$ is a diagonal matrix,  and multiplying \autoref{eq:var_reduced} from the left with $\bm B_0$, we obtain the structural form of the model:
%We can use the Cholesky decomposition for  $\bm \Sigma = \bm B_0^{-1} \tilde{\bm \Sigma} \bm B_0^{-1'}$, where $\bm B_0^{-1}$ denotes a lower-triangular square matrix with ones on the main diagonal and $\tilde{\bm \Sigma} = \text{diag}(\sigma_1^2, \dots, \sigma_M^2)$ is a diagonal matrix containing the error variances for variables $i = 1,\dots,M$. Pre-multiplying equation \autoref{eq:var_reduced} by $\bm B_0$ we can rewrite it in its recursive form:
\begin{equation*}
    \bm B_0 \bm y_t = \bm \beta + \bm B_1 \bm y_t + \dots + \bm B_P \bm y_{t-P} + \bm \varepsilon_t,\quad \bm \varepsilon_t \sim \mathcal{N}(\bm 0, \tilde{\bm \Sigma}),
\end{equation*}
with $\bm \beta = \bm B_0 \bm \alpha$, $\bm B_p = \bm B_0 \bm A_p$ (for $p=1,\dots,P$) and $\bm \varepsilon_t = \bm B_0 \bm u_t$ being a vector of mutually independent error terms. Since $\tilde{\bm \Sigma}$ is diagonal, it is straightforward to derive the representation for equation $i$ only:
\begin{equation*}
    y_{t,i} = \hat{\bm \vartheta}_i' \tilde{\bm y}_{t,i} + \tilde{\bm \theta}_i' \tilde{\bm x}_{t,i} + \varepsilon_{t,i}, \quad \varepsilon_{t,i} \sim \mathcal{N}(0,\sigma_i^2),
\end{equation*}
where $y_{t,i}$ and $\varepsilon_{t,i}$ denote the $i$-th elements of $\bm y_t$ and $\bm \varepsilon_t$, respectively. For later reference, we have introduced two blocks of parameters.  First, the $(i-1)$-dimensional vector $\tilde{\bm y}_{t,i} = (-y_{t,1}, \dots, -y_{t,i-1})'$ contains contemporaneous observations of the variables ordered above variable $i$ in $\bm y_t$ and $\hat{\bm \vartheta}_i$ are the corresponding contemporaneous parameters. Note that  $\tilde{\bm y}_{t,i}$ and $\tilde{\bm \vartheta}_i$ only exist for  $i>1$. Second, $\tilde{\bm x}_{t,i} = (1, \bm y_{t-1}', \dots, \bm y_{t-P}')'$ is a $(MP+1)$-dimensional vector that contains lagged observations and the intercept term. The corresponding regression coefficients are given by $\tilde{\bm \theta_i} = (\beta_i, B_{1,i,1},\dots, B_{1,i,M}, \dots, B_{P,i,1}, \dots B_{P,i,M})'$. 

We can rewrite the regression model for equation $i$ more concisely as:
\begin{equation*}%\label{eq:var_recursive_i}
    y_{t,i} = \bm \theta_i' \bm x_{t,i} + \varepsilon_{t,i},\quad \varepsilon_{t,i} \sim \mathcal{N}(0, \sigma_i^2),
\end{equation*}
with $\bm x_{t,i} = (\tilde{\bm y}_{t,i}, \tilde{\bm x}_{t,i})$ and $\bm \theta_i = (\hat{\bm \vartheta}_i, \hat{\bm \theta}_i)$ being vectors of dimensions $(i+MP)$, respectively. Stacking observations over time yields
\begin{equation*}
    \bm Y_i = \bm X_i \bm \theta_i + \bm \varepsilon_i,\ \bm \varepsilon_i \sim \mathcal{N}(\bm 0, \sigma_i^2 \bm I_T),
\end{equation*}
where $\bm Y_i = (y_{1,i}, \dots, y_{T,i})'$ is a $T$-dimensional vector and $\bm X_i = (\bm x_{1,i}', \dots, \bm x_{T,i}')'$ is a matrix of dimension $T \times (i+MP)$.

Since $\tilde{\bm \Sigma}$ is diagonal, the likelihood of the full system is the product of $M$ conditionally independent Gaussian densities:
\begin{equation*}
    p(\bm Y_1, \dots, \bm Y_M|, \bm \theta_1, \dots, \bm \theta_M, \sigma_1^2, \dots, \sigma_M^2) = \left(\prod_{i=1}^M p(\bm Y_i | \bm \theta_i, \sigma_i^2) \right).
\end{equation*}
%It is important to note that rewriting the VAR in its recursive form does not affect likelihood, i.e.: $p(\bm Y_1, \dots, \bm Y_M | \bm A, \bm \Sigma) = p(\bm Y_1, \dots, \bm Y_M|, \bm \theta_1, \dots, \bm \theta_M, \sigma_1^2, \dots, \sigma_M^2)$. 

We use this form of the likelihood because it effectively allows us to treat each equation independently and hence introduce separate coarsening parameters per series. This is predicated on the fact that certain series in $\bm y_t$ might be more prone to misspecification than other series. Hence, using a single coarsening parameter on the joint likelihood implies a trade-off; we can either control for substantial degrees of misspecification or little/no misspecification. Therefore, we set up the  coarsened likelihood for each equation of the VAR and introduce separate coarsening parameters $\phi_i$:
\begin{equation*}
    \tilde{p}(\bm Y_i|\bm \theta_i, \sigma^2_i, \phi_i) =  \left( \prod_{t=1}^T p( y_{t,i} | \bm \theta_i, \sigma_i^2)^{\phi_i} \right).
\end{equation*}
Rewriting this equation yields the equation-specific coarsened likelihood:
\small
\begin{equation*}
    \tilde{p} (\bm Y_i | \bm \theta_i, \sigma_i^2, \phi_i, ) = (2 \pi \sigma_i^2)^{-\frac{\phi_i T}{2}} \text{exp}\left(-\frac{\phi_i}{2 \sigma_i^2} \sum_{t=1}^T(y_{t,i} - \bm \theta_i' \bm x_{t,i})^2 \right).
\end{equation*}
The joint coarsened likelihood is then given by:
\begin{equation}\label{eq:jointcoarsened_lik}
    \tilde{p}(\bm Y_1, \dots, \bm Y_M|\bm \theta_1, \dots, \bm \theta_M, \sigma_1^2, \dots, \sigma_M^2, \phi_1, \dots, \phi_M, ) = \prod_{i=1}^M  \tilde{p} (\bm Y_i | \bm \theta_i, \sigma_i^2, \phi_i, ).
\end{equation}

For large values of $M$ and $P$, the number of parameters can quickly exceed the number of available observations, and the OLS estimator ceases to exist. Hence, regularization is necessary, and Bayesian approaches that rely on specifying priors on $\bm \theta_1, \dots, \bm \theta_M$ and $\hat {\sigma}_1^2, \dots, \hat{\sigma}_M^2$ are commonly employed. In the next sub-section we will derive the coarsened posterior under the asymmetric conjugate prior.
    
\subsection{Coarsened Bayesian Analysis of the VAR}\label{subsec:posteriors}
The natural conjugate prior \citep[see, e.g.,][]{kadiyala1997numerical, koop2013forecasting, carriero2015bayesian} implies that the amount of shrinkage is proportional across equations (scaled by the corresponding elements of the covariance matrix $\bm \Sigma$). Since we start from the assumption that the amount of misspecification can differ across equations, this assumption is overly restrictive. As a solution, we use the asymmetric conjugate prior proposed in \cite{chan2022asymmetric}. This prior has the convenient property that it leads to analytical posterior results and can be combined with the coarsened likelihood in \autoref{eq:jointcoarsened_lik}. Moreover, it leads to a model that is order-invariant since the prior on the contemporaneous terms translates into an inverted Wishart prior on $\bm \Sigma$.

We assume that the priors for each equation $i = 1, \dots, M$ are mutually independent such that $p(\bm \theta_1, \dots, \bm \theta_M, \sigma_1^2, \dots, \sigma_M^2) = \prod_i^M p(\bm \theta_i, \sigma_i^2)$. The prior on the regression coefficients is normally distributed and conditions on $\sigma_i^2$:
\begin{equation*}
    (\bm \theta_i | \sigma_i^2) \sim \mathcal{N}(\underline{\bm \theta}_i, \sigma_i^2 \underline{\bm V}_i)
\end{equation*}
with $\underline{\bm \theta}_i$ denoting the prior mean and $\underline{\bm V_i}$ a prior variance-covariance matrix of dimension $K_i \times K_i$. 
We follow the \cite{sims1998bayesian} tradition and assume that, a priori, the elements in $\bm y_t$ follow $M$ independent random walks. This is achieved by centering  $\bm B_1$  on $\bm I_M$ (or to $\bm 0_M$ if the data is stationary) and $\bm B_p\ (p\neq1)$ on $\bm 0_{M \times M}$. Let $\underline{\bm \theta}_i = (\underline{\tilde{\bm \vartheta}}_i, \underline{\tilde{\bm \theta}}_i)$, where $\underline{\tilde{\bm \vartheta}}$ and $\underline{\tilde{\bm \theta}}_i$ denote the prior mean for the block of contemporaneous and dynamic coefficients, respectively. For equation $i=1,\dots,M$, we set the prior mean as $\underline{\tilde{\bm \theta}}_i = (\bm 0_{(i-1)\times1}', 1, \bm 0_{(M-i)\times1}')'$.  The prior on the contemporaneous coefficients is centered on zero: $\underline{\tilde{\bm \vartheta}} = \bm 0_{i-1\times1}$.

We assume that the prior variance-covariance matrix $\underline{\bm V}_i = \text{diag}(\underline{\bm v}_{\tilde{\vartheta},i}^2, \underline{\bm v}_{\tilde{\theta},i}^2)$ is diagonal. In this paper, we follow \cite{chan2022asymmetric} and  set $\bm v_{\tilde{\vartheta},i}^2 = (1/s_1^2, \dots, 1/s_{i-1}^2)$, where $s_i^2$ denotes the OLS estimate of the residual error variance of an $AR(P)$ model for  variable $i$. The elements of $\underline{\bm v}_{\tilde{\theta}}^2$ are set using the Minnesota prior, which shrinks coefficients associated to higher-order lags of $\bm y_t$  and makes a distinction between the own and other lags of a particular target variable. In particular, the $k^{th}$ element of $\underline{\bm v}_{\tilde{\theta}}^2$ reads:
\begin{equation*}
    \underline{v}_{\tilde{\theta},i,k}^2 = \begin{cases}
        \frac{\kappa_{1,i}}{p^2 s_i^2}, & \text{for the coefficient of the } p \text{-th lag of variable }i\\
        \frac{\kappa_{2,i}}{p^2 s_j}, & \text{for the coefficient of the } p \text{-th lag of variable } j \neq i\\
        \kappa_3, & \text{for the regression intercept}
    \end{cases}
\end{equation*}
where, for each equation $i$, $\kappa_{1,i}$ controls the amount of shrinkage induced on the coefficients  associated with lags of variable $i$ and $\kappa_{2,i}$ is the shrinkage hyperparameter for coefficients associated with lags of other variables. More information on how we set $\kappa_{1,i}$ and $\kappa_{2,i}$ is provided in  \autoref{subsec:hyperparms}. The hyperparameter $\kappa_3$ is set equal to $10^2$ to effectively introduce no prior information on the intercept term. 

On the error variances we use an inverse Gamma prior:
\begin{equation*}
    \sigma_i^2 \sim \mathcal{IG}(\underline{\nu}_i, \underline{S}_i),
\end{equation*}
where $\nu_i$ and $S_i$ denote the scale and shape parameter, respectively.
We set $\nu_i = 1+i/2$ and $\underline{s}_i = s_i^2/2$. \cite{chan2022asymmetric} shows that this choice, in combination with the prior on $\hat{\vartheta}_i$, leads to a prior on the reduced-form error variance-covariance matrix which follows an inverse Wishart distribution $\bm \Sigma \sim \mathcal{IW}(M+2, \underline{\bm S})$ with $M+2$ degrees of freedom and prior scaling matrix $\underline{\bm S} = \text{diag}(s_1^2, \dots, s_M^2)/(M + 2)$ . %Note, that a non-diagonal prior expectation $ \underline{\bm S}$ can be derived by setting $\underline{\tilde{\bm \vartheta}}_i$ appropriately.

At this point, it is worth stressing that our coarsened likelihood can be combined with any of the priors commonly used in the VAR literature.  For instance,  priors based on the use of structural models to inform parameter estimates \citep[see, e.g.,][]{ingram1994supplanting, del2004priors, de2018debt, loria2022economic}, priors utilizing information on the long-term behavior of the time series under scrutiny \citep{giannone2019priors} or priors that force the VAR towards factor models \citep{huber2023subspace} can be easily incorporated in our framework. 

The joint prior $\prod_{i=1}^M p(\bm \theta_i, \sigma_i^2) = \prod_{i=1}^M p(\bm \theta_i| \sigma_i^2) \times p(\sigma_i^2)$ can be combined with \autoref{eq:jointcoarsened_lik} to obtain the joint coarsened posterior distribution:
\begin{align*}
    p(\bm \theta_1, & \dots, \bm \theta_M, \sigma_1^2, \dots, \sigma_M^2 | \bm Y_1, \dots, \bm Y_M, \phi_1, \dots, \phi_M) \\ 
    =\ & \prod_{i=1}^M c_i (2\pi)^{-\frac{\phi_i T}{2}} (\sigma_i)^{-(\overline{\nu}_i + \frac{i + MP}{2} +1)} \text{exp}\left(-\frac{1}{\sigma_i^2}(\overline{S}_i + \frac{1}{2}(\bm \theta_i - \overline{\bm \theta}_i)'\overline{\bm V}_i (\bm \theta_i - \overline{\bm \theta}_i))\right),
\end{align*}
which is the product of $M$ Gaussian-Inverse-Gamma distributions:
\begin{equation}
    (\bm \theta_i, \sigma_i^2 | \bm Y) \sim \mathcal{NIG}(\overline{\bm \theta}_i, \overline{\bm V}_i, \overline{v}_i, \overline{S}_i).
\end{equation}
We provide an analytical expression for the normalizing constant $c_i$ of the posterior in \autoref{subsec:hyperparms}. The posterior hyperparameters are given by:
\begin{align*}
    \overline{\bm V}_i &= (\underline{\bm V}_i^{-1} + \phi_i \bm X_i' \bm X_i)^{-1}, \\
    \overline{\bm \theta}_i &= \overline{\bm V}_i (\underline{\bm V}_i^{-1} \underline{\bm \theta}_i + \phi_i \bm X_i'\bm Y_i) \\
    \overline{\nu}_i &= \underline{\nu}_i + \frac{\phi_i T}{2}, \\
    \overline{S}_i &= \underline{S}_i + (\phi_ i \bm Y_i' \bm Y_i + \underline{\bm \theta}_i' \underline{\bm V}_i \underline{\bm \theta}_i - \overline{\bm \theta}_i' \overline{\bm V}_i \overline{\bm \theta}_i)/2.
\end{align*}

As discussed in \cite{chan2022asymmetric}, it is straightforward to sample from this distribution by first sampling $\sigma_i^2 \sim \mathcal{IG}(\overline{\nu}_i, \overline{S}_i)$ marginally and then drawing $\bm \theta_i | \sigma_i^2 \sim \mathcal{N}(\overline{\bm \theta}_i, \sigma_i^2 \overline{\bm V}_i)$ conditionally on the current draw of $\sigma_i^2$. 
    
\subsection{Deciding on the degree of coarsening and setting the prior hyperparameters}\label{subsec:hyperparms}

For each equation in the VAR we need to decide on the learning rate as well as the hyperparameters controlling the prior tightness.   There are several ways to choose $\phi_i$. First, one can use prior information on the amount of misspecification. However, for possibly large-dimensional macroeconomic datasets, this turns out to be unfeasible. Second, we could use cross-validation. This essentially implies that one recomputes the model to obtain quantities of interest (such as forecasts or impulse responses) for different values of $\phi_i$ and picks the one that minimizes a loss function. Third, from a Bayesian perspective one could integrate out $\phi_i$. \cite{gruenwald_van-ommen_2017_ba} show, however, that this strategy does not work, since without substantial prior information on $\phi_i$ the model severely underestimates the true degree of misspecification and we would end up setting $\phi_i$ close to one.

In this paper, we use the SafeBayes algorithm of \cite{gruenwald_van-ommen_2017_ba} that is fully automatic and has excellent theoretical and empirical properties. Since our model is written in its recursive form, the SafeBayes algorithm proceeds on an equation-by-equation basis. 

Before starting the algorithm, we need to define a grid of length $R$ of possible learning rates $\bm \Phi_i$. In our empirical work, $\bm \Phi_i = (0.05, 0.1, \dots, 1)'$ for all $i$. Then, conditional on a particular value $\phi_i^{(j)} \in \bm \Phi_i$, we first decide on the optimal shrinkage parameters of the Minnesota prior. This is done by maximizing the coarsened marginal likelihood (ML) with respect to $\kappa_{1,i}, \kappa_{2,i}$ while conditioning on $\phi_i^{(j)}$:
\begin{equation*}
    \kappa^{(j)}_{1,i}, \kappa^{(j)}_{2,i}  = \argmax_{\kappa_{1,i}, \kappa_{2,i}} \tilde{p}(\bm Y_i| \phi_i = \phi_i^{(j)},  \kappa_{1,i}, \kappa_{2,i}),
\end{equation*}
with $\tilde{p}(\bm Y_i| \phi_i = \phi_i^{(j)},  \kappa_{1,i}, \kappa_{2,i})$ denoting the coarsened  ML specific to equation $i$. This is given by:
\begin{equation*}
    \tilde{p}_i(Y_i| \phi_i, \kappa_{1,i}, \kappa_{2,i}) = (2 \pi)^{\frac{\phi_i T}{2}}| \underline{\bm V}_i|^{-\frac{1}{2}} |\overline{\bm V}_i|^{\frac{1}{2}} \frac{\Gamma(\overline{\nu}_i)\underline{s}_i^{\underline{\nu}_i}}{\Gamma(\underline{\nu}_i)\overline{s}_i^{\overline{\nu}_i}}.
\end{equation*}
Here, we let $\Gamma(\bullet)$ denote the Gamma function and $|\bm A|$ denotes the determinant of a matrix $\bm A$.  It should be noted that we optimize the coarsened ML for each value of $\phi_i$, implying that the hyperparameters are set as a function of the degree of misspecification. This is related to \cite{gonzalez2025misspecification} who use misspecification robust loss functions instead of the marginal likelihood to select the hyperparameters of a Bayesian VAR. Contrary to our approach, this is done on a full-system basis.

\begin{algorithm}[t!]
    \caption{SafeBayes \citep{gruenwald_van-ommen_2017_ba}}
    \label{alg:safebayes}
    \begin{algorithmic}
        \Require Data $\bm Y$, learning rate proposals $\bm \Phi$
        \Ensure learning rates $\phi_1, \dots, \phi_M$, prior hyperparameters $\kappa_{1,1},\ \kappa_{2,1}, \dots, \kappa_{1,M},\ \kappa_{2,M}$
        \For{$i$ in $1,\dots,M$}
            \For{all $\phi \in \Phi$}
                \State Select $\kappa_{1,i,\phi}^*, \kappa_{2,i,\phi}^* = \argmax_{\kappa_{1,i}, \kappa_{2,i}} \tilde{p}(Y_i| \kappa_{1,i}, \kappa_{2,i}, \phi_i)$
                %\State Set $s_{i, \phi} = 0$
                \For{$t$ in $1,\dots, T-1$}
                    \State Obtain coarsened posterior expectation conditional on data up to time $t$: 
                    \State $\mathbb{E}_t(\bm \theta_i, \sigma_i^2 | \bm Y_{i,1:t}, \phi, \kappa_{1,i}^*, \kappa_{2,i}^*)$
                    \State Calculate loss function by predicting actual next outcome:
                    \State $l_{i,\phi, t} = -\text{log}\ f(y_{i,t+1}|\bm x_{i,t+1},\mathbb{E}_t(\bm \theta_i, \sigma_i^2 | \bm Y_{i,1:t}, \phi, \kappa_{1,i}^*, \kappa_{2,i}^*))$
                \EndFor
            \EndFor \\
            Choose $\phi_i := \argmin_{\phi \in \Phi} \sum_{t=1}^{T-1} l_{i,\phi,t}$ and corresponding $\kappa_{1,i,\phi_i}^*, \kappa_{2,i,\phi_i}^*$
        \EndFor
    \end{algorithmic}
\end{algorithm}

Conditional on $\phi^{(j)}_i$ and the optimal shrinkage parameters $\kappa_{1,i}^{(j)}, \kappa_{2,i}^{(j)}$,  we then loop through $t = 1,\dots,T-1$ and derive the coarsened posterior expectations for $\mathbb{E}_t[\theta_i]$ and $\mathbb{E}_t[\sigma_i^2]$ using only data up to time $t$. We use these to evaluate the loss function $l_{i,\phi,t} = -\text{log}\ f(y_{i,t+1} | x_{i,t+1}, \mathbb{E}_{t}[\bm \theta_i], \mathbb{E}[\sigma_i^2])$ at time $t$. In our empirical work, the loss function equals the log predictive likelihood (LPL), a commonly used forecast evaluation metric. This implies setting $f(\cdot) = \mathcal{N}(y_{i, t+1}|\mathbb{E}_{t}[\bm \theta_i]'\bm x_{i,t+1}, \mathbb{E}_t[\sigma_i^2])$.
After performing this operation for all $t$, we choose the learning rate out of $\Phi$ that minimizes the sum of loss functions over the entire sample:
\begin{equation*}
    \phi^*_i = \argmin_{\phi \in \Phi} \sum_{t=1}^{T-1} l_{i,\phi,t},
\end{equation*}
and choose the associated prior hyperparameters $\kappa_{i,1} = \kappa_{i,1}^*$ and $\kappa_{i,2} = \kappa_{i,2}^*$.  This choice of the loss function is also motivated in \cite{gruenwald_van-ommen_2017_ba}.

This is repeated for each equation until we end up with  $M$ optimal learning rates and associated prior hyperparameters. Notice that the conjugate nature of our model makes this step computationally tractable. If posterior quantities such as the predictive mean, the predictive variance or the marginal likelihood would not be available in closed form, one would have to use simulation-based methods for each value of $\phi_i^{(j)}$ in our grid. \autoref{alg:safebayes} gives an overview of the algorithm.

\section{Evidence based on synthetic data }\label{sec:sim_study}
In this section, we carry out a structured evaluation of the empirical performance of our coarsened BVAR model using a detailed Monte Carlo exercise. First, we employ a straightforward illustrative example grounded in a DGP with non-Gaussian errors. Subsequently, we examine the predictive capacity of our method in various DGPs.

\subsection{An illustrative example}
To investigate whether the estimated coarsening parameter effectively picks up model misspecification, we simulate $T=500$ observations from a small linear VAR where the errors are Student's t distributed with varying degrees of freedom $\nu$.\footnote{Information on the precise DGP used can be found in \autoref{app:dgps} of the Online Appendix.} In particular, we decrease the degrees of freedom from $20$ (in which case the shape of the Student-t distribution generates a moderate amount of outliers such that the standard BVAR is only lightly misspecified) to $3$ (the lowest value for which the moments of the Student-t distribution exist; implying a maximum amount of misspecification). This leads to 18 different DGPs, from each of which we take $100$ draws to estimate the cBVAR. 

\begin{figure}
    \centering
    \includegraphics[width=0.8\linewidth]{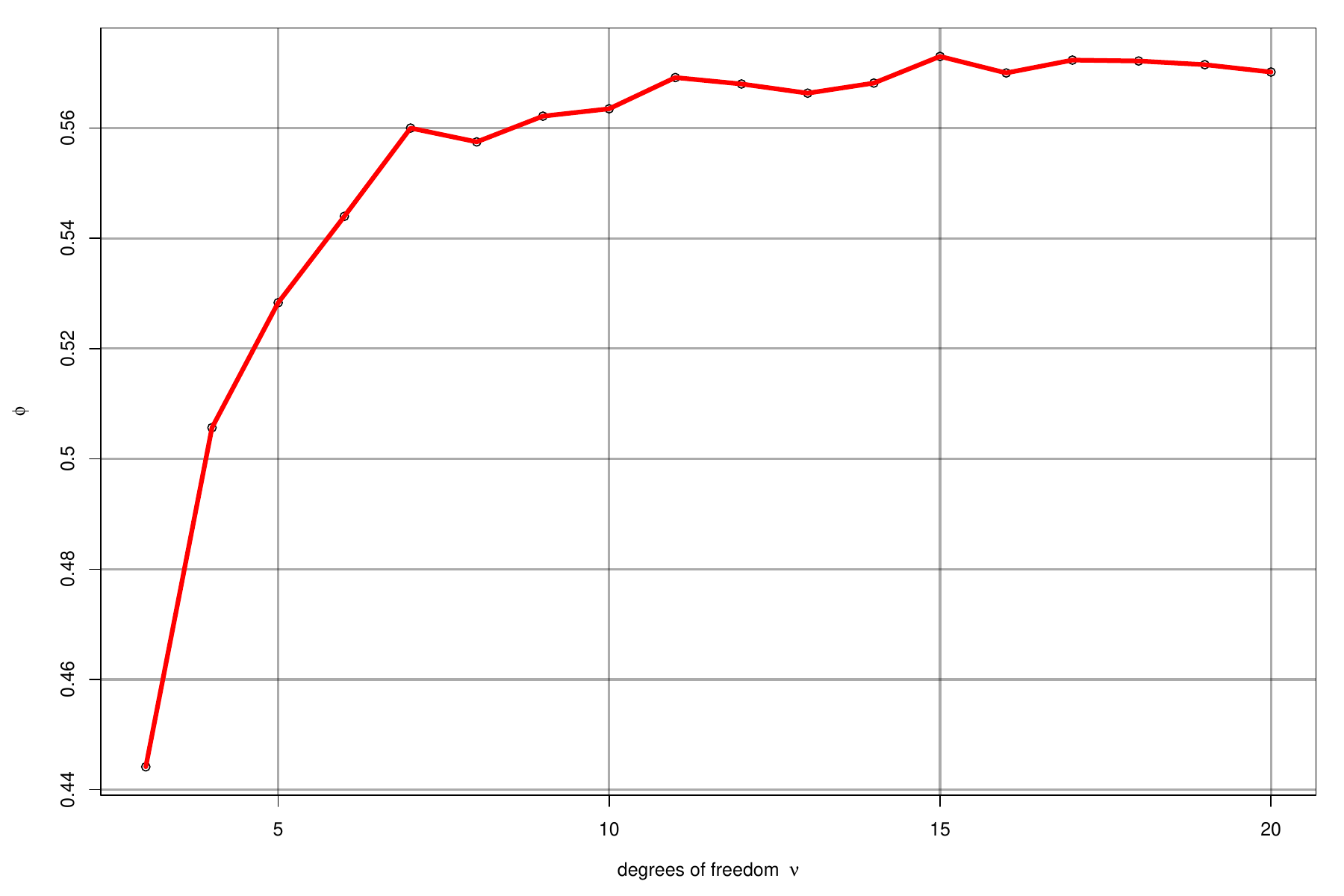}
    \caption{Estimated Coarsening parameter vs. degrees of freedom of the Student-t distribution generating the model errors.\\
    \textbf{Notes:}  Estimates are the mean over 100 draws from each DGP and over all three VAR equations.\\
    \textbf{Legend:} \textcolor{plotred}{$-$} represents a smoothed spline.}
    \label{fig:dof_vs_phi}
\end{figure}

\autoref{fig:dof_vs_phi} shows the mean estimate (averaged over all realizations from each DGP and all VAR equations) for each of these DGPs. Starting with the extreme case of three degrees of freedom, our model selects an average learning rate of around $0.44$, controlling for the fact that the shocks are heavy-tailed and the misspecified model we estimate has much lighter tails. This learning rate results in a significantly wider predictive distribution, suggesting that the cBVAR model is more capable of accommodating outliers compared to the conventional BVAR model.

Considering larger degrees of freedom reveals that the learning rate increases quickly (to around 0.5) before reaching a level of around 0.57 for larger degrees of freedom. This is surprising given that, in this case, the standard BVAR is actually well-specified. However, notice that the number of observations is large (given the size of the model), and outliers are still more likely under the t distribution with moderate degrees of freedom. In this case, the SafeBayes algorithm would set the learning rate below one. This is driven by the loss function that depends on the log predictive score. 

\subsection{Simulation results based on a large universe of data generating processes}
\subsubsection{Design of the Monte Carlo exercise}
We evaluate the performance of our coarsened BVAR against that of a standard Bayesian VAR without likelihood tempering across a range of different data generating processes (DGPs).  We vary these DGPs across two dimensions. Firstly, we allow for different conditional mean specifications. We consider a standard linear VAR (labeled "Linear"), a VAR with structural breaks in the parameters ("Breaks"), a VAR augmented with exogenous variables ("Exo") and a smooth-transition VAR ("Transition"). Secondly, for the error term we consider standard, Gaussian errors ("Gaussian"), errors arising from a Student-t distribution ("Student"), heteroskedastic errors whose variance arises from a stochastic volatility process ("SV") and errors featuring a moving average component ("MA"). The latter specification allows us to test whether coarsening helps for dynamically misspecified models along the lines of \cite{schorfheide2005var}  and \cite{gonzalez2025misspecification}. Details on the specifications of these DGPs can be found in \autoref{app:dgps} of the Online Appendix.

Moreover, to investigate the relationship between model performance and size, we consider three different model sizes. The first is a small VAR with three variables ("small"), the second is a medium-sized VAR containing nine variables ("medium") and the third is a large 27-variable VAR ("large"). 

These combinations of conditional mean and variance specifications reflect a large share of models commonly applied by empirical macroeconomists. Practitioners are also often interested in estimating larger models, motivating our wish to understand how coarsening performs for models of different sizes. Each of the DGPs is then calibrated by estimating the respective model parameters on US macro data taken from the FRED-MD data base \citep{mccracken2016fred}. We estimate all models with $P=2$ lags.

From each DGP, we draw $100$ different realizations of $\bm y_t$ of size $T=200$ to estimate the coarsened and standard BVAR. For each of these runs, we then use the DGP to simulate $1,000$ out-of-sample datasets with $H=12$ observations each.  This yields a total of $100 \cdot 1,000 = 100,000$ evaluation points at each forecast horizon. For each evaluation point in the holdout, we compute the LPL of a particular cBVAR vis-\'{a}-vis the  BVAR (in differences such that numbers above zero mean better performance of the cBVAR).

\subsubsection{Predictive results}

\autoref{tab:lpls_sim} displays the mean differences in LPLs averaged across all test sets and all draws from each DGP.\footnote{Results for point forecasts are provided in the Online Appendix.} \autoref{tab:lpls_sim} is divided into three blocks containing results for the small, medium and large models, respectively. Each of these, in turn, consists of three blocks for the different forecast horizons. We also report the results of a \cite{diebold1995comparing} test with the null hypothesis of equal forecast performance.  

In general, the coarsened VAR exhibits a superior performance relative to the traditional Bayesian VAR. We observe a clear and steady pattern: the benefits of using the coarsened likelihood as a substitute for the standard version grow in parallel with the forecast horizon. As VAR predictive distributions are generated iteratively, errors due to model misspecification tend to accumulate and intensify the more extended the forecast period is \citep[see, e.g.,][]{marcellino2006comparison}.

Another salient observation is that the forecast gains increase with model size. As outlined above, the misspecified standard VAR fits a model based on data that did not arise from the postulated DGP. Since larger models tend to fit the data more strongly, the negative implications of misspecification convolute and have a deleterious impact on  forecasting performance that increases with model size. 

\begin{table}[!htbp]
\centering
\begin{threeparttable}
\caption{\label{tab:lpls_sim}Differences in joint LPL scores for predictive distributions.}
\centering
\fontsize{9}{11}\selectfont
\begin{tabular}[t]{l>{}l>{\raggedleft\arraybackslash}p{1.5cm}>{\raggedright\arraybackslash}p{.5cm}>{\raggedleft\arraybackslash}p{1.5cm}>{\raggedright\arraybackslash}p{.5cm}>{\raggedleft\arraybackslash}p{1.5cm}>{\raggedright\arraybackslash}p{.5cm}>{\raggedleft\arraybackslash}p{1.5cm}>{\raggedright\arraybackslash}p{.5cm}}
\toprule\toprule
\multicolumn{1}{c}{\textbf{}} & \multicolumn{1}{c}{\textbf{}} & \multicolumn{2}{c}{\textbf{Gaussian}} & \multicolumn{2}{c}{\textbf{Student}} & \multicolumn{2}{c}{\textbf{SV}} & \multicolumn{2}{c}{\textbf{MA}} \\
\cmidrule(l{3pt}r{3pt}){3-4} \cmidrule(l{3pt}r{3pt}){5-6} \cmidrule(l{3pt}r{3pt}){7-8} \cmidrule(l{3pt}r{3pt}){9-10}
 &  & $\Delta$ LPLS &  & $\Delta$ LPLS &  & $\Delta$ LPLS &  & $\Delta$ LPLS & \\
\midrule
\addlinespace[.1em]
\multicolumn{10}{c}{\textit{One-month-ahead predictions}}\\
\multirow{12}{*}{\rotatebox[origin=c]{90}{\normalsize \textbf{Small-sized}}} & \textbf{Linear} & -0.01 &  & -0.02 &  & -0.02 &  & -0.01 & \\
 & \textbf{Break} & 0.03 & *** & 0.02 & *** & 0.03 & *** & 0.04 & ***\\
 & \textbf{Exo} & 0.03 & *** & -0.04 &  & 0.01 & *** & 0.01 & ***\\
 & \textbf{Transition} & -0.03 &  & -0.05 &  & -0.03 &  & -0.02 & \\
\cmidrule{2-10}
\addlinespace[.1em]
\multicolumn{10}{c}{\textit{One-quarter-ahead predictions}}\\
 & \textbf{Linear} & 0.02 & *** & 0.11 & *** & 0.04 & *** & 0.43 & ***\\
 & \textbf{Break} & 0.38 & *** & 0.54 & *** & 0.39 & *** & 0.92 & ***\\
 & \textbf{Exo} & 0.91 & *** & 0.15 &  & 0.89 & *** & 0.31 & ***\\
 & \textbf{Transition} & 0.10 & *** & 0.01 &  & 0.12 & *** & 0.46 & ***\\
\cmidrule{2-10}
\addlinespace[.1em]
\multicolumn{10}{c}{\textit{One-year-ahead predictions}}\\
 & \textbf{Linear} & -0.46 &  & -0.03 &  & -0.58 &  & 6.50 & ***\\
 & \textbf{Break} & 2.67 & *** & 2.66 & *** & 2.16 & *** & 9.61 & ***\\
 & \textbf{Exo} & 29.94 & *** & 10.08 & *** & 42.86 & *** & 7.84 & ***\\
 & \textbf{Transition} & 1.32 & *** & -1.28 &  & 1.41 & *** & 6.44 & ***\\
\midrule
\addlinespace[.1em]
\multicolumn{10}{c}{\textit{One-month-ahead predictions}}\\
\multirow{12}{*}{\rotatebox[origin=c]{90}{\normalsize \textbf{Medium-sized}}} & \textbf{Linear} & 0.05 & *** & 0.06 & *** & 0.06 & *** & -0.13 & \\
 & \textbf{Break} & 0.63 & *** & 0.89 & *** & 0.74 & *** & 0.58 & ***\\
 & \textbf{Exo} & 0.79 & *** & 0.65 & *** & 1.01 & *** & 1.24 & ***\\
 & \textbf{Transition} & 0.30 & *** & 0.18 & *** & 0.25 & *** & 0.40 & ***\\
\cmidrule{2-10}
\addlinespace[.1em]
\multicolumn{10}{c}{\textit{One-quarter-ahead predictions}}\\
 & \textbf{Linear} & 1.47 & *** & 1.80 & *** & 1.54 & *** & 4.65 & ***\\
 & \textbf{Break} & 5.13 & *** & 7.02 & *** & 6.25 & *** & 5.35 & ***\\
 & \textbf{Exo} & 24.95 & *** & 17.04 & *** & 23.81 & *** & 30.37 & ***\\
 & \textbf{Transition} & 2.30 & *** & 1.42 & *** & 2.00 & *** & 3.10 & ***\\
\cmidrule{2-10}
\addlinespace[.1em]
\multicolumn{10}{c}{\textit{One-year-ahead predictions}}\\
 & \textbf{Linear} & 6.69 & *** & 5.74 & *** & 6.15 & *** & 28.52 & ***\\
 & \textbf{Break} & 20.46 & *** & 19.39 & *** & 22.95 & *** & 22.22 & ***\\
 & \textbf{Exo} & 955.96 & *** & 425.01 & *** & 772.48 & *** & 1030.49 & ***\\
 & \textbf{Transition} & 14.12 & *** & 0.72 &  & 12.78 & *** & 20.77 & ***\\
\midrule
\addlinespace[.1em]
\multicolumn{10}{c}{\textit{One-month-ahead predictions}}\\
\multirow{12}{*}{\rotatebox[origin=c]{90}{\normalsize \textbf{Large-sized}}} & \textbf{Linear} & 3.91 & *** & 4.94 & *** & 3.94 & *** & 5.76 & ***\\
 & \textbf{Break} & 5.96 & *** & 7.23 & *** & 6.56 & *** & 8.98 & ***\\
 & \textbf{Exo} & 21.30 & *** & 12.98 & *** & 17.42 & *** & 31.54 & ***\\
 & \textbf{Transition} & 0.53 & *** & 0.96 & *** & 0.77 & *** & 0.76 & ***\\
\cmidrule{2-10}
\addlinespace[.1em]
\multicolumn{10}{c}{\textit{One-quarter-ahead predictions}}\\
 & \textbf{Linear} & 22.55 & *** & 26.31 & *** & 23.83 & *** & 42.96 & ***\\
 & \textbf{Break} & 16.98 & *** & 20.25 & *** & 18.59 & *** & 23.89 & ***\\
 & \textbf{Exo} & 103.53 & *** & 66.36 & *** & 92.34 & *** & 157.14 & ***\\
 & \textbf{Transition} & 4.49 & *** & 5.54 & *** & 5.10 & *** & 8.69 & ***\\
\cmidrule{2-10}
\addlinespace[.1em]
\multicolumn{10}{c}{\textit{One-year-ahead predictions}}\\
 & \textbf{Linear} & 69.99 & *** & 85.33 & *** & 74.48 & *** & 121.29 & ***\\
 & \textbf{Break} & 52.76 & *** & 58.37 & *** & 54.51 & *** & 71.30 & ***\\
 & \textbf{Exo} & 495.35 & *** & 284.06 & *** & 471.56 & *** & 713.78 & ***\\
 & \textbf{Transition} & 15.80 & *** & 17.71 & *** & 17.53 & *** & 26.73 & ***\\
\bottomrule\bottomrule
\end{tabular}
\begin{tablenotes}
\small
\item [] \footnotesize \textbf{Notes:} Differences are averaged
    over all draws and test sets for each DGP. Stars indicate p-values from a
    one-sided t-test with alternative hypothesis that cBVAR produces larger LPL scores.
\item [] \footnotesize \textbf{Legend}: $.\sim p<0.16$; $^{*}\sim p<0.1$; $^{**}
    \sim p<0.05$; $^{***}\sim p<0.01$
\end{tablenotes}
\end{threeparttable}
\end{table}

When we focus on specific selections for model size, conditional mean, and variance, some noteworthy patterns emerge. Firstly, in the case of correct model specification (Linear + Gaussian), it appears that coarsening only adversely affects predictive accuracy when the model size is small. However, these effects are minimal and the predictive distributions remain similar to those of the standard BVAR. Conversely, if the model continues to be correctly specified but increases in size, coarsening can enhance predictive performance. This unexpected outcome aligns with the simulation findings of \cite{gruenwald_van-ommen_2017_ba}, who, while examining Bayesian ridge regressions, discover that SafeBayes significantly reduces risk even in the correctly specified case. 

For combinations of the conditional mean and variance that imply a misspecified model, we find substantial gains, but only for forecast horizons greater than one-step-ahead. The largest gains can be found in the case of omitted variables ("Exo"). In this case, the cBVAR produces much more precise density forecasts than the benchmark specification for forecast horizons greater than one-step-ahead. Interestingly, these gains do not always increase if we allow for misspecification in the error terms: the largest gains can be found if we ignore a MA structure in the shocks while the second-largest gains are observed if the shocks are Gaussian. This pattern is consistent across model sizes.

For the other DGPs which imply non-linearities in the conditional mean, we also observe gains from coarsening. However, these are less pronounced than in the case of (neglecting) exogenous factors. Interestingly, we find a different pattern if we allow for distributional misspecification. If we consider structural breaks or smooth transition DGPs, we find that solid gains can already be obtained for the one-step-ahead forecast if the errors feature stochastic volatility or a MA structure. For larger horizons (and model sizes), a pattern similar to the DGP featuring exogenous covariates shows up. In most cases, we find the largest improvements in forecast accuracy if the DGP has MA shocks or homoskedastic Gaussian disturbances.

In conclusion, we find strong evidence in favor of our proposed approach. In the vast majority of cases the cBVAR significantly outperforms its non-coarsened competitor. The only case where the results are somewhat mixed are one-step-ahead  LPLs arising from the small model. For larger model sizes, improvements increase further and tend to be present for short-term forecasts as well.

\section{Macroeconomic forecasting with coarsened VARs}\label{sec:forecasting}

\subsection{Data, model specification and design of the forecasting exercise}
We employ our cBVAR to forecast a range of US macroeconomic variables. Again, we focus on three model sizes. First, we consider a small VAR in three variables. These three variables are the unemployment rate (UNRATE), CPI inflation (CPIAUCSL) and the short-term interest rate (FEDFUNDS). Second, the three series of the small model are complemented with an additional six series. These are average weekly working hours in manufacturing (AWHMAN), the real M2 money stock (M2REAL), the S\&P500 stock price index (S\&P 500), the industrial production index (INDPRO), the spread between the federal funds rate and 10-year treasury rate (T10YFFM) as well as the consumer price index for commodities (CUSR0000SAC). Finally, we consider a large model that builds on the medium-scale model and adds additional 18 series such that the large model comprices 27 variables in total. The data is sourced from the FRED-MD database \citep{mccracken2016fred} and transformed to stationarity using the transformation codes suggested in \cite{mccracken2016fred}. Precise information on the series and transformations is provided in \autoref{tab:data_app} of the Online Appendix. All models we consider include $P=2$ lags.

Model comparison is done using LPLs and mean squared forecast errors (MSFEs). This is done in two ways. We consider two types of LPLs. First, we focus on univariate LPLs for three target series (the unemployment rate, UNRATE; CPI inflation, CPIAUCSL; and short-term interest rates, FEDFUNDS). This gives us information on how well coarsening helps when predicting each of the series individually. Second, we consider the joint LPL over all series in $\bm y_t$. This gives an impression on the overall predictive performance of a model. All loss functions are computed for the one-month-ahead, one-quarter-ahead and one-year-ahead forecast horizon.

The forecast design is as follows. Our sample starts in 1967:M07 and ends in 2023:12. We use 1967:M07 to 1999:M12 as our initial training sample. After estimating the models, we produce the predictive distributions for 2000:M01 up to 2000:M12. We then add one additional observation to the initial estimation window and compute the corresponding predictive distributions (2000:M02 to 2001:M01). This procedure is repeated until we arrive at the end of the sample.

\subsection{Forecasting results}

\subsubsection{Point forecasting accuracy}
We start our discussion with the point forecasting performance for the three focus variables. As a point forecast, we use the posterior median of the respective predictive distributions. The MSFEs for the cBVAR, in ratios relative to the BVAR, are shown in \autoref{tab:fc_mse}. Numbers greater than one imply that the cBVAR produces less precise point forecasts, whereas numbers smaller than one indicate the opposite. Raw MSFEs for the BVARs are shown in the columns labeled 'BVAR'.

\begin{table}[!tbh]

\centering \begin{threeparttable}
\caption{\label{tab:fc_mse}Mean squared error over the entire hold-out period.}
\centering
\begin{tabular}[t]{l>{\raggedleft\arraybackslash}p{10mm}>{\raggedright\arraybackslash}p{1mm}>{\raggedleft\arraybackslash}p{10mm}>{\raggedleft\arraybackslash}p{10mm}l>{\raggedleft\arraybackslash}p{10mm}>{\raggedleft\arraybackslash}p{10mm}l>{\raggedleft\arraybackslash}p{10mm}}
\toprule\toprule
\multicolumn{1}{c}{\textbf{Variables}} & \multicolumn{3}{c}{\textbf{Small}} & \multicolumn{3}{c}{\textbf{Medium}} & \multicolumn{3}{c}{\textbf{Large}} \\
\cmidrule(l{3pt}r{3pt}){1-1} \cmidrule(l{3pt}r{3pt}){2-4} \cmidrule(l{3pt}r{3pt}){5-7} \cmidrule(l{3pt}r{3pt}){8-10}
 & cBVAR &  & BVAR & cBVAR &  & BVAR & cBVAR &  & BVAR\\
\midrule
\addlinespace[0.3em]
\multicolumn{10}{c}{\textit{One-month-ahead predictions}}\\
UNRATE & 0.86 &  & 0.12 & 0.88 & . & 0.12 & 0.93 &  & 0.1\\
CPIAUCSL & 1 &  & 0.02 & 1.02 &  & 0.02 & 1.05 &  & 0.02\\
FEDFUNDS & 0.33 & . & 0.01 & 0.71 & ** & 0 & 0.65 & *** & 0.01\\
\midrule
\addlinespace[0.3em]
\multicolumn{10}{c}{\textit{One-quarter-ahead predictions}}\\
UNRATE & 0.8 & . & 0.43 & 0.79 &  & 0.42 & 0.85 & . & 0.34\\
CPIAUCSL & 0.99 &  & 0.13 & 0.98 &  & 0.13 & 0.97 &  & 0.13\\
FEDFUNDS & 0.41 &  & 0.04 & 0.84 & . & 0.04 & 0.69 & *** & 0.05\\
\midrule
\addlinespace[0.3em]
\multicolumn{10}{c}{\textit{One-year-ahead predictions}}\\
UNRATE & 0.93 & * & 1.07 & 0.91 &  & 0.98 & 0.85 & . & 1.06\\
CPIAUCSL & 1.05 &  & 0.72 & 0.95 &  & 0.75 & 1.2 &  & 0.63\\
FEDFUNDS & 0.91 &  & 0.17 & 0.89 & * & 0.23 & 0.78 &  & 0.38\\
\bottomrule\bottomrule
\end{tabular}
\begin{tablenotes}
\small
\item [] \footnotesize \textbf{Notes: } Results for the cBVAR are the
    ratio relative to the BVAR. Stars indicate p-values from a one-sided DM-test with
    alternative hypothesis that cBVAR produces smaller forecast errors.
\item [] \footnotesize \textbf{Legend:} $.\sim p<0.16$; $^{*}\sim p<0.1$; $^{**}\sim p<0.05$; $^{***}\sim p<0.01$
\end{tablenotes}
\end{threeparttable}
\end{table}

When assessing point forecasts, we find that coarsening leads to more accurate point forecasts for all variables except inflation. For the unemployment rate, we find improvements of around 14 percentage points (PPs) at the one-month-ahead horizon. For one-quarter-ahead, these improvements reach about 20 PPs before deteriorating appreciably when one-year-ahead predictions and the small dataset are being considered. These gains remain similar when we consider the medium-sized dataset (with very small improvements for one-quarter- and one-year-ahead forecasts). When we consider the large model, we find somewhat weaker gains from coarsening for forecasts up to one-quarter-ahead. However, for longer-run predictions (one year in advance), we observe improvements of around 15 PPs. These somewhat smaller improvements for short-run unemployment rate forecasts are most likely driven by the fact that larger models mitigate misspecification issues by trading unobserved heterogeneity with observed heterogeneity (through the inclusion of many more series that soak up patterns a more flexible econometric model would otherwise pick up).

Turning to inflation forecasts gives rise to a mixed picture. For one-quarter-ahead predictions (and irrespective of the model size), we find MSFE ratios close to (but mostly above) one. Increasing the forecast horizon to one-quarter-ahead slightly changes this picture, suggesting small improvements in forecast performance for the medium and large models. But these improvements are minor and reach around three PPs (for the large model). When we consider one-year-ahead forecasts we find that coarsening hurts point forecast accuracy for small and large models, with pronounced losses of around 20 PPs for the biggest model we consider. Only for the medium model, we observe gains of around 5 PPs.

The time series that profits most from coarsening is the Federal Funds rate. In this case, improvements for the critical one-month-ahead horizon reach staggering 67 PPs for the small model and still sizable improvements of 29 to 35 PPs for the medium and large models, respectively. These strong gains are driven by the zero lower bound (ZLB). Econometrically, the ZLB is a non-linear feature in the FEDFUNDS time series that should be controlled for through, e.g., regime-switching models \citep{liu2019changing} or models that allow for estimating the shadow interest rate \citep{carriero2025shadowrateforecasting}. The BVAR ignores this, leading to misspecification. This results in predictions that display substantial amounts of high-frequency variation even if the short-term interest rate is stuck close to zero. By contrast, the coarsened model produces a predictive median that displays much less high-frequency variation (see \autoref{fig:pred_dist_FEDFUNDS_small_1} to \autoref{fig:pred_dist_FEDFUNDS_small_12} of the Online Appendix), and this leads to substantial improvements in MSFEs. Once we increase the forecast horizon, these gains become slightly smaller, but even for the one-year-ahead horizon, remain sizable.

\subsubsection{Density forecasting accuracy}
We now turn to discussing whether coarsening also helps for higher-order moments by considering LPL differences to the uncoarsened VAR. \autoref{tab:fc_scores} shows average LPL differences between the cBVAR and the BVAR as well as absolute LPLs for the BVAR benchmark. Numbers exceeding zero indicate outperformance of the cBVAR while negative numbers indicate the opposite. Unlike \autoref{tab:fc_mse}, we also have rows that show the performance of the joint density forecast.

 In terms of density forecasts, we find an overall picture that is closely related to the one observed for point forecasting performance. Starting with the joint density forecast performance, we find improvements relative to the BVAR for all three model sizes and forecast horizons. Notice that the general pattern that performance improvements increase with the predictive horizon is visible for the medium- and large-scale models. For the small model, we find strong improvements for one-quarter-ahead LPLs and less pronounced gains for one-month- and one-year-ahead LPLs.
 
\begin{table}[!tbh]

\centering \begin{threeparttable}
\caption{\label{tab:fc_scores}Average LPL scores over the entire hold-out period.}
\centering
\begin{tabular}[t]{l>{\raggedleft\arraybackslash}p{14mm}>{\raggedright\arraybackslash}p{1mm}>{\raggedleft\arraybackslash}p{14mm}>{\raggedleft\arraybackslash}p{14mm}l>{\raggedleft\arraybackslash}p{14mm}>{\raggedleft\arraybackslash}p{14mm}l>{\raggedleft\arraybackslash}p{14mm}}
\toprule\toprule
\multicolumn{1}{c}{\textbf{Variables}} & \multicolumn{3}{c}{\textbf{Small}} & \multicolumn{3}{c}{\textbf{Medium}} & \multicolumn{3}{c}{\textbf{Large}} \\
\cmidrule(l{3pt}r{3pt}){1-1} \cmidrule(l{3pt}r{3pt}){2-4} \cmidrule(l{3pt}r{3pt}){5-7} \cmidrule(l{3pt}r{3pt}){8-10}
 & cBVAR &  & BVAR & cBVAR &  & BVAR & cBVAR &  & BVAR\\
\midrule
\addlinespace[0.3em]
\multicolumn{10}{c}{\textit{One-month-ahead predictions}}\\
Joint & 0.28 & * & -2.99 & 0.26 & * & -3.69 & 0.57 & ** & -3.88\\
UNRATE & 0.01 & *** & 0.42 & -0.01 & *** & 0.44 & -0.01 & *** & 0.44\\
CPIAUCSL & 0.11 & . & 0.90 & 0.00 & *** & 1.01 & 0.02 & *** & 1.01\\
FEDFUNDS & 0.39 & * & -1.77 & 0.42 & *** & -1.93 & 2.64 & *** & -16.03\\
\midrule
\addlinespace[0.3em]
\multicolumn{10}{c}{\textit{One-quarter-ahead predictions}}\\
Joint & 1.26 & . & -13.57 & 1.47 & . & -15.71 & 3.75 & * & -18.95\\
UNRATE & 0.13 & *** & -3.32 & 0.27 & *** & -3.42 & 0.80 & *** & -4.11\\
CPIAUCSL & 0.72 & . & -0.07 & 0.26 & *** & -0.11 & 0.66 & *** & -0.65\\
FEDFUNDS & 1.86 & * & -16.73 & 2.89 & ** & -37.45 & 16.81 & *** & -141.52\\
\midrule
\addlinespace[0.3em]
\multicolumn{10}{c}{\textit{One-year-ahead predictions}}\\
Joint & 1.60 & * & -39.09 & 2.21 & * & -41.11 & 15.12 & ** & -61.13\\
UNRATE & -0.74 & *** & -23.32 & 2.36 & *** & -25.66 & -0.59 & ** & -24.96\\
CPIAUCSL & 0.65 & * & -3.63 & 1.46 & *** & -6.14 & 4.29 & ** & -12.02\\
FEDFUNDS & 0.85 & ** & -63.07 & 11.17 & *** & -175.01 & 58.64 & *** & -607.74\\
\bottomrule\bottomrule
\end{tabular}
\begin{tablenotes}
\small
\item [] \footnotesize \textbf{Note:} Results for the cBVAR are in
    differences relative to the BVAR. Stars indicate p-values from a one-sided
    DM-test with alternative hypothesis that cBVAR produces larger LPL scores.
\item [] \footnotesize \textbf{Legend:} $.\sim p<0.16$; $^{*}\sim p<0.1$;
    $^{**}\sim p<0.05$; $^{***}\sim p<0.01$
\end{tablenotes}
\end{threeparttable}
\end{table}

 When we hone in on the variable-specific performance a story similar to the one we have seen in \autoref{tab:fc_mse} emerges. For unemployment rate forecasts, we find muted gains for one-month-ahead predictions. These increase with the forecast horizon. Notice that, for all three forecast horizons, the large model produces the weakest forecasts in absolute terms. 

 For CPI inflation and, in difference to the point forecasts, we find that the cBVAR improves upon the BVAR for the one-quarter-ahead forecast horizon and for all three model sizes considered. For one-month-ahead and one-year-ahead predictions we find either negligible improvements (in the case of the small model) or losses.

 Finally, for short-term interest rate forecasts we find improvements for forecast horizons greater than one-month-ahead. For the one-month-ahead horizon, modest gains are visible if the small model is adopted. For larger model sizes, improvements are minor.

In summary, the good performance for point forecasts is accompanied by more accurate density predictions. This holds not only for the three focus variables but also when we consider  the joint density forecasts. 

Similarly to our Monte Carlo exercise, we observe that coarsening gains increase  with model size, because larger versions of the standard BVAR suffer more heavily from misspecification. In particular, extreme events such as the global Covid-19 pandemic affect many series at once, leading to large losses in forecast gains if not adequately controlled for. 

\subsubsection{Density forecast performance over time}
The discussion thus far has focused on average forecast performance. To analyze whether the performance gains from coarsening vary over time, we now analyze the evolution of the cumulative LPLs over time. The figure shows the variable-specific and joint cumulative LPLs across all three model sizes and forecast horizons. The blue lines refer to the one-month-ahead, the red lines to the one-quarter-ahead and the green lines to the one-year-ahead forecast horizons. 

\begin{figure}[!th]
    \centering
    % HEADER
    \begin{subfigure}[b]{0.3\textwidth}
        \centering
        \textbf{Small-sized model}
    \end{subfigure}
    \hfill
    \begin{subfigure}[b]{0.3\textwidth}
        \centering
        \textbf{Medium-sized model}
    \end{subfigure}
    \hfill
    \begin{subfigure}[b]{0.3\textwidth}
        \centering
        \textbf{Large-sized model}
    \end{subfigure}
    % Joint 
    \begin{subfigure}[!htbp]{\textwidth}
         \centering
         \includegraphics[width=.3\textwidth]{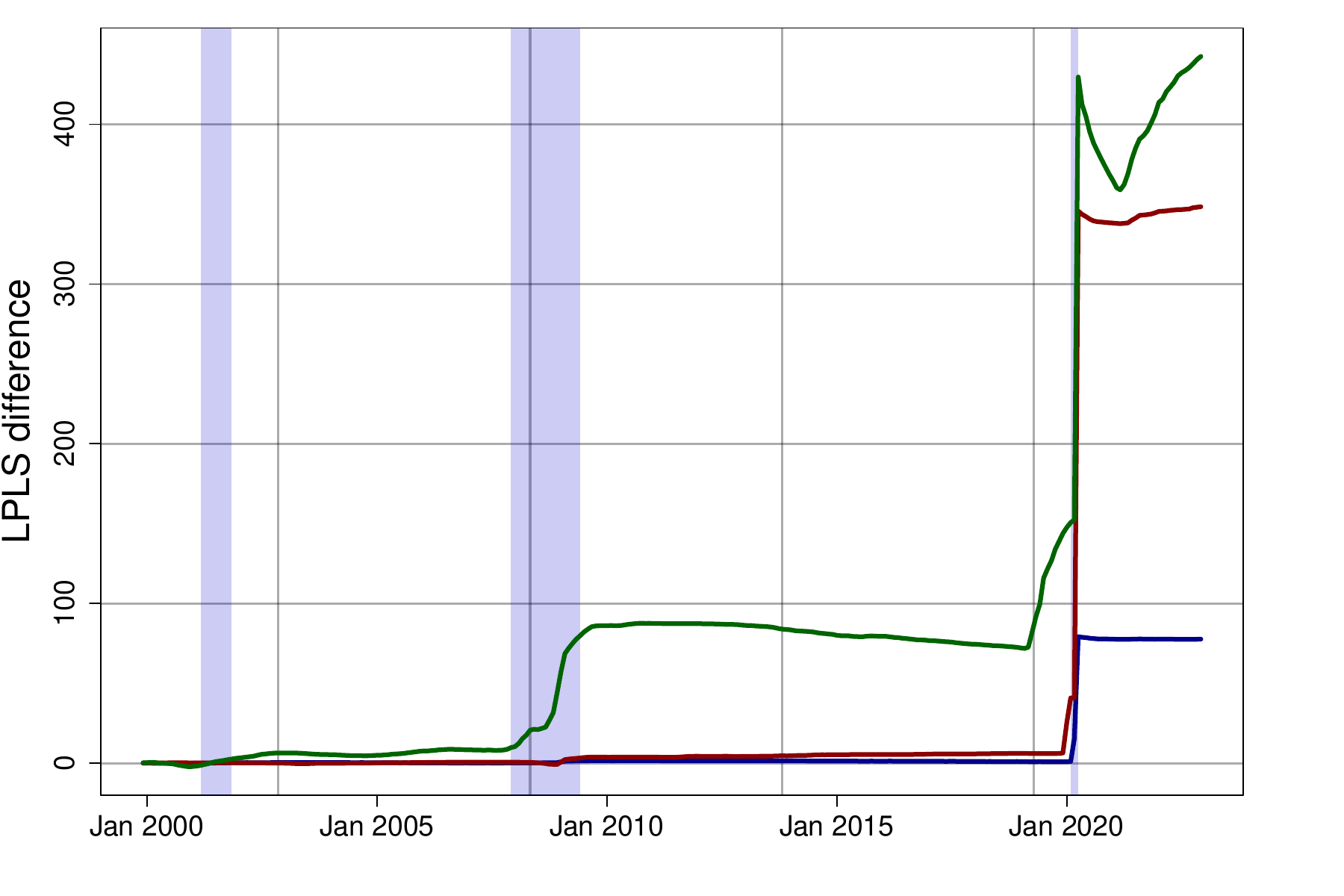} \hfill
         \includegraphics[width=.3\textwidth]{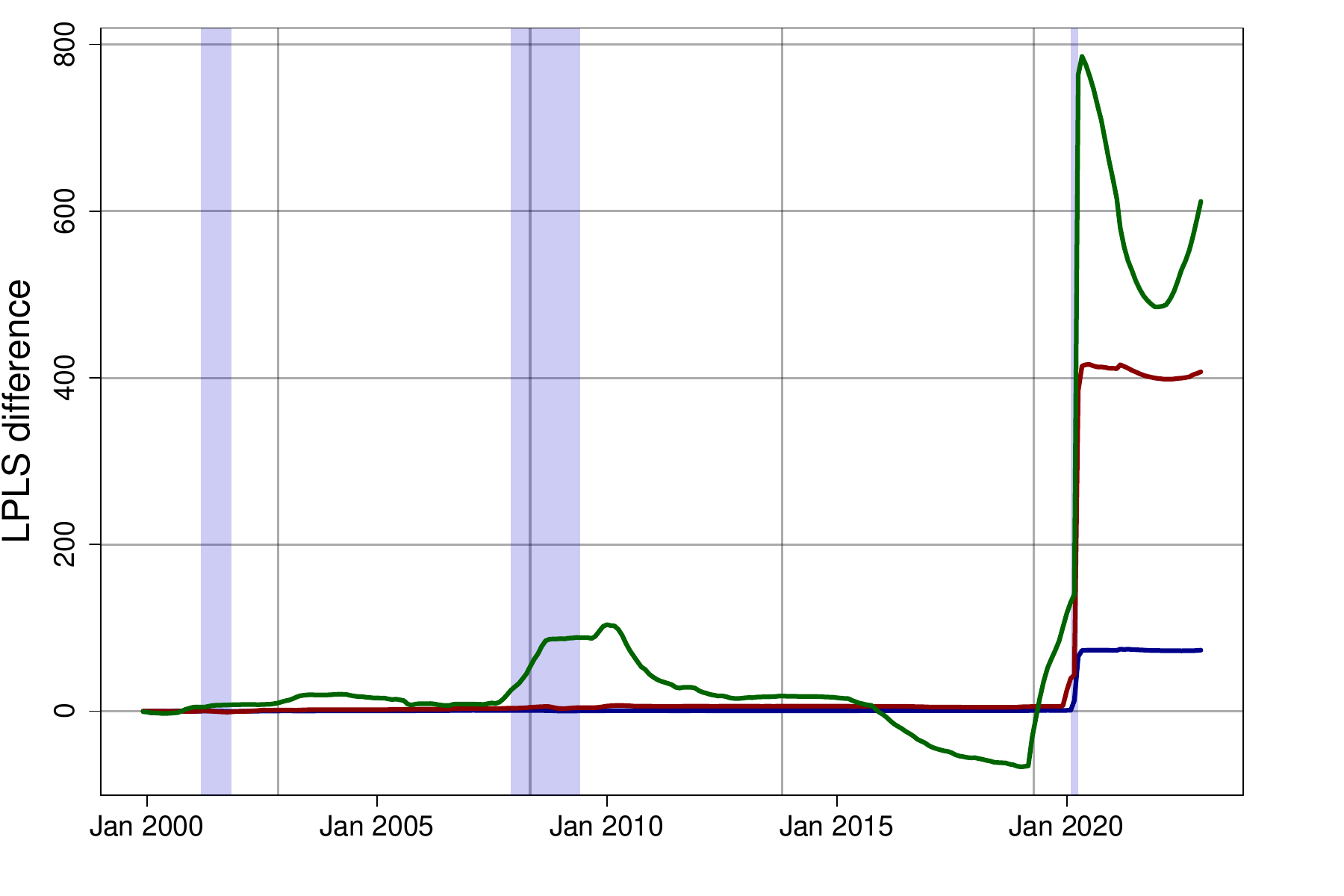} \hfill
         \includegraphics[width=.3\textwidth]{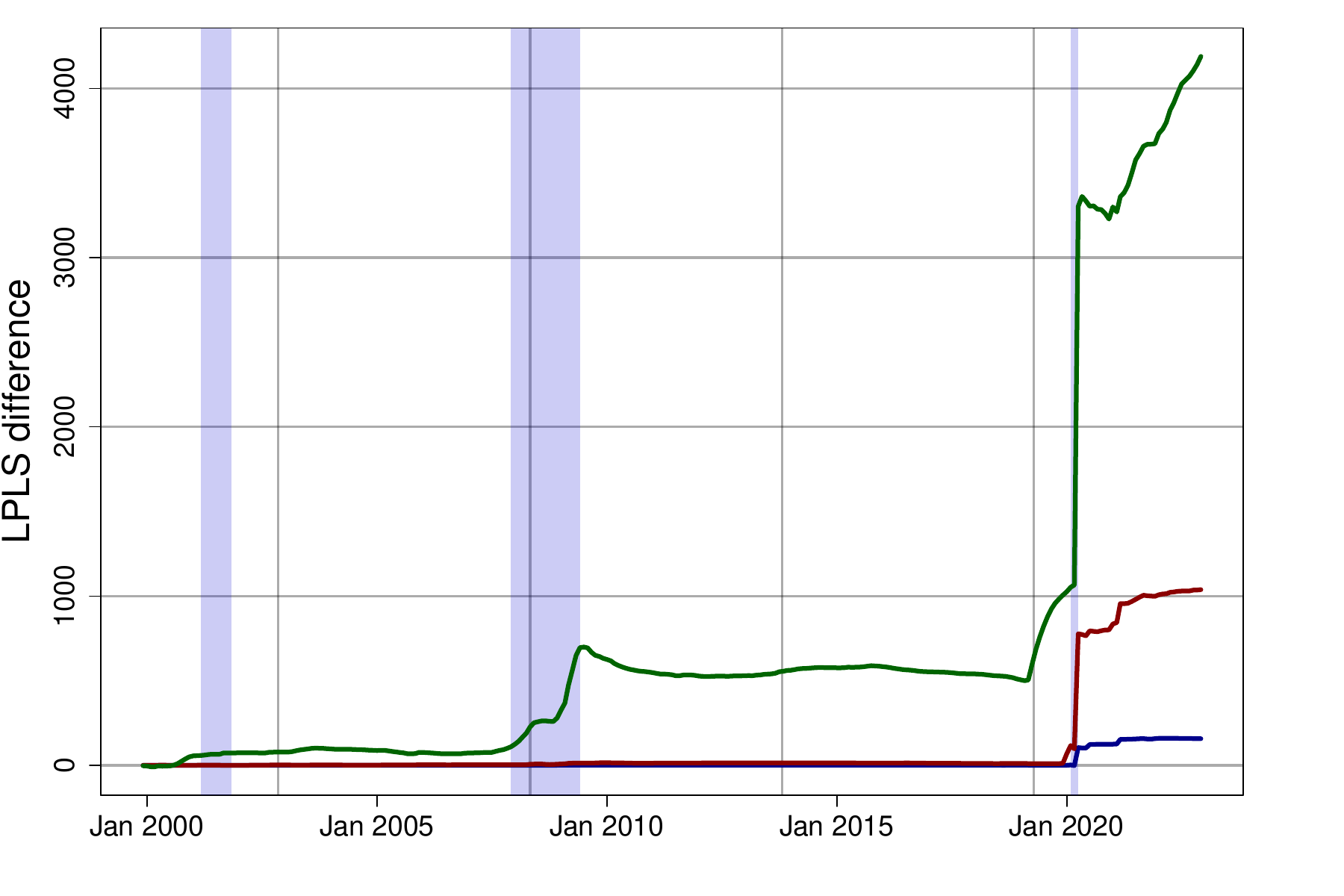}
         \caption{Joint}
         \label{fig:lpls_Joint}
     \end{subfigure}
     \vfill
     
     % UNRATE
     \begin{subfigure}[!htbp]{\textwidth}
         \centering
         \includegraphics[width=.3\textwidth]{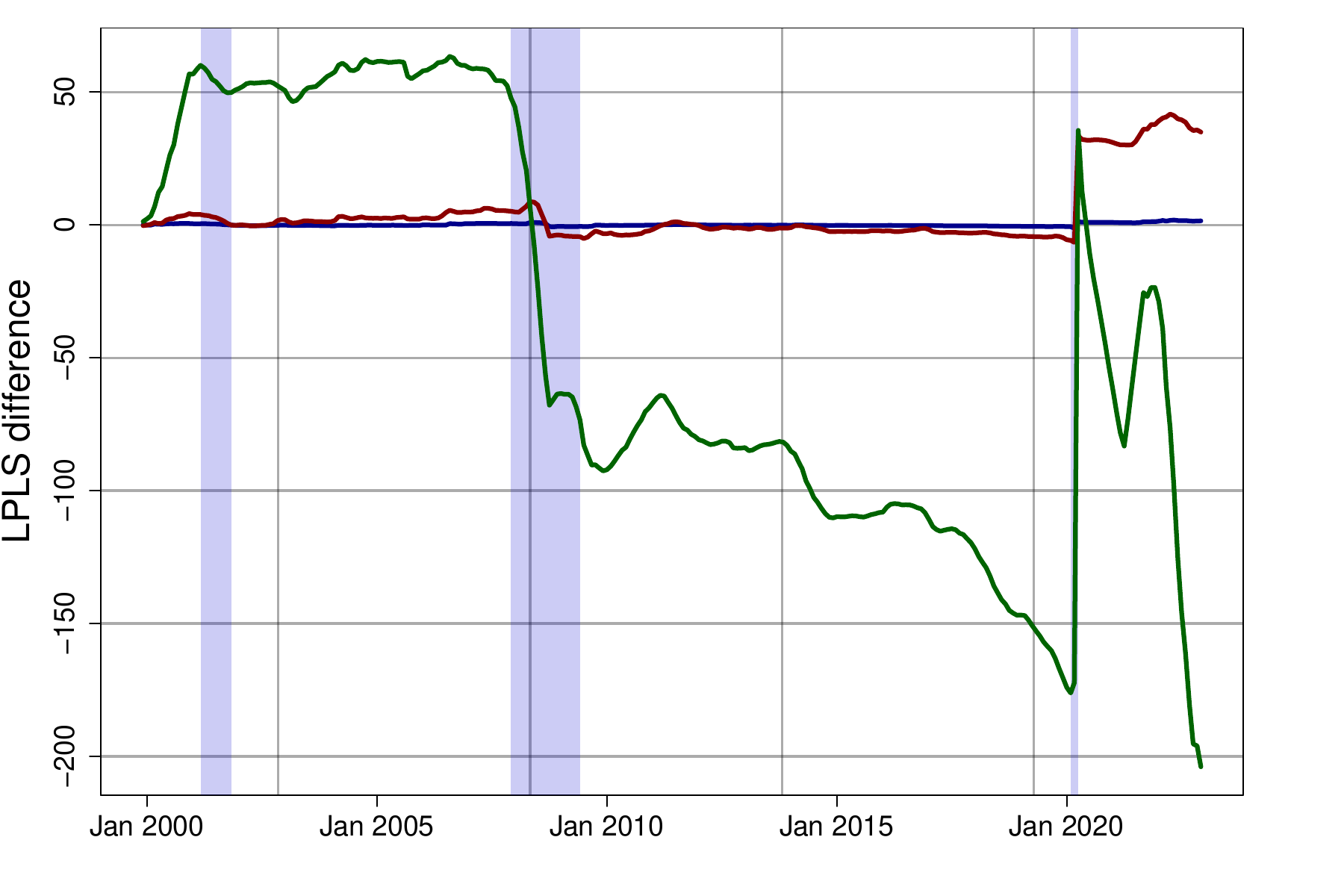}\hfill
         \includegraphics[width=.3\textwidth]{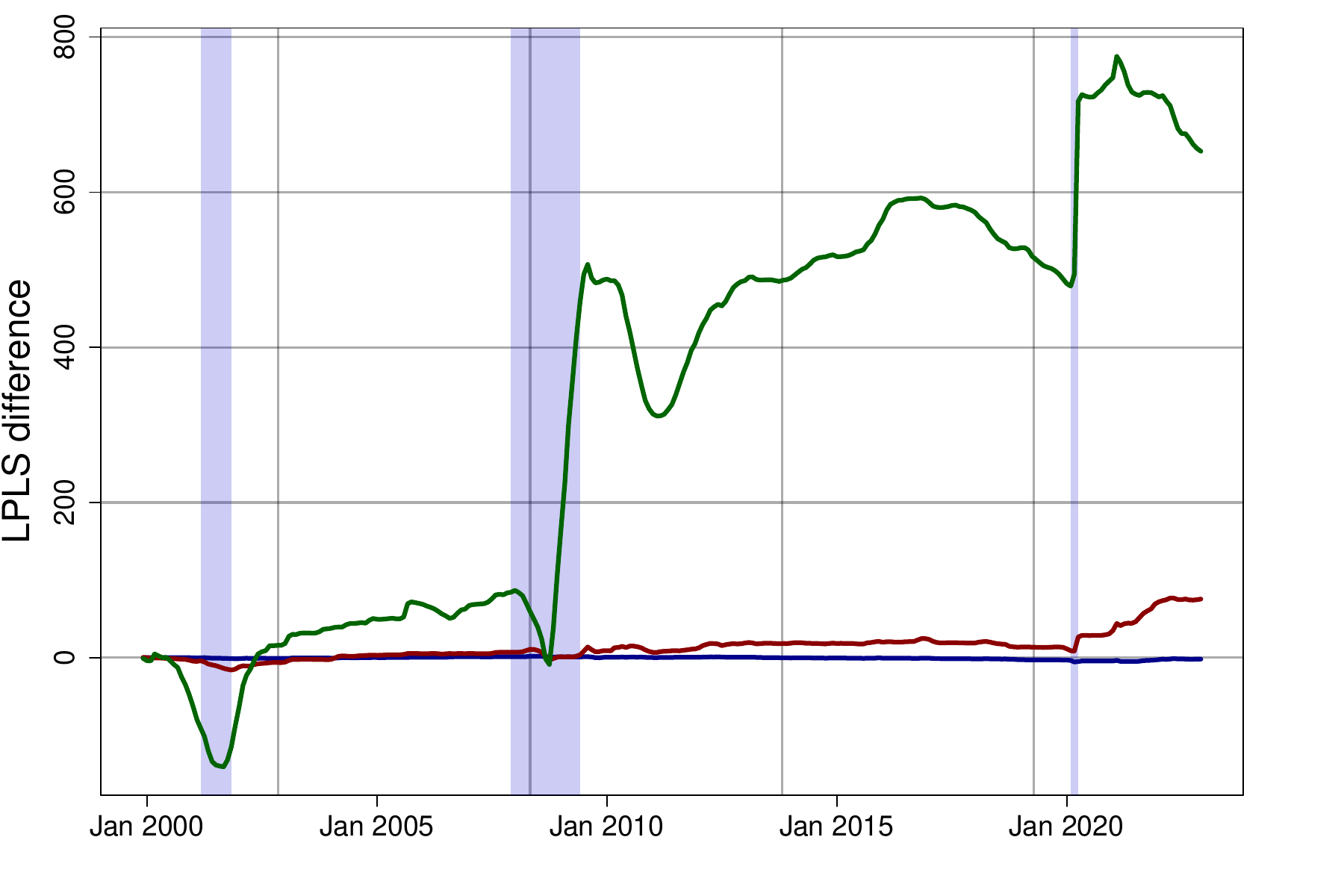}\hfill
         \includegraphics[width=.3\textwidth]{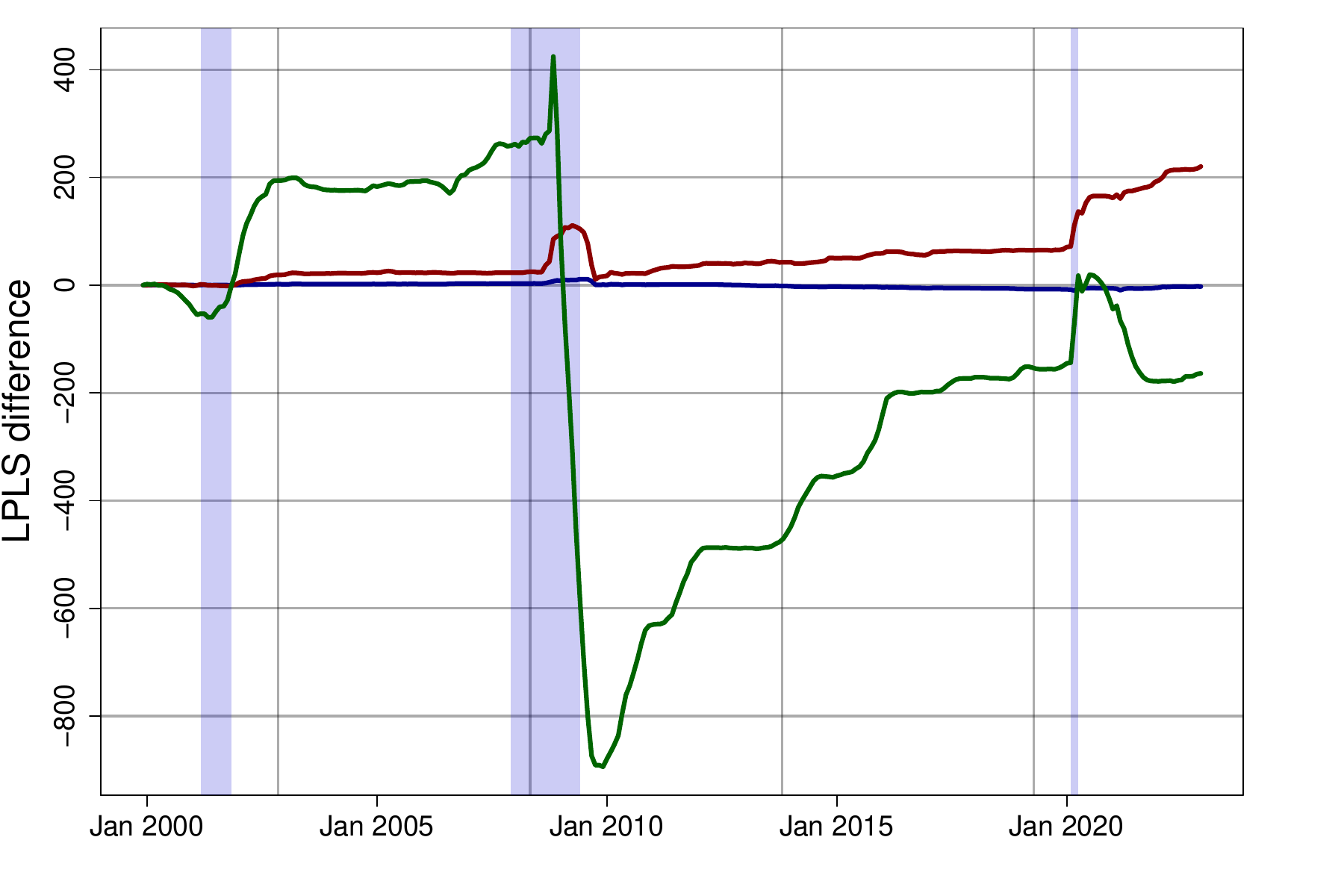}
         \caption{UNRATE}
         \label{fig:lpls_UNRATE}
     \end{subfigure}
     
     % CPIAUCSL
     \begin{subfigure}[!htbp]{\textwidth}
         \centering
         \includegraphics[width=.3\textwidth]{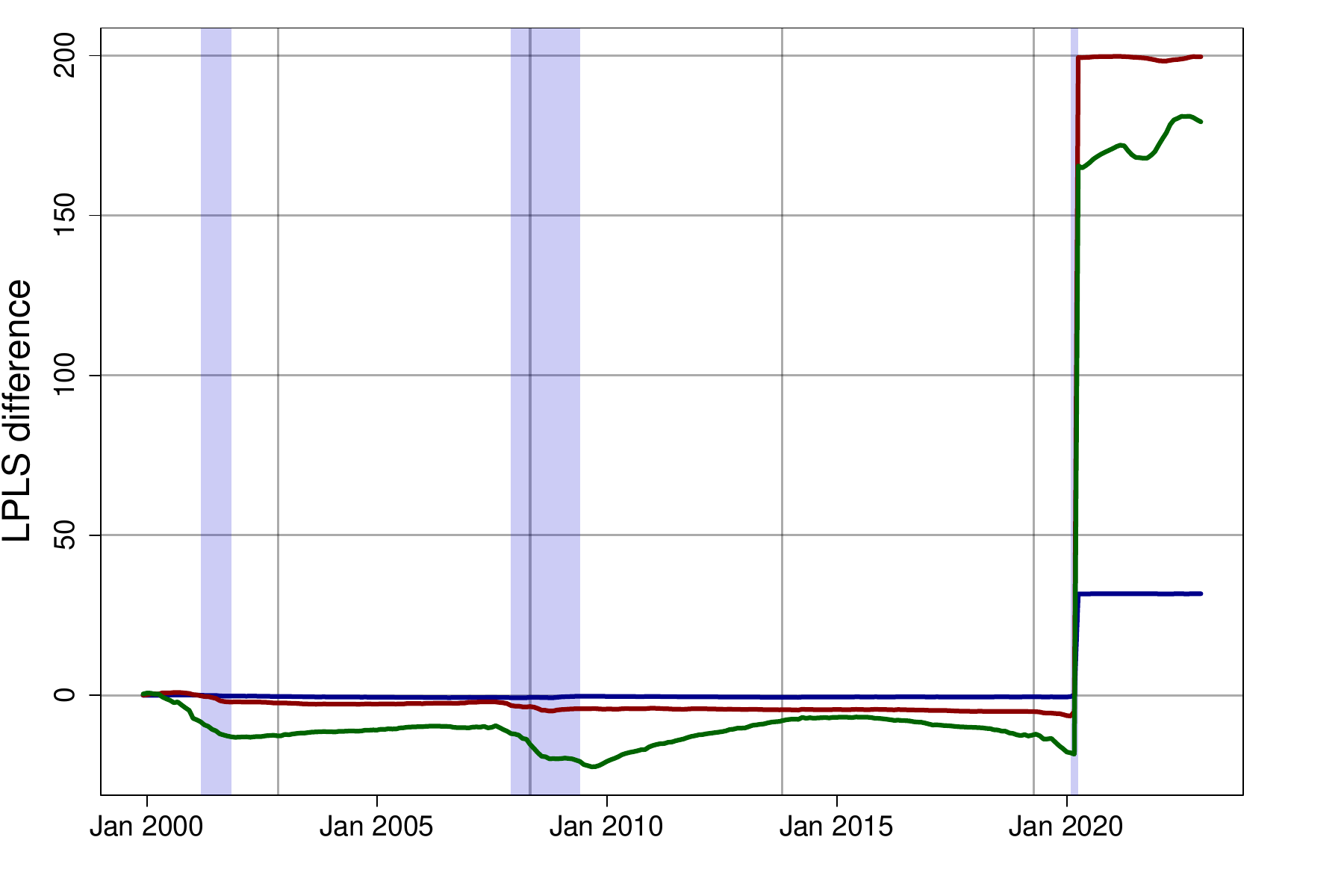} \hfill
         \includegraphics[width=.3\textwidth]{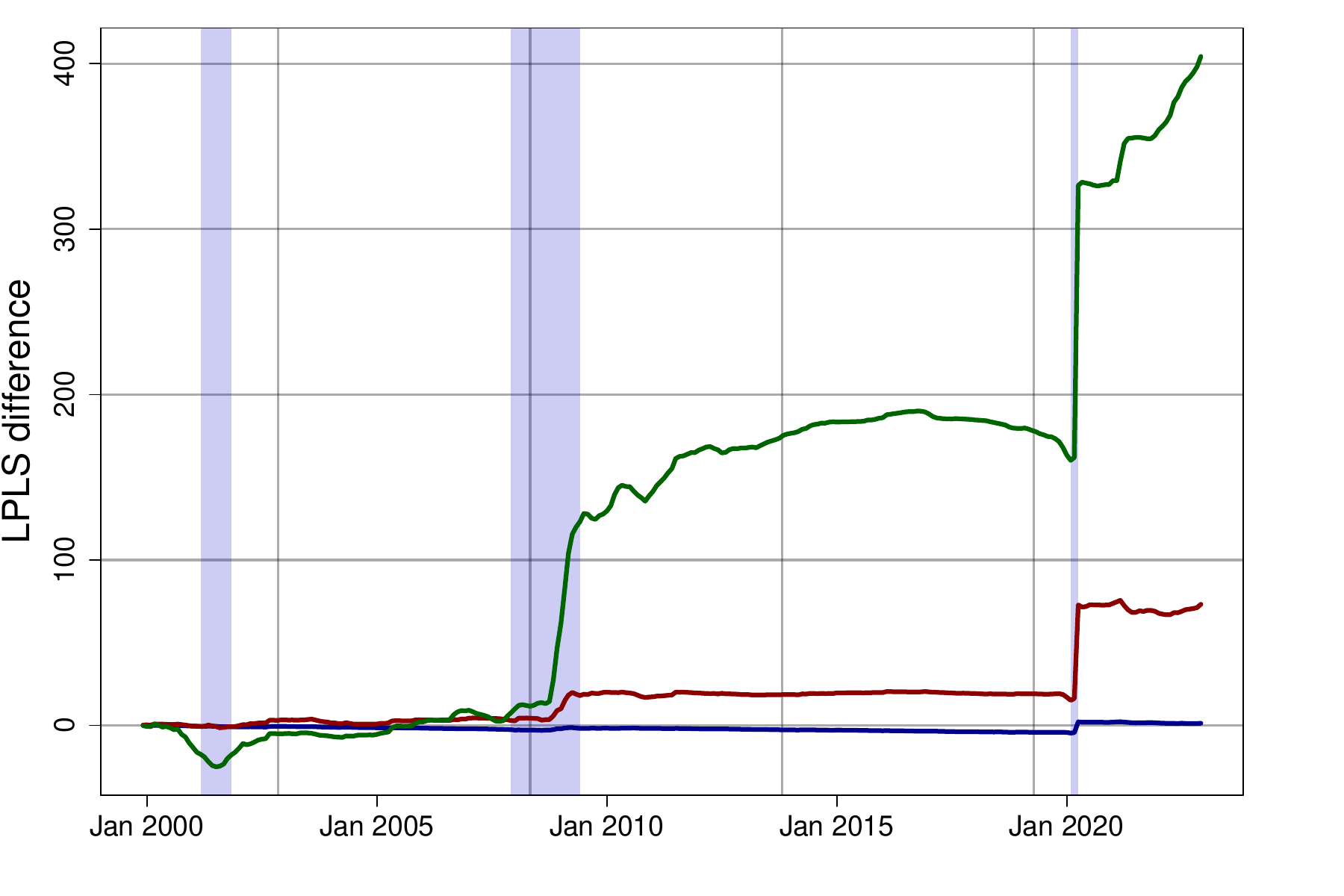} \hfill
         \includegraphics[width=.3\textwidth]{figures/figures_fc/lpls_CPIAUCSL_small.pdf}
         \caption{CPIAUCSL}
         \label{fig:lpls_CPIAUCSL}
     \end{subfigure}
     
     % FEDFUNDS
     \begin{subfigure}[!htbp]{\textwidth}
         \centering
         \includegraphics[width=.3\textwidth]{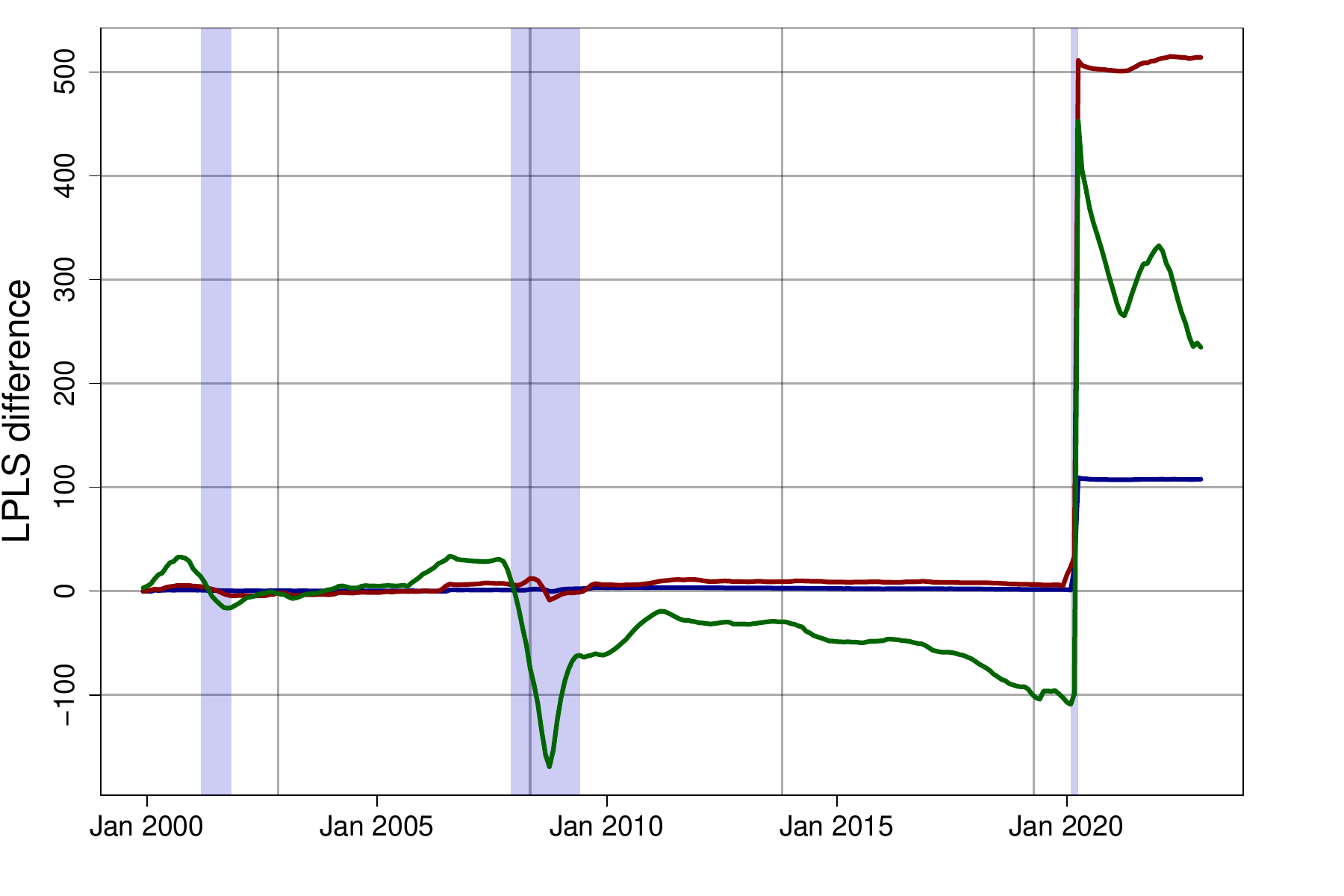} \hfill
         \includegraphics[width=.3\textwidth]{figures/figures_fc/lpls_FEDFUNDS_small.pdf} \hfill
         \includegraphics[width=.3\textwidth]{figures/figures_fc/lpls_FEDFUNDS_small.pdf}
         \caption{FEDFUNDS}
         \label{fig:lpls_FEDFUNDS}
     \end{subfigure}
     \caption{Cumulative LPLS differences over time.\\
     \textbf{Legend:} \textcolor{plotblue}{$-$} refers to the one-month-ahead, \textcolor{plotred}{$-$} to one-quarter-ahead and \textcolor{plotgreen}{$-$} to one-year-ahead forecasts. Shaded areas indicate the NBER recession dates.}
    \label{fig:lpls_1}
\end{figure}

We again start with the overall density forecast performance, measured in terms of the joint LPLs over time and displayed in \autoref{fig:lpls_Joint}. For small datasets, we find little differences for short-term forecasts (one-month- and one-quarter-ahead) until the onset of the pandemic. In this case, we observe that coarsening substantially improves predictive performance. This finding carries over to medium- and large-scale models. During the pandemic, some of the time series (such as the unemployment rate) displayed substantial outliers that are inconsistent with a Gaussian homoskedastic model. Coarsening, by contrast, leads to more conservative credible intervals, attributing a higher probability to these extreme observations and thus effectively improving density forecasting performance. 

However, it should be noted that the joint density forecasting performance is not exclusively driven by the pandemic. When we consider the medium and large models, we observe that for higher-order forecasts, improvements relative to the BVAR arose much earlier and appear to increase during (or shortly after) turbulent periods.

Next, we consider variable-specific LPLs. For the unemployment rate (see \autoref{fig:lpls_UNRATE}), we find a picture similar to the one of the joint LPLs: smaller differences for one-month- and one-quarter-ahead LPLs up to the pandemic and then substantial gains during the pandemic. This holds for the small and medium dataset. For one-year-ahead LPLs we find that coarsening already helps shortly after the global financial crisis (GFC), approximately coinciding with the period where the unemployment rate peaked. This pattern holds for all three model sizes. Only for the large-sized model and one-year-ahead forecasts we find a much weaker performance of the cBVAR between 2010:01 and the pandemic. 

For inflation (\autoref{fig:lpls_CPIAUCSL}), we again find small differences between one-month-ahead and one-quarter-ahead forecast accuracy and all three model sizes. There is again some evidence that the inflation forecasts arising from the cBVAR become more accurate after the Covid-19 pandemic, a period characterized by a sustained increase in inflation. This pattern is visible for all forecast horizons except for the one-year-ahead forecast and the small model. In this case, we find that cBVAR performs better than the benchmark in the first part of the hold-out period (2020:01 to around 2009:06), is inferior during the second part of the hold-out (2009:07 to 2020:05) before recovering. When we consider the medium-sized model and one-year-ahead predictions, we find that after the GFC, the forecast accuracy premium from coarsening increases, and the cBVAR consistently returns more accurate density predictions than the BVAR. These accuracy improvements also increase during the pandemic and post-pandemic part of the sample. For the large model, the pattern resembles that of the small model.

The LPLs for the short-term interest rate (\autoref{fig:lpls_FEDFUNDS}) evolve similarly to the other two series. Small differences in the pre-pandemic period are followed by substantial variation during the pandemic. Strikingly, we find that there is an appreciable improvement in LPLs during the zero lower bound period only for the medium-sized model. This is not visible for the small and large information sets we consider, and, at first glance, appears to be inconsistent with our findings for the point forecasts. Notice, however, that for LPLs it does not only matter how well we fit the first moment of the predictive density, but higher order moments also play an important role. 

\subsection{How does misspecification change over time?}
The previous sections emphasized that coarsening leads to improvements in predictive performance, particularly  at longer forecast horizons. In this section, we ask how much weight is actually put on the likelihood function. To do so, we plot the learning rates for the three focus variables over the hold-out. Doing so gives us a general impression on the degree of misspecification and whether this is subject to time-variation.

\begin{figure}[!t]
    \begin{subfigure}[b]{0.32\textwidth}
         \centering
         \includegraphics[width=\textwidth]{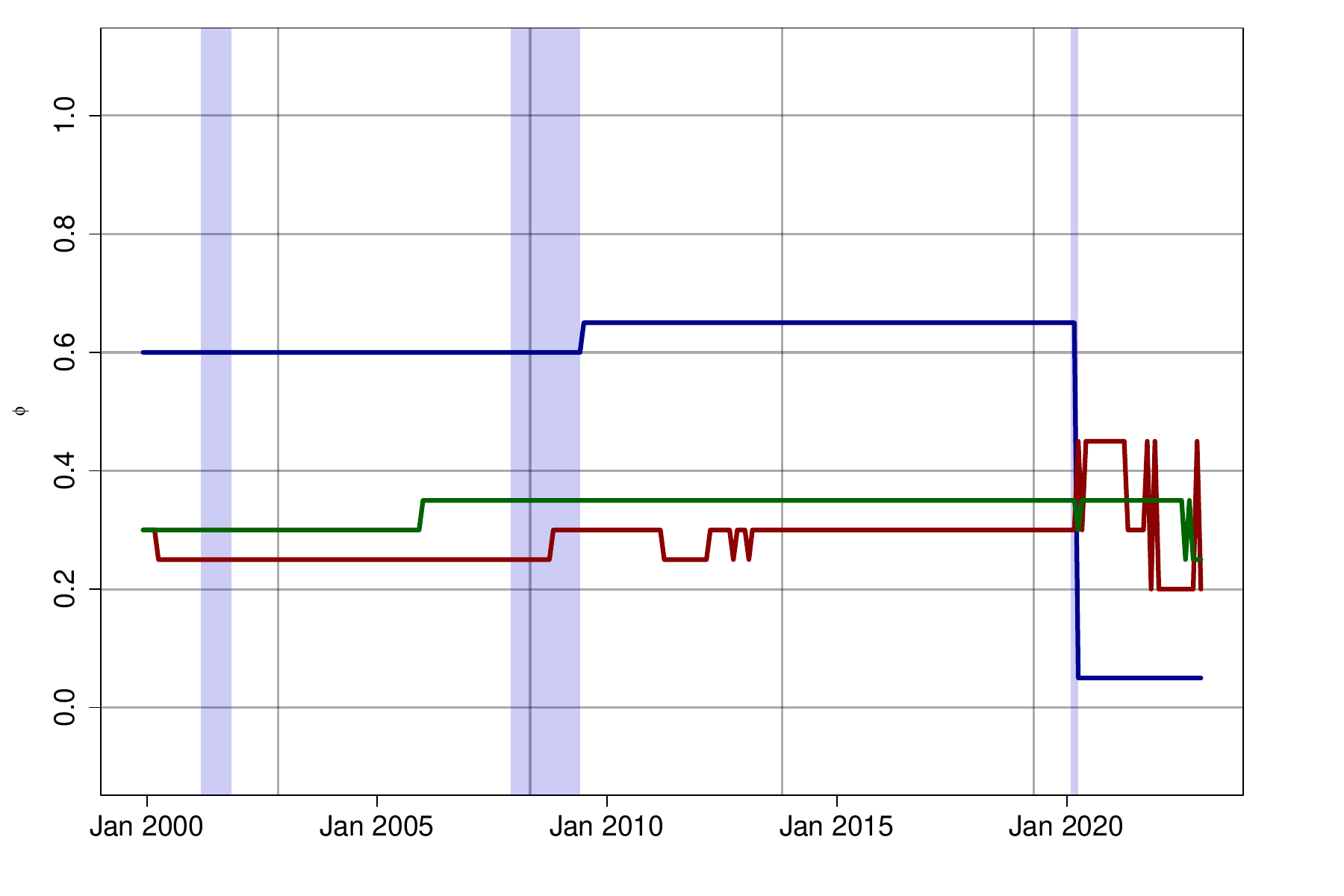}
         \caption{Small-sized VAR}
         \label{fig:phi_over_time_small}
     \end{subfigure}
     \hfill
     \begin{subfigure}[b]{0.32\textwidth}
         \centering
         \includegraphics[width=\textwidth]{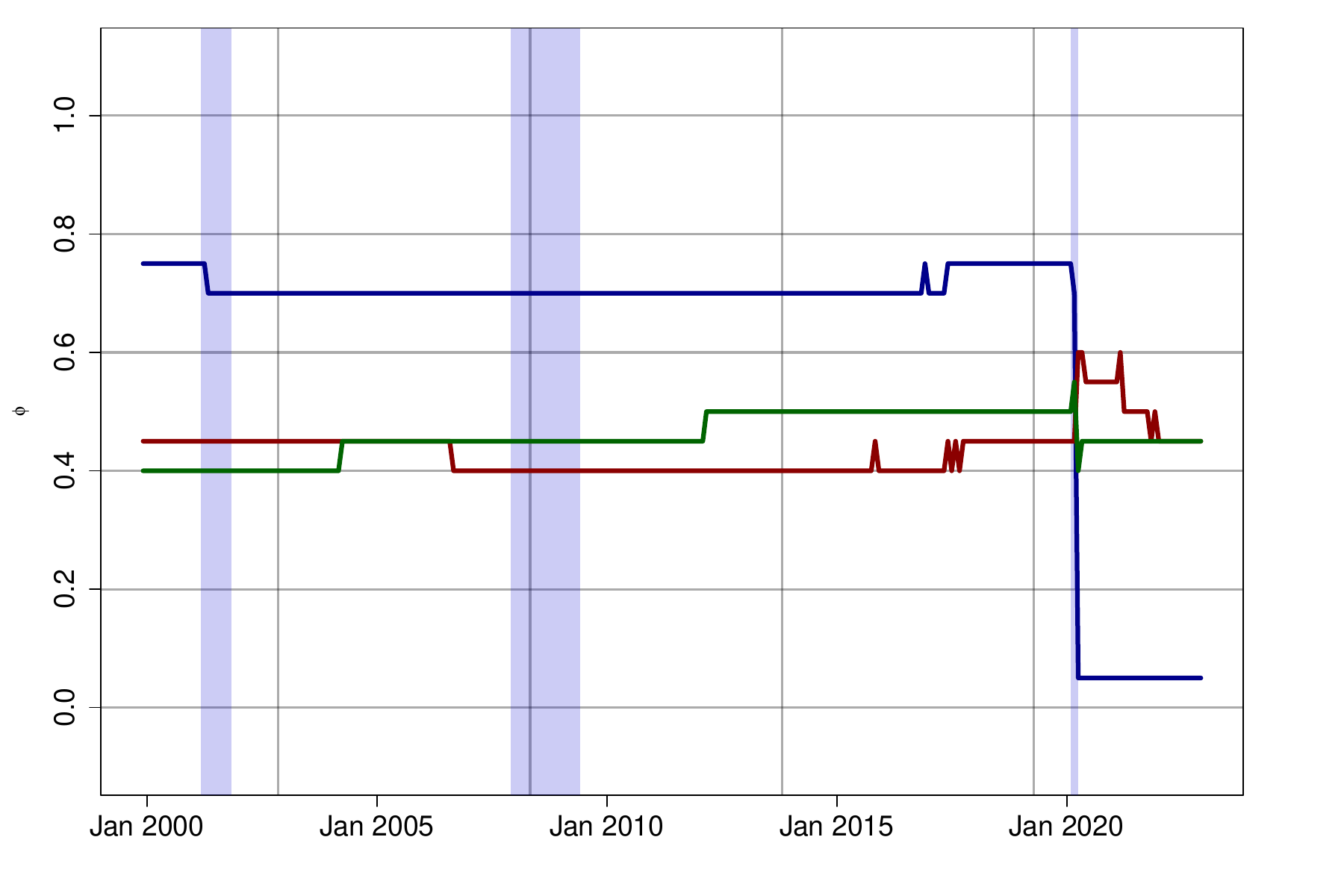}
         \caption{Medium-sized VAR}
         \label{fig:phi_over_time_medium}
     \end{subfigure}
     \hfill
     \begin{subfigure}[b]{0.32\textwidth}
         \centering
         \includegraphics[width=\textwidth]{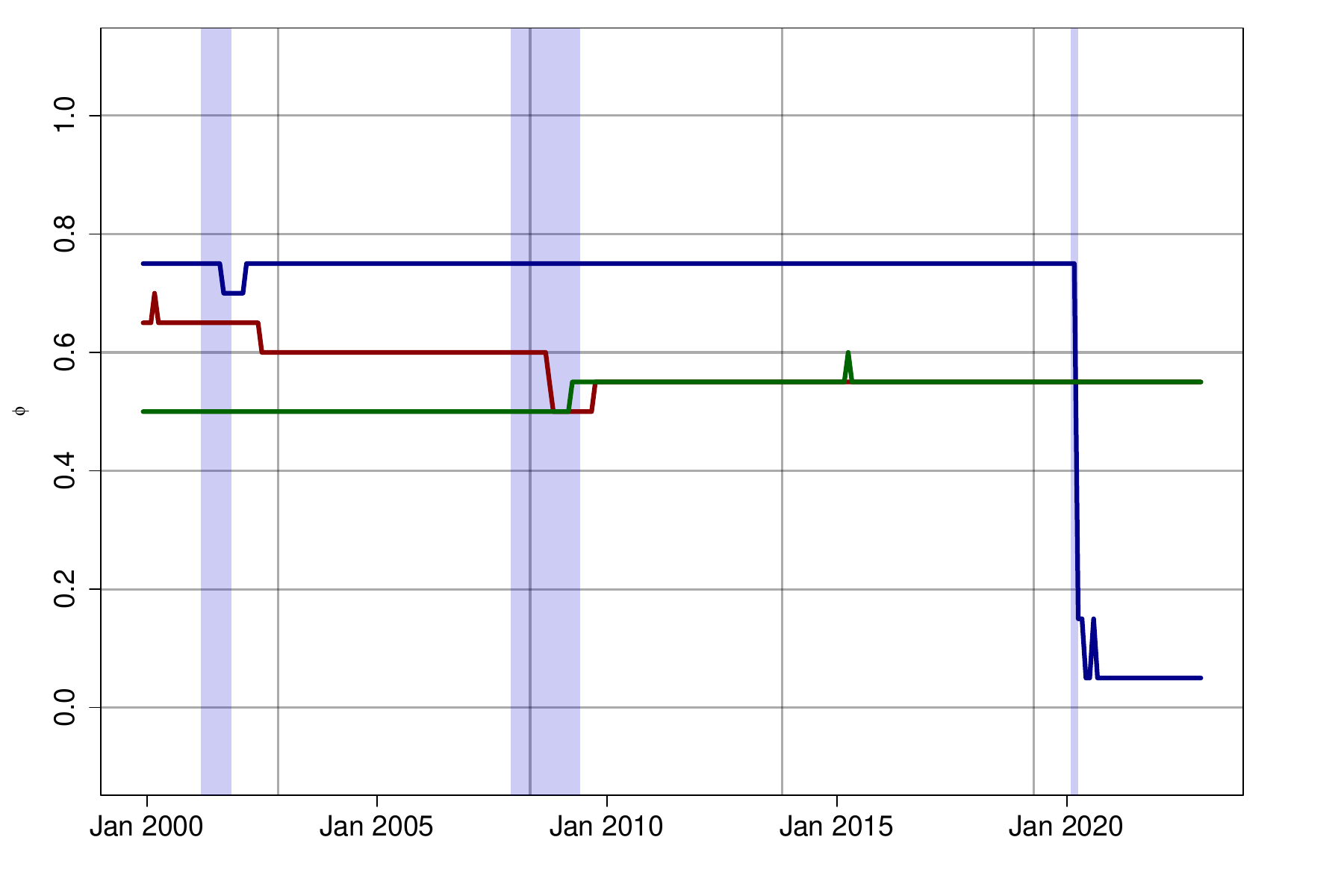}
         \caption{Large-sized VAR}
         \label{fig:phi_over_time_large}
     \end{subfigure}
    \caption{Estimated likelihood weights over the different training samples for the focus variables.\\
    \textbf{Legend:} \textcolor{plotblue}{$-$} refers to UNRATE, \textcolor{plotred}{$-$} to CPIAUCSL and \textcolor{plotgreen}{$-$} to FEDFUNDS. Blue shaded areas indicate the NBER recession dates.}
    \label{fig:phi_over_time}
\end{figure}

\autoref{fig:phi_over_time} shows the estimated $\phi_i's$ for the unemployment rate (in blue), inflation (in red), and the short-term interest rate (in green) over the hold-out sample. Across the three model sizes, we observe that $\phi_i$ is much smaller than one throughout our hold-out period, pointing towards a substantial degree of misspecification. Among the focus series, we find that CPI inflation and short-term interest rates have the lowest learning rates whereas the one of the unemployment rate is much higher. An interesting finding is that, for all three target series, adding information to the model seems to increase the estimated learning rates. This finding is driven by the reduced risk of omitted variables in larger models. 

In terms of time variation, we find some rather unsystematic movements in $\phi_i$ for all three variables prior to the pandemic. In some cases, the learning rate slightly increases while in other cases, we have declining learning rates. However, these movements are relatively small and should not affect the predictive distribution too much. 

However, when we focus on the pandemic period,  we find a common pattern across all models considered: the unemployment rate equation suddenly requires a much lower learning rate. Again, and consistent with the findings for LPLs over time, this indicates that the sharp increase in unemployment rates (and quick recoveries after the lockdowns subsided) requires a more flexible econometric model. Our cBVAR reflects this necessity by producing a much lower learning rate during these periods. 

\section{Conclusion}
Common practice in empirical macroeconomics is to estimate simple and interpretable models to derive stylized facts and inform decision making. However, these simple models often suffer from misspecification of various, and unknown, types, which has a deleterious effect on estimation, inference, and predictive performance. In this paper, we offer a simple solution to obtain more robust results. Our proposal is to modify standard Bayesian multivariate econometric models. Instead of conditioning on the observed data, we propose to condition on the event that there is a distance between the sampling distribution of the observed data and a hypothetical idealized sampling distribution. Using relative entropy as a loss metric gives rise to a simple approximation that amounts to tempering the likelihood by a learning rate, connecting this approach to the literature on power posteriors.  

Since different endogenous series in a VAR can feature different degrees of misspecification, we allow for different learning rates across equations. The resulting likelihood is then coupled with an asymmetric prior, allowing for conjugate updating. To select the learning rates, we use a fully automatic approach that selects the prior shrinkage parameters and the learning rates jointly.

We consider an extensive Monte Carlo exercise with a wide range of different DGPs that are inspired by models actually used by empirical macroeconomists. These differ in terms of the conditional mean and variance specifications, giving rise to 16 different DGPs.  We show that standard BVARs in the presence of misspecification of different forms produce predictive distributions that are substantially less accurate than those produced by the coarsened counterpart. 

Using actual US monthly data, we show that the cBVAR produces more accurate point and density forecasts than the standard BVAR, with larger gains at longer horizons and when using small models, and it does not suffer much from the substantial outliers observed during the pandemic. This suggests that our approach can also be used to obtain more robust inference in the presence of large outliers. 

%Finally, we show that a coarsened model can help also for more structural applications. When we consider the effects of uncertainty shocks, we find that the cBVAR produces no evidence of a real activity overshoot, contrary to the literature that relies on observed proxies of uncertainty, with negative real activity responses that are milder but more persistent than with a standard BVAR.
\clearpage
% \singlespacing
% \small{\setstretch{0.85}
\addcontentsline{toc}{section}{References}
\bibliography{lit}
\bibliographystyle{frbcle.bst}%}

\clearpage
\begin{appendices}
\setcounter{figure}{0}    
\setcounter{table}{0}
\renewcommand{\thetable}{\thesection.\arabic{table}}
\renewcommand{\thefigure}{\thesection.\arabic{figure}}

% \section{Codes}

% Computations were performed using the programming languages Julia \citep{julia} and R \citep{R}. Replication codes are available in a dedicated \hyperlink{https://github.com/tscheckel/Coarsened_VAR}{Github repository}.

\section{Technical Appendix}
\setcounter{figure}{0}    
\setcounter{table}{0}
\renewcommand{\thetable}{\thesection.\arabic{table}}
\renewcommand{\thefigure}{\thesection.\arabic{figure}}

\subsection{Posterior Distribution}

Combining the joint prior and likelihood functions, we can derive the joint posterior distribution as:
\begin{equation*}
\begin{split}
    p(\bm \theta_1& , \dots, \bm \theta_M, \sigma_1^2 \dots, \sigma_M^2 | \bm Y_1, \dots, \bm Y_M, \phi_1, \dots, \phi_M)\\
    \propto &\ p(\bm \theta_1, \dots, \bm \theta_M, \sigma_1^2, \dots, \sigma_M^2)\ \tilde{p}(\bm Y_1, \dots, \bm Y_M|\bm \theta_1, \dots, \theta_M, \sigma_1^2, \dots, \sigma_M^2, \phi_1, \dots, \phi_M) \\
    &\ = \prod_{i=1}^M p(\bm \theta_i, \sigma_i^2)\ \hat{p}(\bm Y_i| \bm \theta_i, \sigma_i^2, \phi_i) \\
    &\ = \prod_{i=1}^M c_i (\sigma_i^2) ^{-(\underline{\nu}_i + \frac{i+MP}{2} + 1)}\text{exp}\left(-\frac{1}{\sigma_i^2}(\underline{S}_i + \frac{1}{2}(\bm \theta_i - \underline{\bm \theta}_i)'\underline{\bm V}_i^{-1}(\bm \theta_i - \underline{\bm \theta}_i))\right) \\
    &\ \ \times (2\pi \sigma_i^2)^{- \frac{\phi_i T}{2}} \text{exp}\left(-\frac{\phi_i}{2\sigma_i^2}(\bm Y_i - \bm X_i \bm \theta_i)'(\bm Y_i - \bm X_i \bm \theta_i)\right) \\
    &\ = \prod_{i=1}^Mc_i (2\pi)^{-\frac{\phi_i T}{2}} (\sigma_i^2)^{-(\underline{\nu}_i + \frac{\phi_i T + i+MP}{2})+1)} \\
     &\ \ \times \text{exp}\left(-\frac{1}{\sigma_i^2}(\underline{S}_i + \frac{1}{2}(\bm \theta_i' (\underline{\bm V}_i^{-1} + \phi_i X_i' X_i)\bm \theta_i - 2 \bm \theta_i'(\underline{\bm V}_i^{-1}\underline{\bm \theta}_i + \phi_i \bm X_i' \bm y_i) + \underline{\bm \theta}_i' \underline{\bm V}_i^{-1} \underline{\bm \theta}_i + \phi_i \bm Y_i' \bm Y_i))\right) \\
     &\ = \prod_{i=1}^M c_i (2\pi)^{-\frac{\phi_i T}{2}} (\sigma_i)^{-(\overline{\nu}_i + \frac{i + MP}{2} +1)} \text{exp}(-\frac{1}{\sigma_i^2}(\overline{S}_i + \frac{1}{2}(\bm \theta_i - \overline{\bm \theta}_i)'\overline{\bm V}_i (\bm \theta_i - \overline{\bm \theta}_i)))
     \end{split}
\end{equation*}

\subsection{Simulation of Data Generating Processes}\label{app:dgps}
Unless specified otherwise specified, the parameters below are obtained by estimating the corresponding specification on US data (using the datasets outlined in \autoref{tab:data_app}). 

All DGPs have the general form:
\begin{equation}
    \bm y_t = \bm A_{1t} \bm y_{t-1} + \bm A_{2t} \bm y_{t-2} + \bm B \bm X_t + \bm u_t + \bm \theta \bm u_{t-1},\ \bm u_t \sim \text{p}(\bm u_t| \bm 0_M, \bm \Sigma_t),
\end{equation}
where $\bm y_t$ is an $M$-dimensional vector of endogenous variables at time $t$ and $\bm X_t$ is a $K=5$-dimensional vector of additional exogenous regressors at time $t$. The associated parameter matrices are given by $\bm A_{1,t}, \bm A_{2,t}, \bm B$. The reduced- form shocks at time $t$ are denoted $\bm u_t$. It follows some distribution $\text{p}(\cdot | \bm 0_M, \bm \Sigma_t)$ with mean $\bm 0_M$ and variance-covariance matrix $\bm \Sigma_t = \bm A_0 \bm D_t \bm A_0$. Here, $\bm A_0$ is a lower triangular matrix with ones on the main diagonal and $\bm D_t$ a diagonal matrix containing the error variances at time $t$. Lastly, $\bm \theta$ is a moving average coefficient matrix. 

We consider four different kinds of DGPs with respect to modeling the conditional mean and the error structure. These are outlined in further detail below.
\subsubsection{Variations with respect to the conditional mean}
\begin{enumerate}
    \item Our baseline is a linear VAR ("Linear") where we set $\bm A_{1t} = \bm A_1, \bm A_{2t} = \bm A_2,\ \forall t=1,\dots,T$ and $\bm B = 0_{M\times K}$. The parameter matrices $\bm A_1$ and $\bm A_2$ are obtained by estimating a VAR(2) following \cite{giannone2015prior}. This model has been applied extensively in the empirical literature, a recent example is \cite{antolindiaz-surico_2025_aer}.
    \item In the second case, the DGP arises from a smooth transition model ("Transition") and we set $\bm A_{pt} = s_t \bm A_{pa} + (1-s_t) \bm A_{pb}$, with $\bm A_{pa}, \bm A_{pb}$ being the baseline regimes for $p = 1,2$. The mixture variable $s_t \in [0,1]$ is given by the following equation:
        \begin{equation*}
             s_t = \frac{1}{1+\text{exp}((-\gamma) z_t)},
        \end{equation*}
    where we set the speed of transition parameter $\gamma = 3$ and simulate $z_t$ arise from a random walk with unit variance. We set $\bm B = \bm 0_{M\times K}$ and, conditional on $s_t$ (which is known given $\{z_t\}_{t=1}^T$), estimate the parameter matrices again using BVAR(2) models on the corresponding macro time series. An recent application of this approach is \cite{caggiano-etal_2022_jae}.
    \item The third DGP follows a break-point model ("Break") such that the VAR coefficients are given by $\bm A_{pt} = \bm A_{pk}, p = 1,2$, where $\bm A_{pk}$ are changing every 50 observations. We again set $\bm B = \bm 0_{M\times K}$ and obtain the coefficients through BVAR(2) models estimated over the respective sub-samples. A model of this kind has recently been applied by \cite{check-piger_2021_jmcb}.
    \item In the case of the DGP featuring exogenous regressors ("Exo"), we let $\bm A_{1t} = \bm A_1, \bm A_{2t} = \bm A_2,\ \forall t=1,\dots,T$ (as in the linear baseline model) but allow $\bm B$ to be a non-zero coefficient matrix which each of its elements drawn from a normal distribution with mean zero and variance $\frac{1}{M^2}$. We simulate each element of $X_t$ from an AR(1) model with persistence parameter $0.95$ and error variance $0.0625$. An example of this is \cite{forni-etal_2014_ej}.
\end{enumerate}

\subsubsection{Variations with respect to the error structure}
\begin{enumerate}
    \item For the DGP labeled "Gaussian" we let the errors arise from a homoskedastic Gaussian distribution $\text{p}(\cdot) = \mathcal{N}(\bm 0_M, \bm D)$ (such that $\bm D_t = \bm D\ \forall t=1,\dots,T$). $\bm D$ has been calibrated by estimating a VAR(2) model following \cite{giannone2015prior}. We set $\bm \theta = 0_{M\times M}$. This simple model has been extensively applied in the empirical literature, a recent example is again \cite{antolindiaz-surico_2025_aer}.
    \item Additionally, we explore a similar setup (labeled "Student") with $\text{p}(\cdot) = t_3(\bm 0_M, \bm D)$ being the Student-t distribution with three degrees of freedom and again $\bm D_t = \bm D\ (\forall t=1,\dots,T)$ and $\bm \theta = \bm 0_{M\times M}$ ("Student"). Recent work by \cite{karlsson2023vector} emphasizes the empirical importance of allowing for heavy tails in the error distribution.
    \item The DGP considering stochastic volatility ("SV") sets assumes that the natural logarithm of the error variances evolves according to the following AR(1) model:
    \begin{equation*}
        \text{ln}(\bm D_t) = \text{ln} (\bm D) + 0.95 \cdot \left(\text{ln}(\bm D_{t-1}) - \text{ln}(\bm D)\right) + \bm \nu_t
    \end{equation*}
    where each element of $\bm \nu_t$ is normally distributed with mean zero and variance $\eta^2 = 0.05$. We let $\text{p}(\cdot) = \mathcal{N}(\bm 0_M, \bm D_t)$ and $\bm \theta = \bm 0_{M \times M}$. A recent example of Vector autoregressions featuring stochastic volatility is \cite{carriero_etal_2018_restat}.
    \item For the DGP with a moving-average term ("MA") we set $\text{p}(\cdot) = \mathcal{N}(\bm 0_M, \bm D),\ (\bm D_t = \bm D\ \forall t=1,\dots,T)$. We let $\bm \theta$ be non-zero and draw each of its elements from a normal distribution with mean zero and variance $\frac{1}{M^2}$. A DGP along these lines has recently been considered in \cite{gonzalez2025misspecification}.
\end{enumerate}

\section{Data appendix}

We provide information on the variables comprising the small-, medium- and larged sized models in \autoref{tab:data_app}. The table also contains their respective FRED-MD variable codes and information on the transformations we applied. To preserve their cointegrating relationship, all variables (expect price series) are typically in (log-)levels

\begin{landscape}\begin{table}[!htbp]

\caption{\label{tab:data_app}Variables used in the small, medium and larged models alongside their FRED-MD codes and transformations. Transformation codes refer to the following: 1 = level, 4 = log-level, 5 = log-first differences.}
\centering
\begin{tabular}[t]{lccccc}
\toprule\toprule
FRED-MD Code & Variable & Transformation & Small & Medium & Large\\
\midrule
UNRATE & Civilian Unemployment Rate & 4 & X & X & X\\
CPIAUCSL & CPI : All Items & 5 & X & X & X\\
FEDFUNDS & E?ective Federal Funds Rate & 1 & X & X & X\\
AWHMAN & Avg Weekly Hours : Manufacturing & 4 &  & X & X\\
M2REAL & Real M2 Money Stock & 5 &  & X & X\\
S\&P 500 & S\&P s Common Stock Price Index: Composite & 5 &  & X & X\\
INDPRO & IP Index & 5 &  & X & X\\
T10YFFM & 10-Year Treasury C Minus FEDFUNDS & 1 &  & X & X\\
CUSR0000SAC & CPI : Commodities & 5 &  & X & X\\
CUMFNS & Capacity Utilization: Manufacturing & 4 &  &  & X\\
CE16OV & Civilian Employment & 5 &  &  & X\\
CLAIMSx & Initial Claims & 4 &  &  & X\\
RPI & Real Personal Income & 5 &  &  & X\\
CES3000000008 & Avg Hourly Earnings : Manufacturing & 5 &  &  & X\\
HOUST & Housing Starts: Total New Privately Owned & 5 &  &  & X\\
S\&P PE ratio & S\&P s Composite Common Stock: Price-Earnings Ratio & 4 &  &  & X\\
DPCERA3M086SBEA & Real personal consumption expenditures & 5 &  &  & X\\
RETAILx & Retail and Food Services Sales & 5 &  &  & X\\
TOTRESNS & Total Reserves of Depository Institutions & 5 &  &  & X\\
NONREVSL & Total Nonrevolving Credit & 5 &  &  & X\\
GS10 & 10-Year Treasury Rate & 1 &  &  & X\\
BAAFFM & Moody s Baa Corporate Bond Minus FEDFUNDS & 1 &  &  & X\\
EXJPUSx & Japan / U.S. Foreign Exchange Rate & 1 &  &  & X\\
EXUSUKx & U.S. / U.K. Foreign Exchange Rate & 1 &  &  & X\\
WPSFD49207 & PPI: Finished Goods & 5 &  &  & X\\
OILPRICEx & Crude Oil, spliced WTI and Cushing & 5 &  &  & X\\
DSERRG3M086SBEA & Personal Cons. Exp: Services & 5 &  &  & X\\
\bottomrule\bottomrule
\end{tabular}
\end{table}
\end{landscape}

\section{Additional Empirical Results}

\begin{table}[!t]
\centering
\begin{threeparttable}
\caption{\label{tab:mse_sim}Relative MSE scores for predictive distributions.}
\centering
\fontsize{9}{11}\selectfont
\begin{tabular}[t]{l>{}l>{\raggedleft\arraybackslash}p{1.5cm}>{\raggedright\arraybackslash}p{.5cm}>{\raggedleft\arraybackslash}p{1.5cm}>{\raggedright\arraybackslash}p{.5cm}>{\raggedleft\arraybackslash}p{1.5cm}>{\raggedright\arraybackslash}p{.5cm}>{\raggedleft\arraybackslash}p{1.5cm}>{\raggedright\arraybackslash}p{.5cm}}
\toprule\toprule
\multicolumn{1}{c}{\textbf{}} & \multicolumn{1}{c}{\textbf{}} & \multicolumn{2}{c}{\textbf{Gaussian}} & \multicolumn{2}{c}{\textbf{Student}} & \multicolumn{2}{c}{\textbf{SV}} & \multicolumn{2}{c}{\textbf{MA}} \\
\cmidrule(l{3pt}r{3pt}){3-4} \cmidrule(l{3pt}r{3pt}){5-6} \cmidrule(l{3pt}r{3pt}){7-8} \cmidrule(l{3pt}r{3pt}){9-10}
 &  & rel. MSE &  & rel. MSE &  & rel. MSE &  & rel. MSE & \\
\midrule
\addlinespace[.1em]
\multicolumn{10}{c}{\textit{One-month-ahead predictions}}\\
\multirow{12}{*}{\rotatebox[origin=c]{90}{\normalsize \textbf{Small-sized}}} & \textbf{Linear} & 1.01 &  & 1.02 &  & 1.01 &  & 1.01 & \\
 & \textbf{Break} & 0.99 & *** & 1.00 &  & 0.99 & *** & 0.99 & ***\\
 & \textbf{Exo} & 1.00 &  & 1.03 &  & 1.01 &  & 1.01 & \\
 & \textbf{Transition} & 1.03 &  & 1.05 &  & 1.03 &  & 1.01 & \\
\cmidrule{2-10}
\addlinespace[.1em]
\multicolumn{10}{c}{\textit{One-quarter-ahead predictions}}\\
 & \textbf{Linear} & 1.01 &  & 1.01 &  & 1.01 &  & 1.01 & \\
 & \textbf{Break} & 0.98 & *** & 0.98 & *** & 0.98 & *** & 0.98 & ***\\
 & \textbf{Exo} & 0.98 & ** & 1.02 &  & 0.99 & ** & 1.00 & \\
 & \textbf{Transition} & 1.03 &  & 1.06 &  & 1.03 &  & 1.02 & \\
\cmidrule{2-10}
\addlinespace[.1em]
\multicolumn{10}{c}{\textit{One-year-ahead predictions}}\\
 & \textbf{Linear} & 1.04 &  & 1.05 &  & 1.05 &  & 1.00 & \\
 & \textbf{Break} & 0.98 & *** & 0.99 &  & 0.99 & *** & 0.99 & ***\\
 & \textbf{Exo} & 0.96 & *** & 1.00 &  & 0.96 & *** & 0.99 & *\\
 & \textbf{Transition} & 1.03 &  & 1.11 &  & 1.03 &  & 0.99 & ***\\
\midrule
\addlinespace[.1em]
\multicolumn{10}{c}{\textit{One-month-ahead predictions}}\\
\multirow{12}{*}{\rotatebox[origin=c]{90}{\normalsize \textbf{Medium-sized}}} & \textbf{Linear} & 0.99 & *** & 1.00 &  & 1.00 &  & 1.02 & \\
 & \textbf{Break} & 0.92 & *** & 0.91 & *** & 0.92 & *** & 0.97 & ***\\
 & \textbf{Exo} & 0.99 & ** & 0.98 & *** & 0.98 & *** & 1.04 & \\
 & \textbf{Transition} & 1.01 &  & 1.04 &  & 1.01 &  & 1.02 & \\
\cmidrule{2-10}
\addlinespace[.1em]
\multicolumn{10}{c}{\textit{One-quarter-ahead predictions}}\\
 & \textbf{Linear} & 0.98 & *** & 0.98 & *** & 0.99 & *** & 0.98 & ***\\
 & \textbf{Break} & 0.89 & *** & 0.88 & *** & 0.89 & *** & 0.93 & ***\\
 & \textbf{Exo} & 0.95 & *** & 0.95 & *** & 0.95 & *** & 0.99 & \\
 & \textbf{Transition} & 1.01 &  & 1.07 &  & 1.02 &  & 1.02 & \\
\cmidrule{2-10}
\addlinespace[.1em]
\multicolumn{10}{c}{\textit{One-year-ahead predictions}}\\
 & \textbf{Linear} & 1.02 &  & 1.05 &  & 1.03 &  & 0.98 & ***\\
 & \textbf{Break} & 0.97 & *** & 1.02 &  & 1.00 &  & 0.98 & ***\\
 & \textbf{Exo} & 0.92 & *** & 0.98 &  & 0.95 & ** & 0.92 & ***\\
 & \textbf{Transition} & 1.02 &  & 1.18 &  & 1.04 &  & 1.00 & *\\
\midrule
\addlinespace[.1em]
\multicolumn{10}{c}{\textit{One-month-ahead predictions}}\\
\multirow{12}{*}{\rotatebox[origin=c]{90}{\normalsize \textbf{Large-sized}}} & \textbf{Linear} & 0.96 & *** & 0.96 & *** & 0.96 & *** & 0.96 & ***\\
 & \textbf{Break} & 0.94 & *** & 0.94 & *** & 0.94 & *** & 0.92 & ***\\
 & \textbf{Exo} & 0.92 & *** & 0.93 & *** & 0.92 & *** & 0.90 & ***\\
 & \textbf{Transition} & 1.01 &  & 1.02 &  & 1.01 &  & 1.02 & \\
\cmidrule{2-10}
\addlinespace[.1em]
\multicolumn{10}{c}{\textit{One-quarter-ahead predictions}}\\
 & \textbf{Linear} & 0.90 & *** & 0.90 & *** & 0.90 & *** & 0.87 & ***\\
 & \textbf{Break} & 0.93 & *** & 0.94 & *** & 0.92 & *** & 0.91 & ***\\
 & \textbf{Exo} & 0.88 & *** & 0.87 & *** & 0.88 & *** & 0.86 & ***\\
 & \textbf{Transition} & 1.01 &  & 1.02 &  & 1.01 &  & 1.00 & ***\\
\cmidrule{2-10}
\addlinespace[.1em]
\multicolumn{10}{c}{\textit{One-year-ahead predictions}}\\
 & \textbf{Linear} & 0.96 & *** & 0.96 & *** & 0.97 & *** & 0.93 & ***\\
 & \textbf{Break} & 0.99 & ** & 1.01 &  & 0.99 & *** & 0.98 & ***\\
 & \textbf{Exo} & 0.91 & *** & 0.92 & *** & 0.92 & *** & 0.88 & ***\\
 & \textbf{Transition} & 1.00 & *** & 1.03 &  & 1.00 &  & 0.99 & ***\\
\bottomrule\bottomrule
\end{tabular}
\begin{tablenotes}
\small
\item [] \footnotesize \textbf{Notes:} Results are averaged over all draws and test sets for each DGP.
  Stars indicate p-values from a one-sided t-test with alternative hypothesis
  that cBVAR produces smaller MSE scores.
\item [] \footnotesize \textbf{Legend:} $.\sim p<0.16$; $^{*}\sim p<0.1$; $^{**}\sim p<0.05$;
  $^{***}\sim p<0.01$
\end{tablenotes}
\end{threeparttable}
\end{table}

\autoref{tab:mse_sim} contains mean squared errors (based on the posterior median) for the Monte Carlo exercise. Again, results for the cBVAR are in relative terms such that scores smaller than one reflect forecast gains.

\begin{figure}[!t]
    \centering
    \begin{subfigure}[b]{0.3\textwidth}
        \centering
        \textbf{Small-sized model}
    \end{subfigure}
    \hfill
    \begin{subfigure}[b]{0.3\textwidth}
        \centering
        \textbf{Medium-sized model}
    \end{subfigure}
    \hfill
    \begin{subfigure}[b]{0.3\textwidth}
        \centering
        \textbf{Large-sized model}
    \end{subfigure}
     % UNRATE
     \begin{subfigure}[b]{0.3\textwidth}
         \centering
         \includegraphics[width=\textwidth]{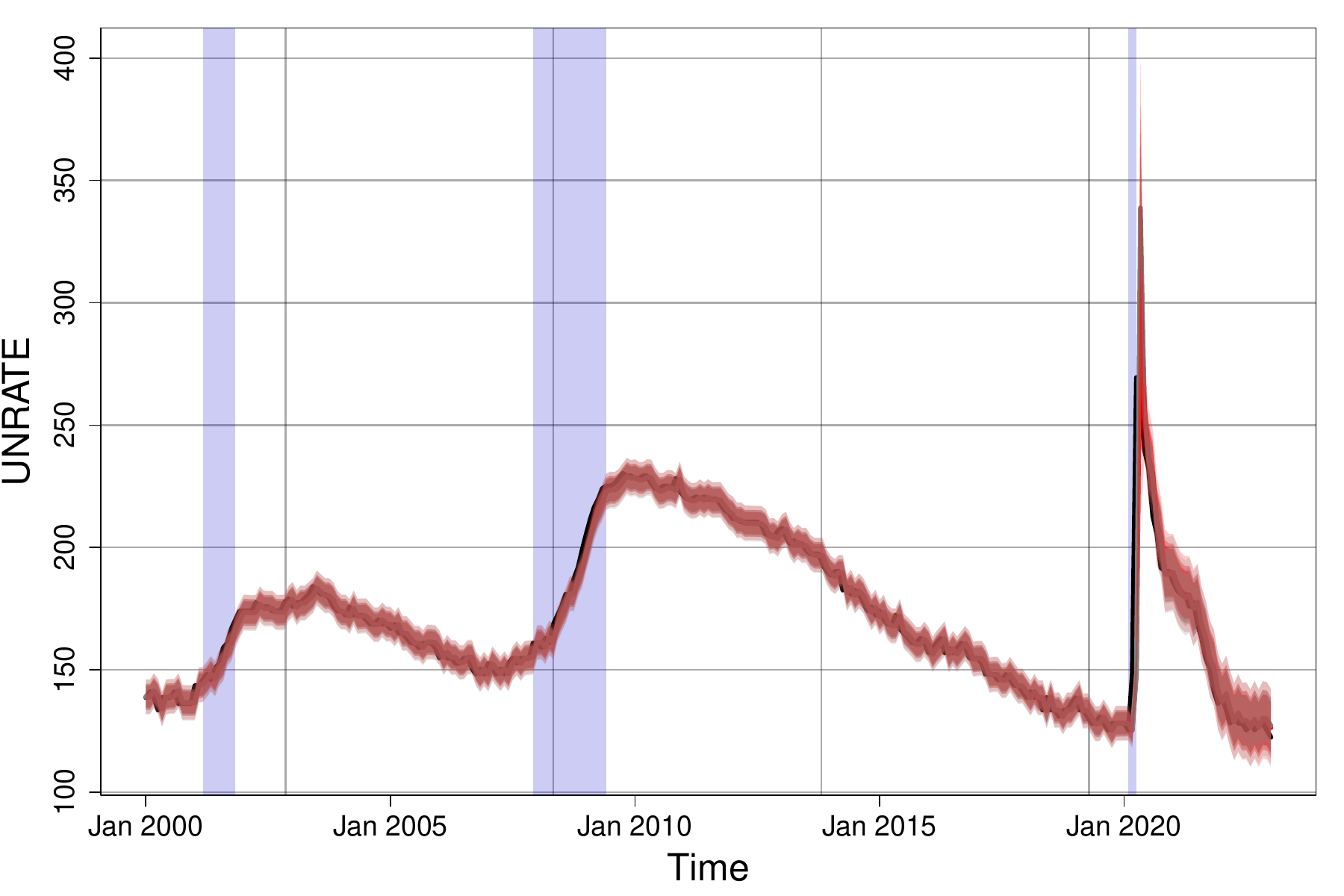}
         \caption{UNRATE}
         \label{fig:pred_dist_UNRATE_small_1}
     \end{subfigure}
     \hfill
     \begin{subfigure}[b]{0.3\textwidth}
         \centering
         \includegraphics[width=\textwidth]{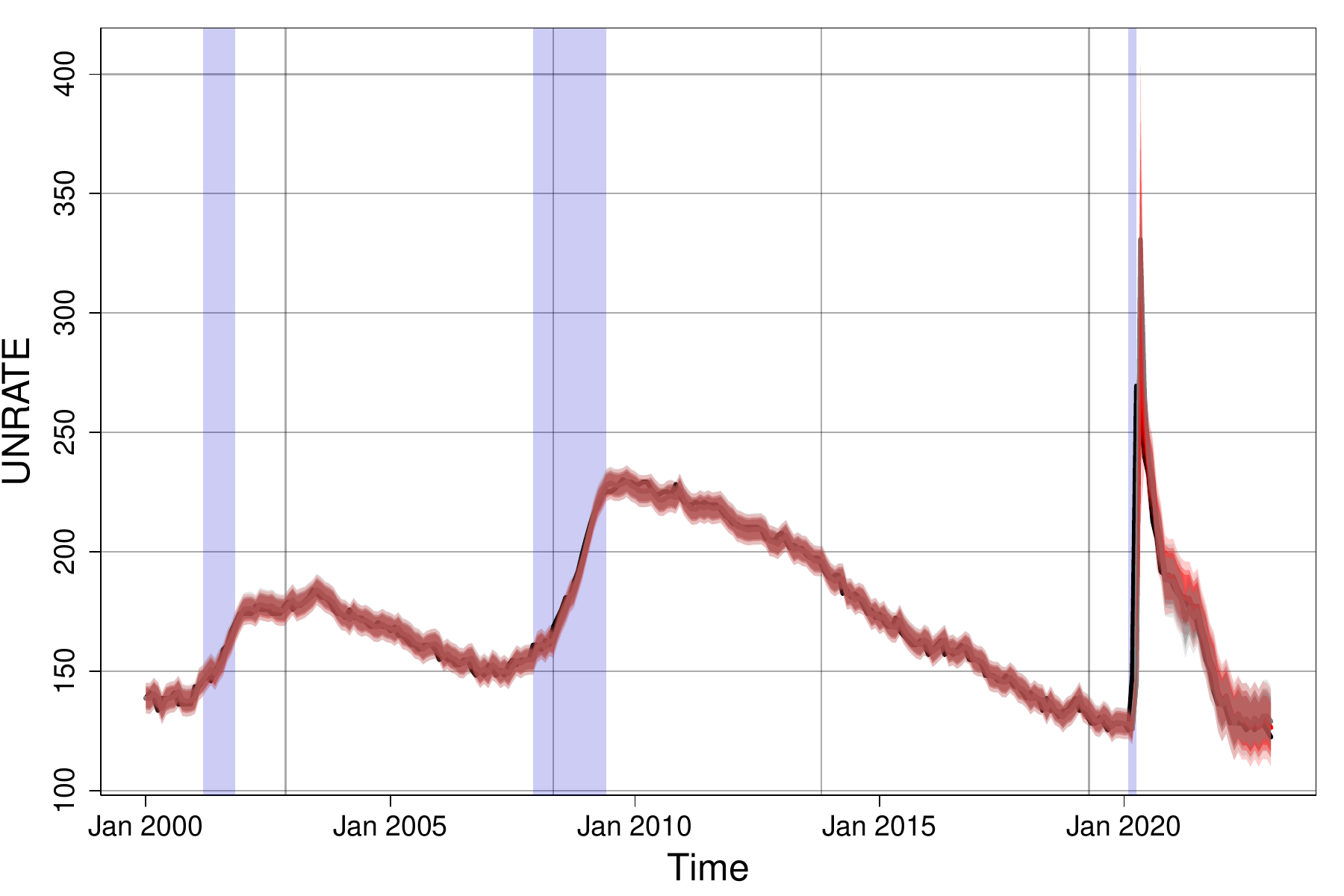}
         \caption{UNRATE}
         \label{fig:pred_dist_UNRATE_medium_1}
     \end{subfigure}
     \hfill
     \begin{subfigure}[b]{0.3\textwidth}
         \centering
         \includegraphics[width=\textwidth]{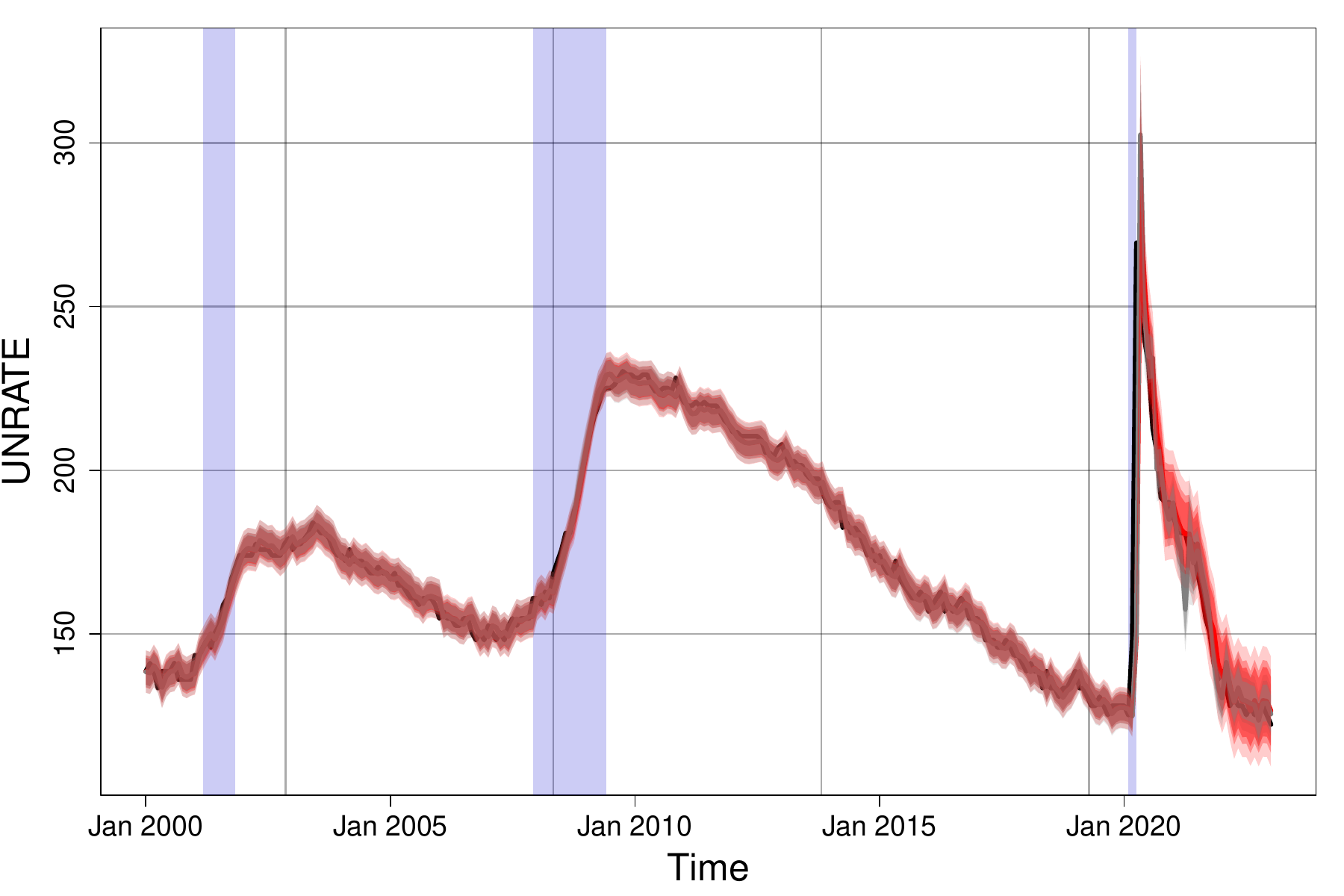}
         \caption{UNRATE}
         \label{fig:pred_dist_UNRATE_large_1}
     \end{subfigure}
     % CPIAUCSL
     \begin{subfigure}[b]{0.3\textwidth}
         \centering
         \includegraphics[width=\textwidth]{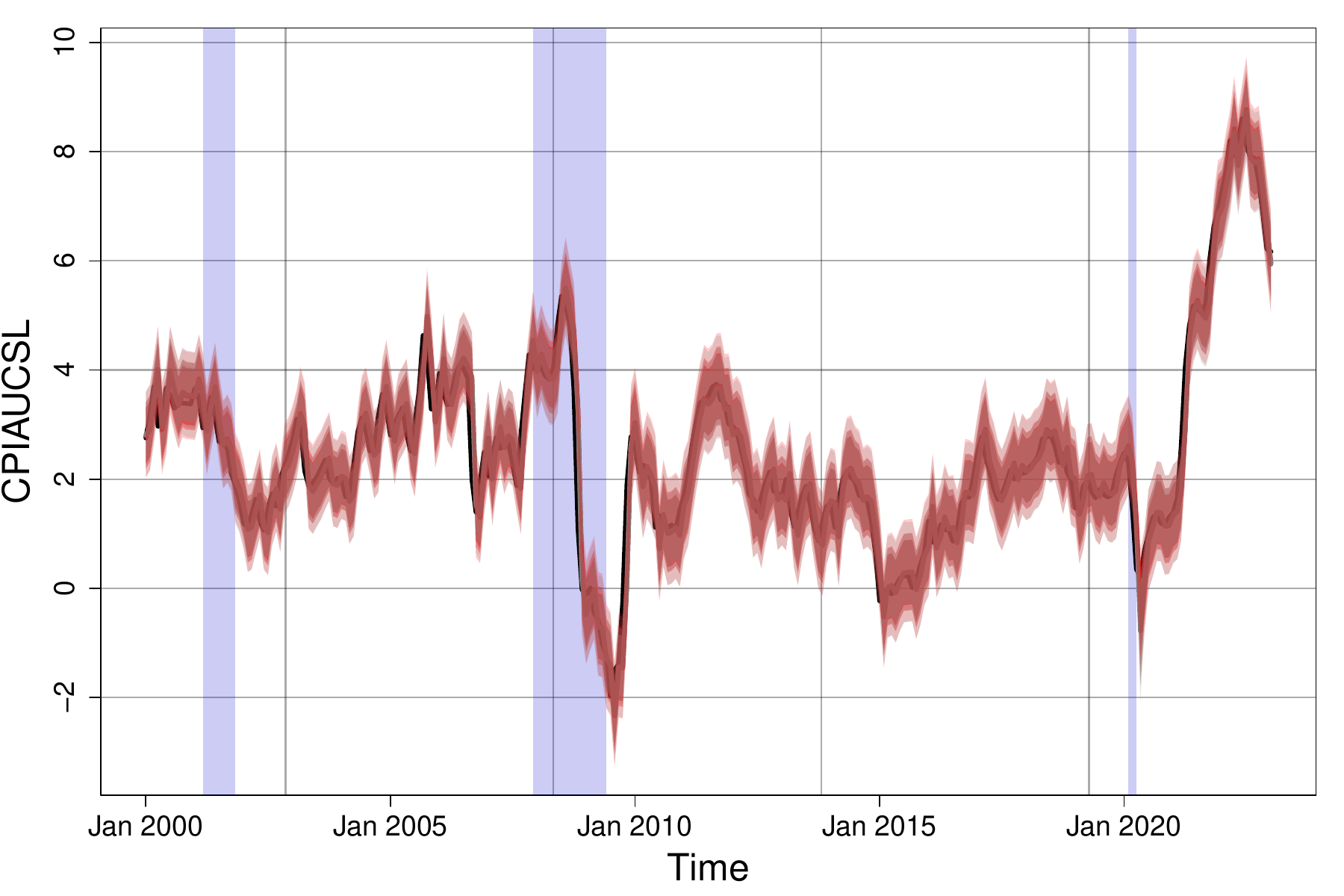}
         \caption{CPIAUCSL}
         \label{fig:pred_dist_CPIAUCSL_small_1}
     \end{subfigure}
     \hfill
     \begin{subfigure}[b]{0.3\textwidth}
         \centering
         \includegraphics[width=\textwidth]{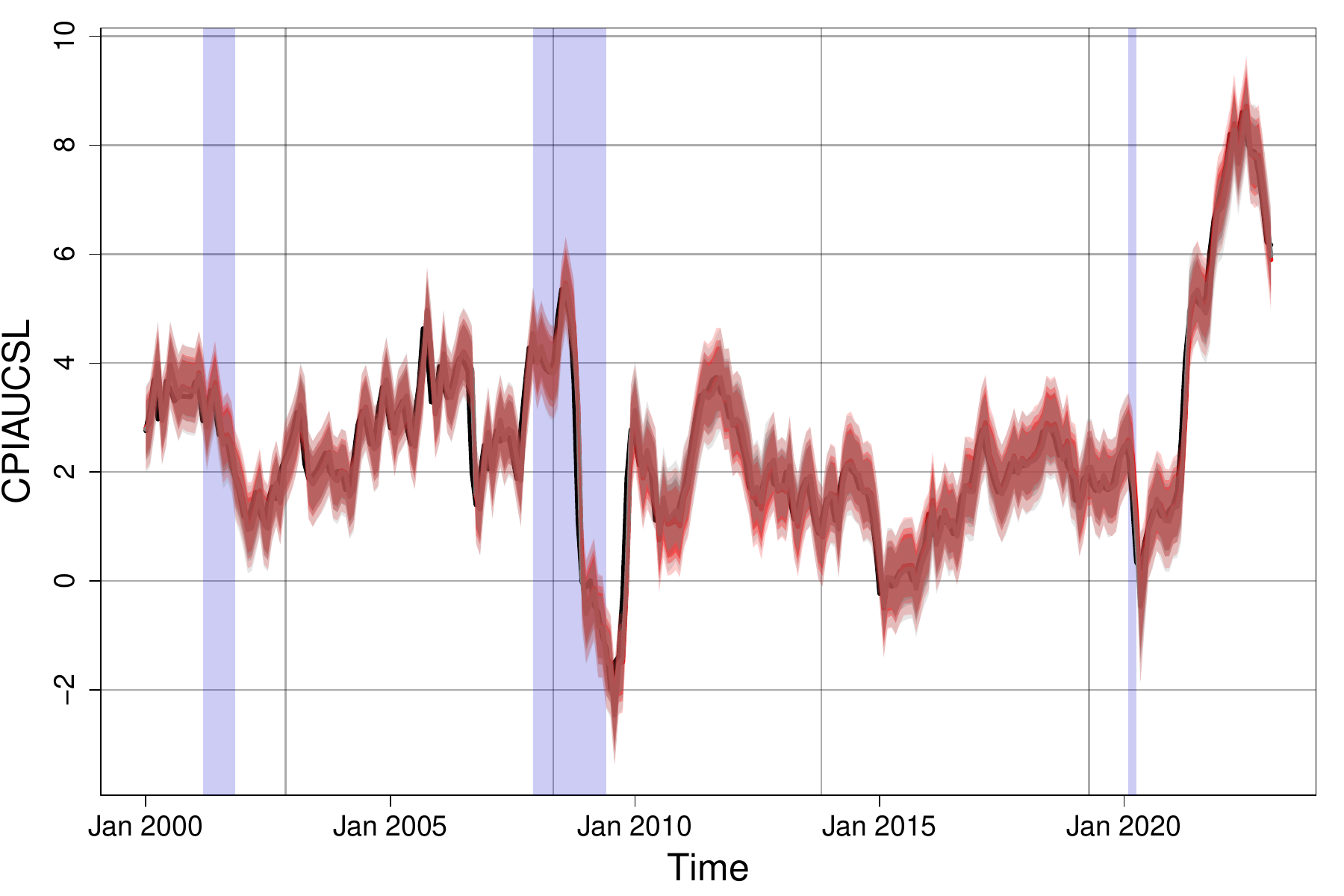}
         \caption{CPIAUCSL}
         \label{fig:pred_dist_CPIAUCSL_medium_1}
     \end{subfigure}
     \hfill
     \begin{subfigure}[b]{0.3\textwidth}
         \centering
         \includegraphics[width=\textwidth]{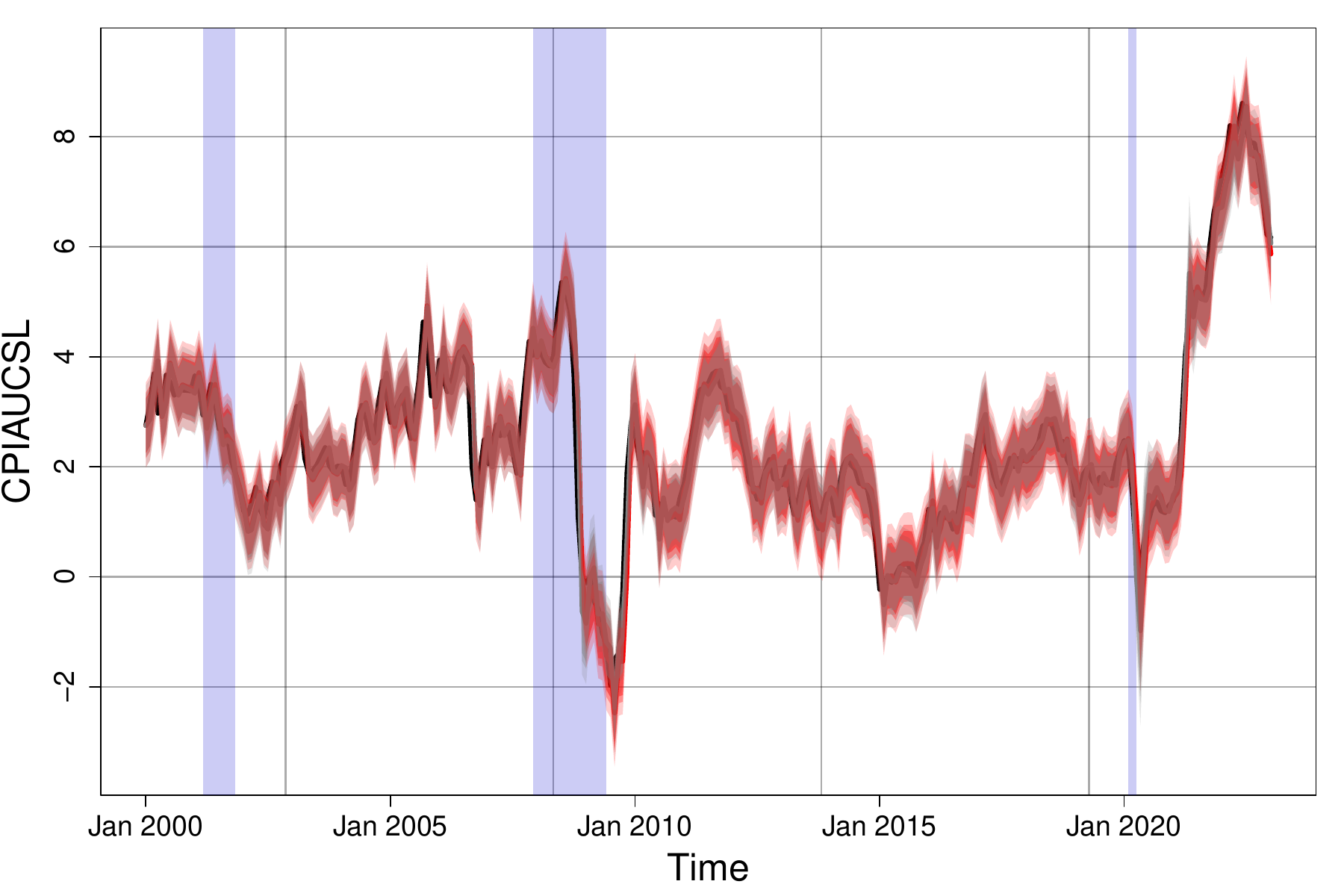} 
         \caption{CPIAUCSL}
         \label{fig:pred_dist_CPIAUCSL_large_1}
     \end{subfigure}
     % FEDFUNDS
     \begin{subfigure}[b]{0.3\textwidth}
         \centering
         \includegraphics[width=\textwidth]{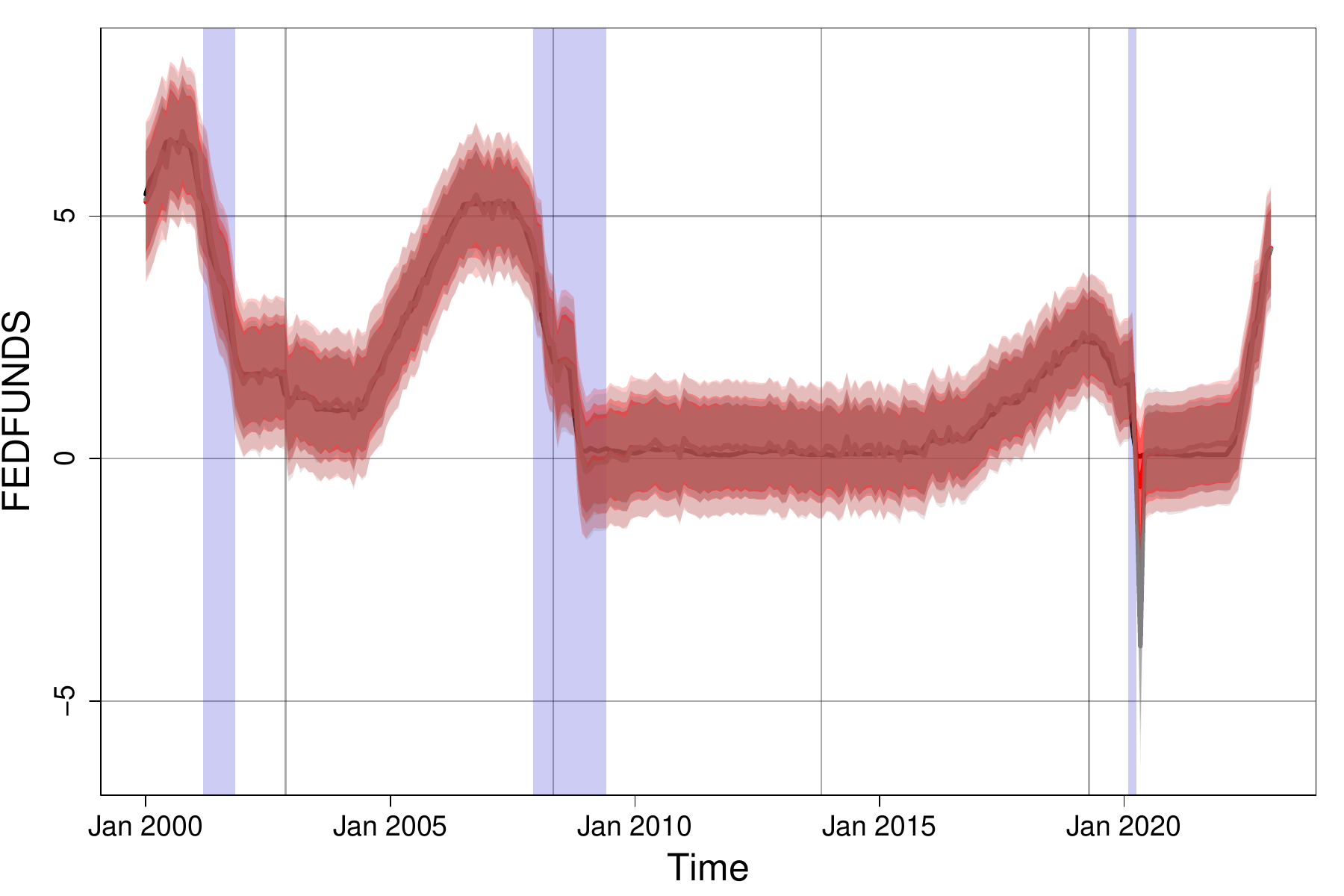}
         \caption{FEDFUNDS}
         \label{fig:pred_dist_FEDFUNDS_small_1}
     \end{subfigure}
     \hfill
     \begin{subfigure}[b]{0.3\textwidth}
         \centering
         \includegraphics[width=\textwidth]{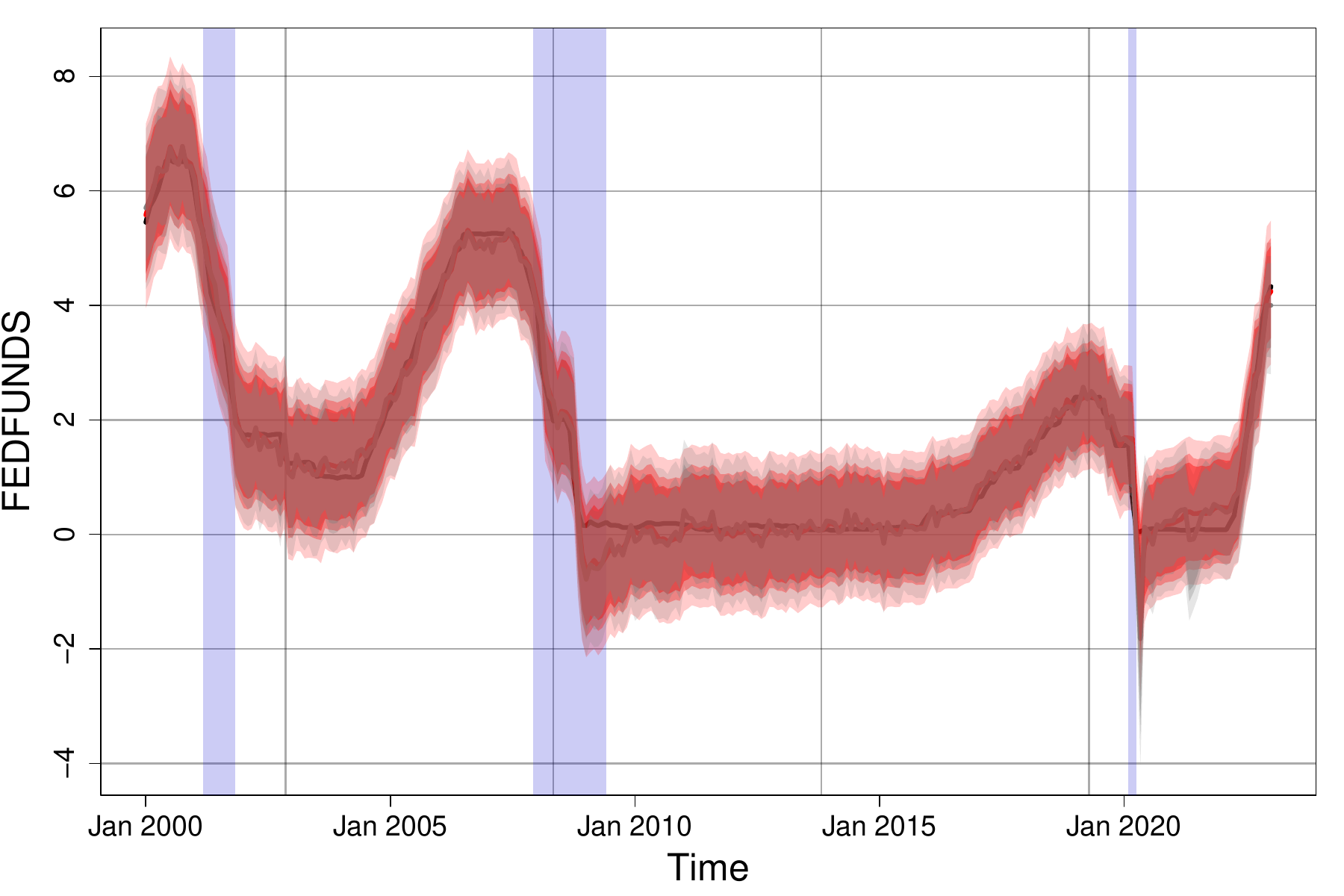}
         \caption{FEDFUNDS}
         \label{fig:pred_dist_FEDFUNDS_medium_1}
     \end{subfigure}
     \hfill
     \begin{subfigure}[b]{0.3\textwidth}
         \centering
         \includegraphics[width=\textwidth]{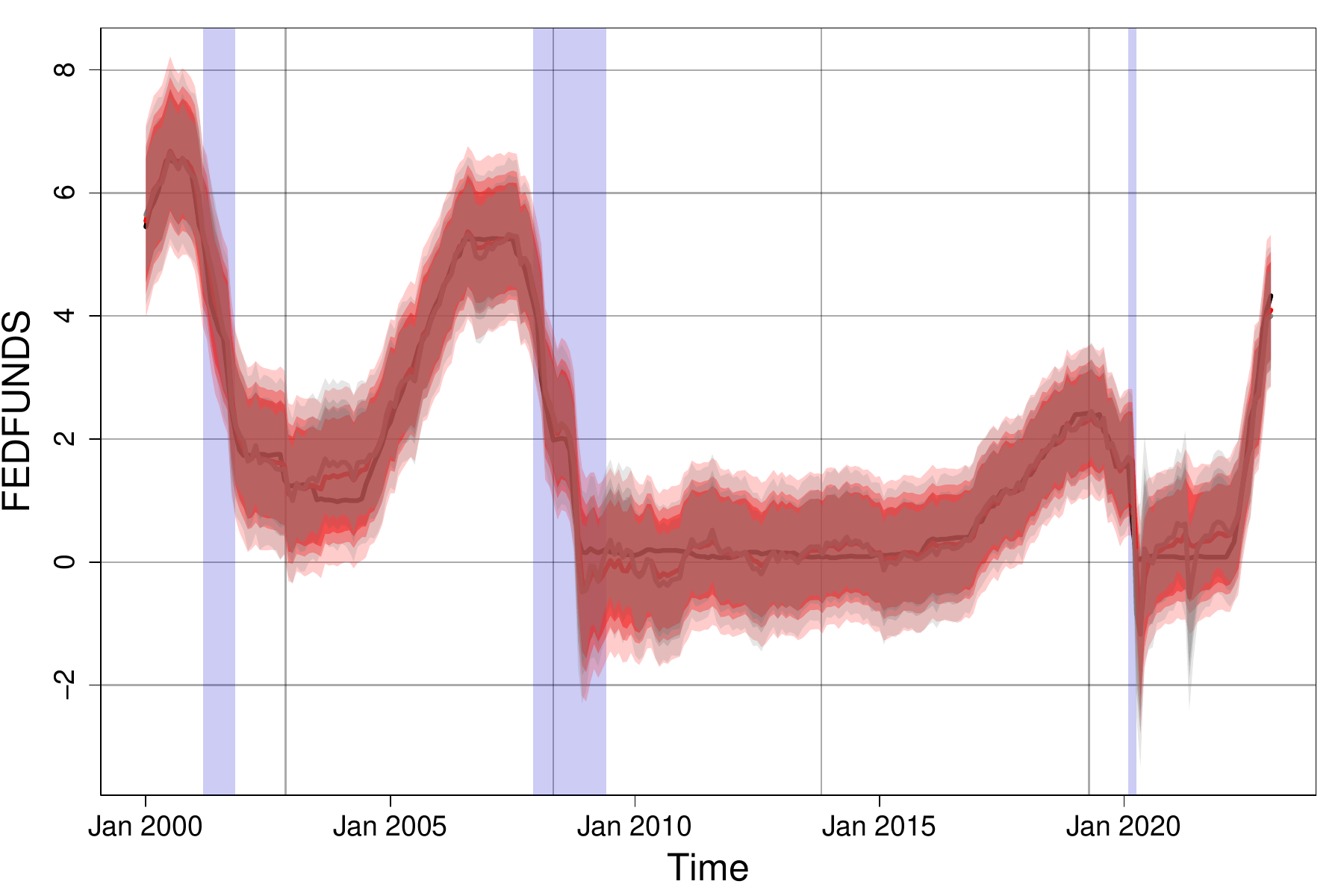}
         \caption{FEDFUNDS}
         \label{fig:pred_dist_FEDFUNDS_medium_1}
     \end{subfigure}
    \caption{Observed path versus one-month-ahead predictive distributions for each of the focus variables.\\
    \textbf{Legend:} \textcolor{black}{$-$} are the observed values, \textcolor{plotred}{$-$} indicates predictions of the cBVAR and \textcolor{plotgrey}{$-$} those of the standard BVAR. Lines denote the posterior median and shaded areas the 90\%, 95\% and 99\% credible intervals (from dark to light). Blue shaded areas indicate the NBER recession dates.}
    \label{fig:pred_dist_1}
\end{figure}

\begin{figure}[!t]
    \centering
    \begin{subfigure}[b]{0.3\textwidth}
        \centering
        \textbf{Small-sized model}
    \end{subfigure}
    \hfill
    \begin{subfigure}[b]{0.3\textwidth}
        \centering
        \textbf{Medium-sized model}
    \end{subfigure}
    \hfill
    \begin{subfigure}[b]{0.3\textwidth}
        \centering
        \textbf{Large-sized model}
    \end{subfigure}
     % UNRATE
     \begin{subfigure}[b]{0.3\textwidth}
         \centering
         \includegraphics[width=\textwidth]{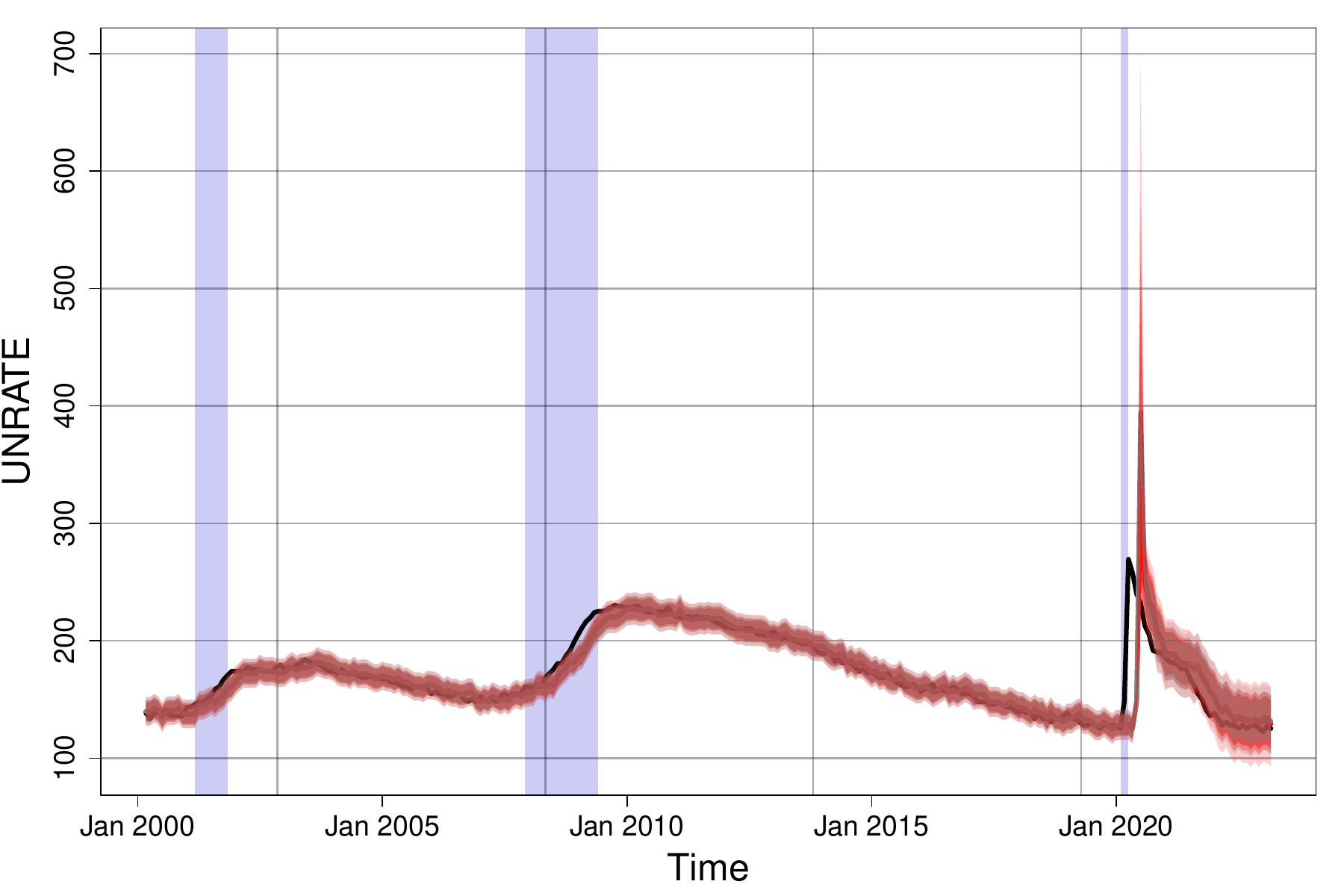}
         \caption{UNRATE}
         \label{fig:pred_dist_UNRATE_small_3}
     \end{subfigure}
     \hfill
     \begin{subfigure}[b]{0.3\textwidth}
         \centering
         \includegraphics[width=\textwidth]{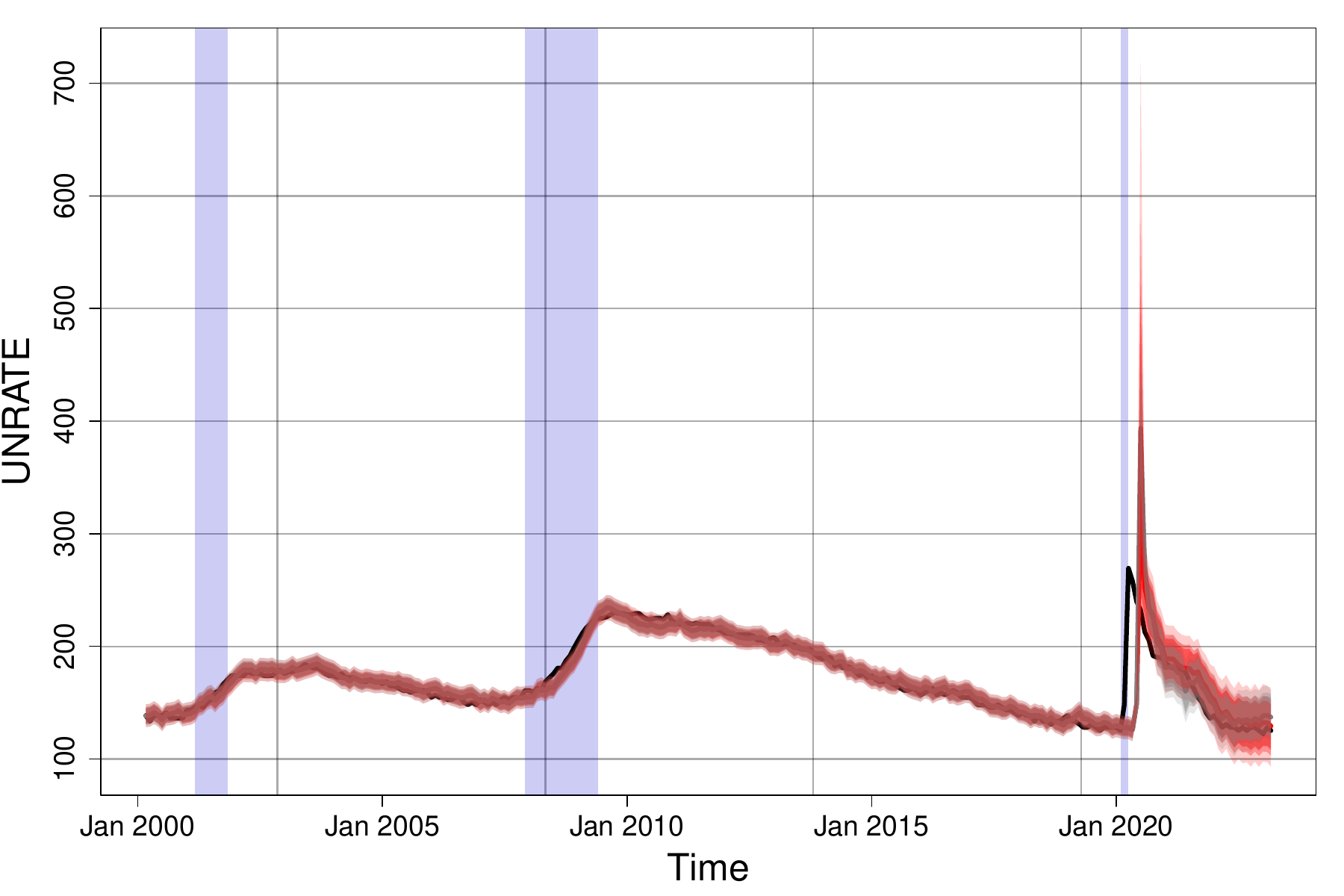}
         \caption{UNRATE}
         \label{fig:pred_dist_UNRATE_medium_3}
     \end{subfigure}
     \hfill
     \begin{subfigure}[b]{0.3\textwidth}
         \centering
         \includegraphics[width=\textwidth]{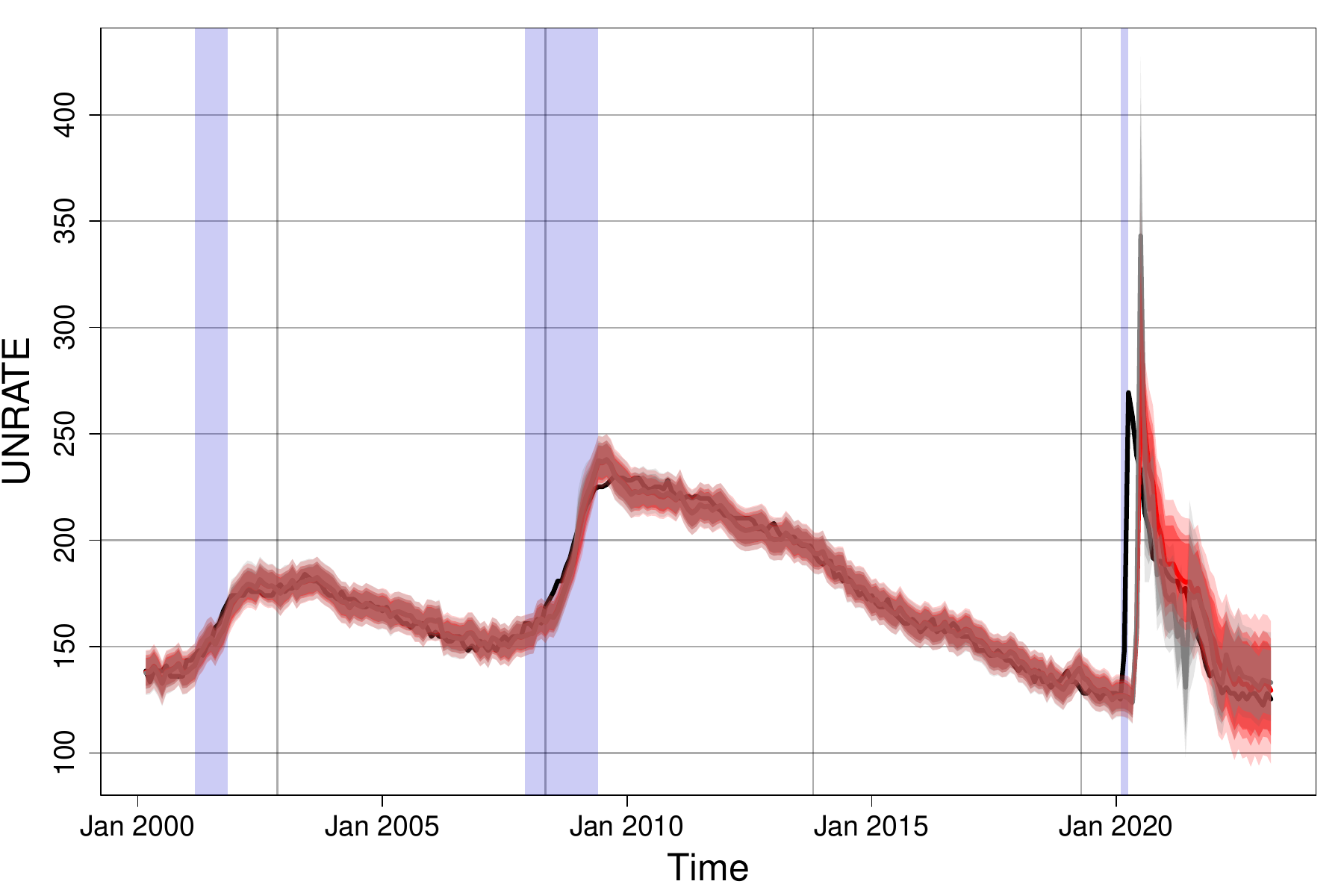}
         \caption{UNRATE}
         \label{fig:pred_dist_UNRATE_large_3}
     \end{subfigure}
     % CPIAUCSL
     \begin{subfigure}[b]{0.3\textwidth}
         \centering
         \includegraphics[width=\textwidth]{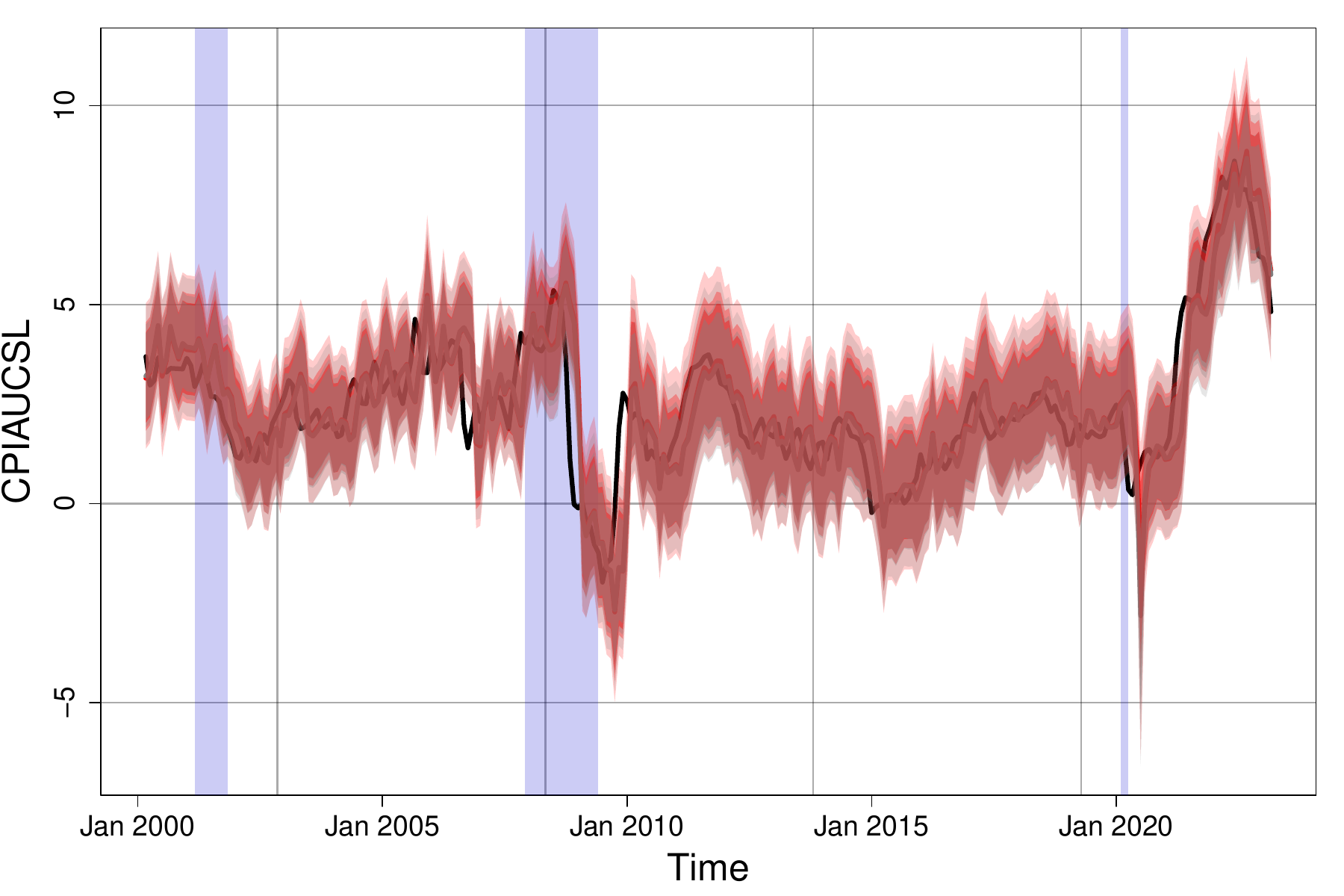}
         \caption{CPIAUCSL}
         \label{fig:pred_dist_CPIAUCSL_small_3}
     \end{subfigure}
     \hfill
     \begin{subfigure}[b]{0.3\textwidth}
         \centering
         \includegraphics[width=\textwidth]{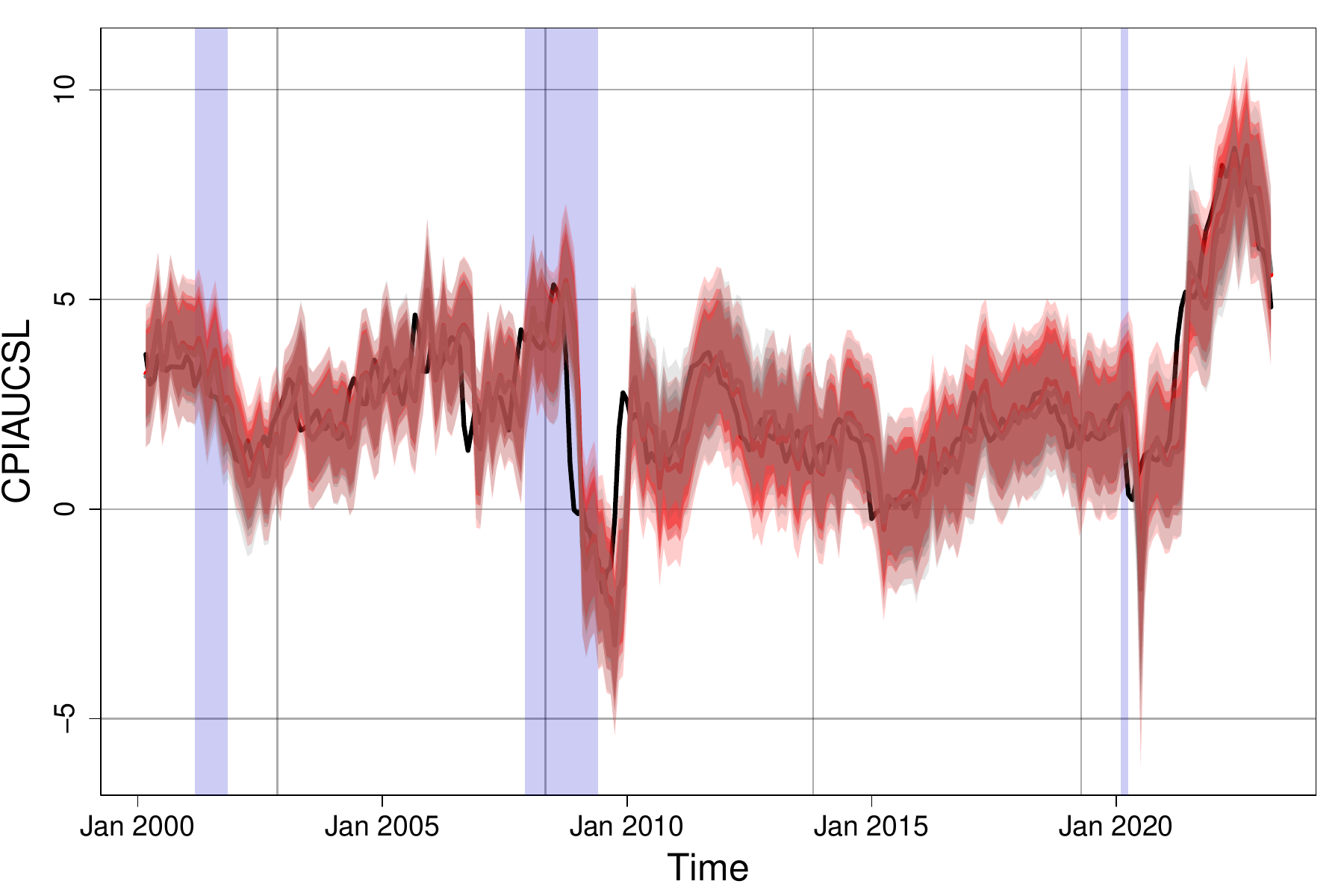}
         \caption{CPIAUCSL}
         \label{fig:pred_dist_CPIAUCSL_medium_3}
     \end{subfigure}
     \hfill
     \begin{subfigure}[b]{0.3\textwidth}
         \centering
         \includegraphics[width=\textwidth]{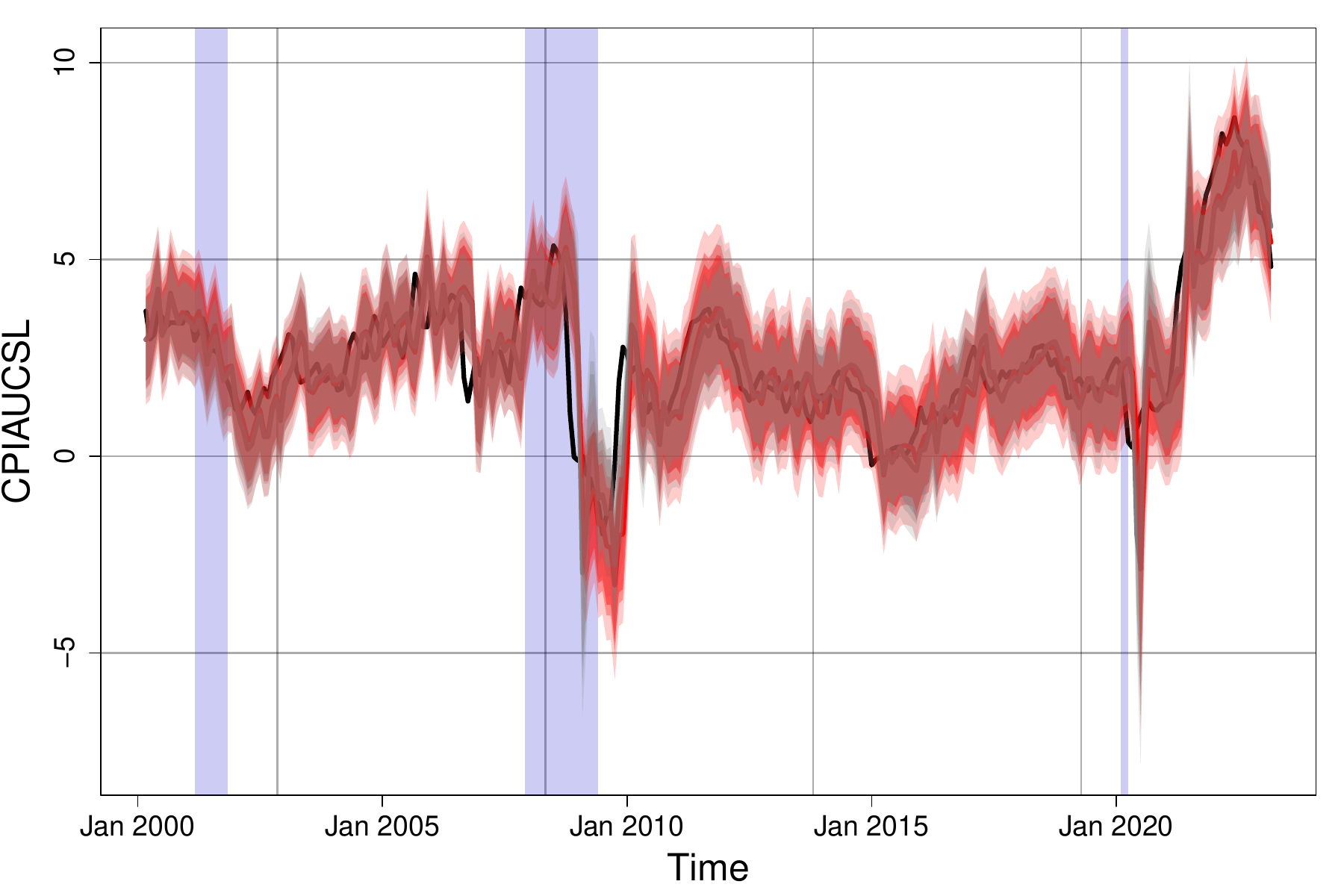}
         \caption{CPIAUCSL}
         \label{fig:pred_dist_CPIAUCSL_large_3}
     \end{subfigure}
     % FEDFUNDS
     \begin{subfigure}[b]{0.3\textwidth}
         \centering
         \includegraphics[width=\textwidth]{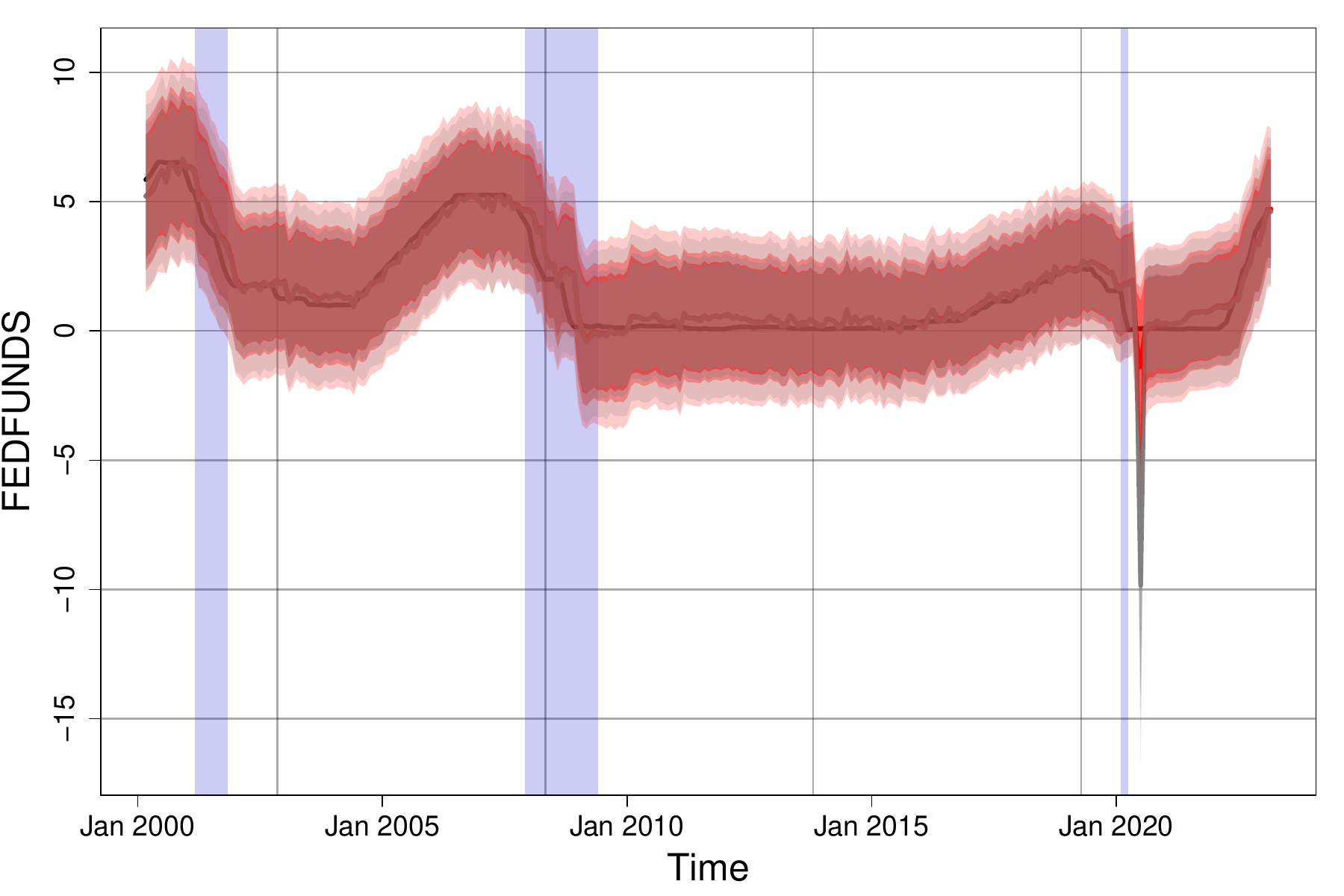}
         \caption{FEDFUNDS}
         \label{fig:pred_dist_FEDFUNDS_small_3}
     \end{subfigure}
     \hfill
     \begin{subfigure}[b]{0.3\textwidth}
         \centering
         \includegraphics[width=\textwidth]{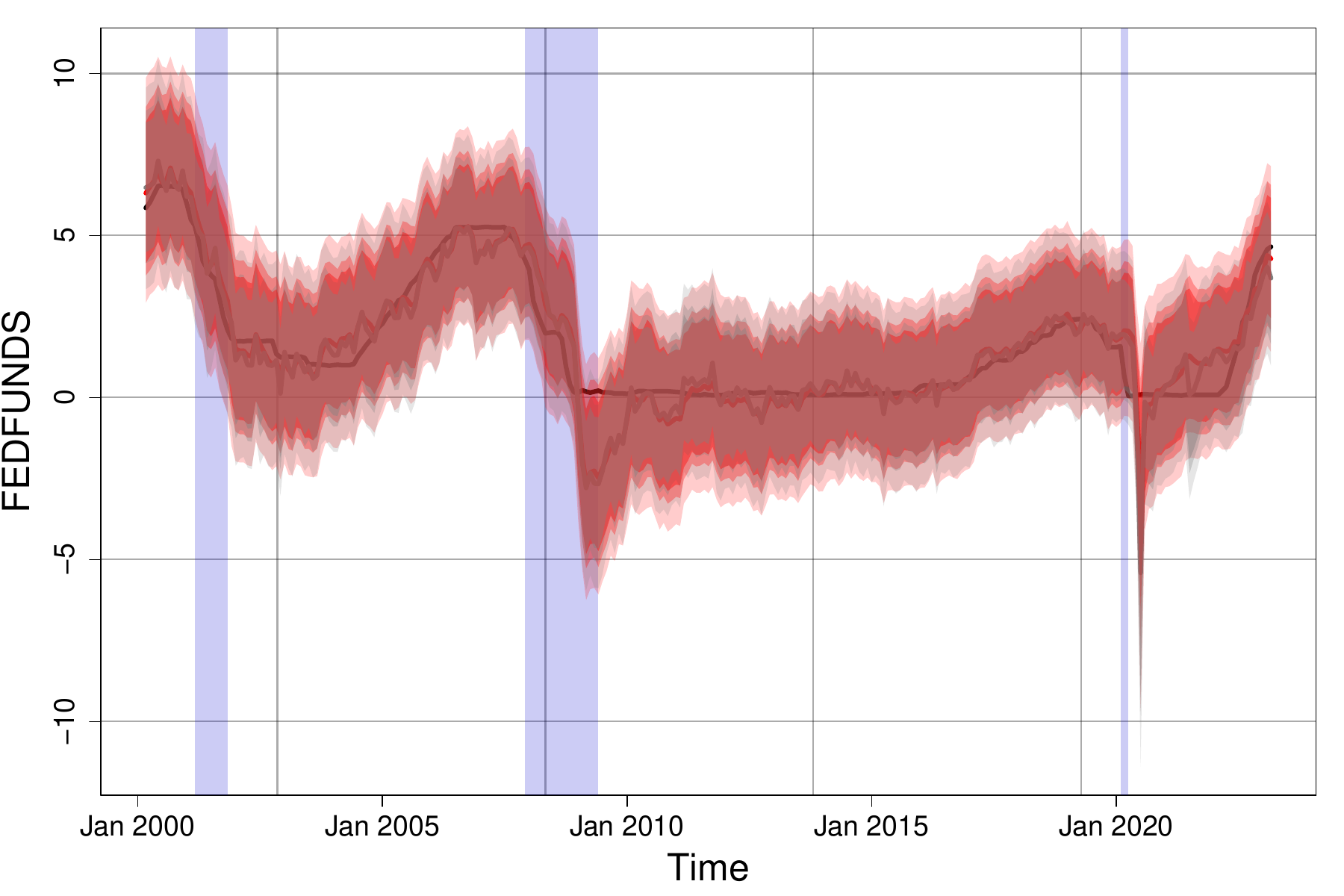}
         \caption{FEDFUNDS}
         \label{fig:pred_dist_FEDFUNDS_medium_3}
     \end{subfigure}
     \hfill
     \begin{subfigure}[b]{0.3\textwidth}
         \centering
         \includegraphics[width=\textwidth]{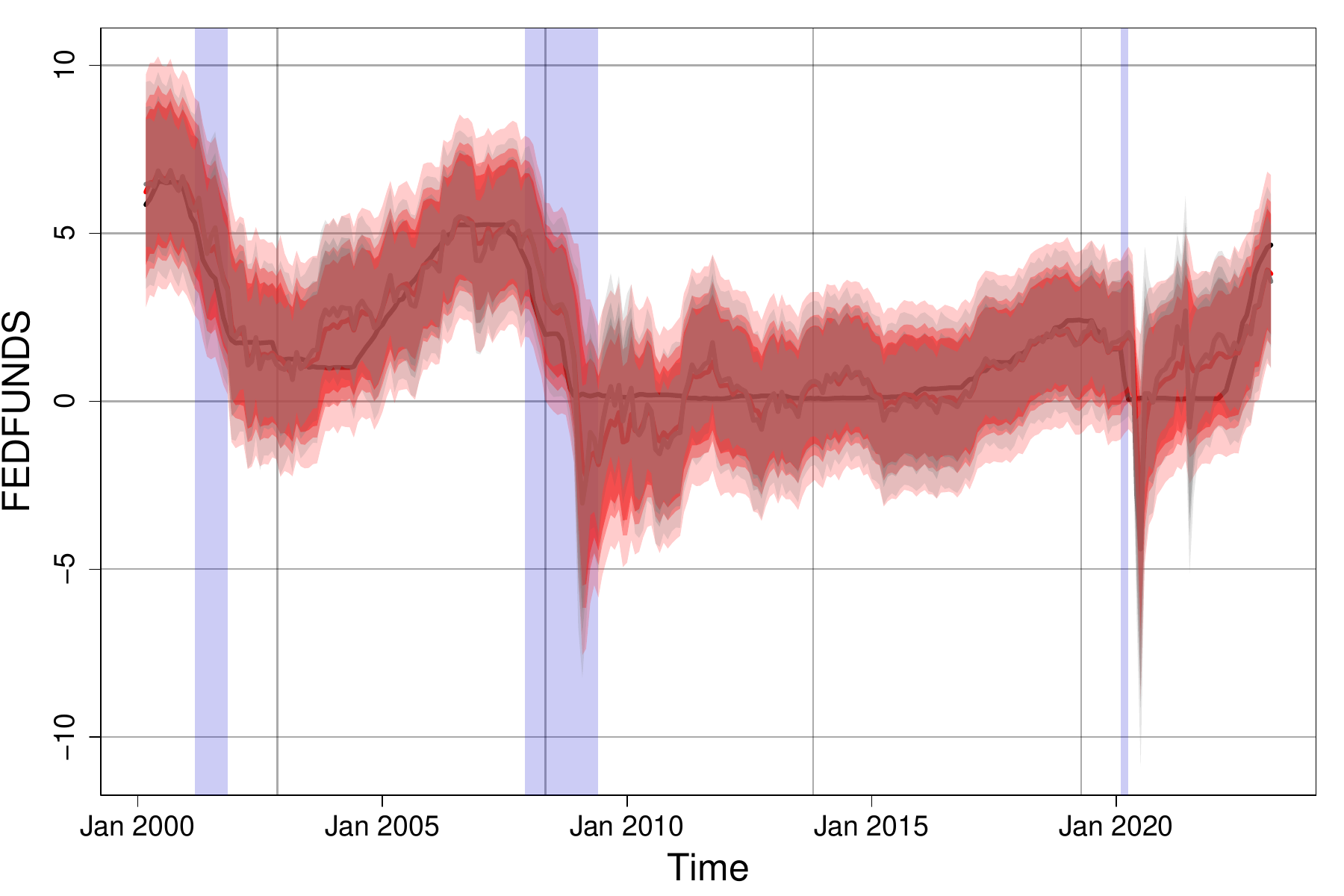}
         \caption{FEDFUNDS}
         \label{fig:pred_dist_FEDFUNDS_medium_3}
     \end{subfigure}
    \caption{Observed path versus one-quarter-ahead predictive distributions for each of the focus variables.\\
    \textbf{Legend:} \textcolor{red}{-} indicates predictions of the cBVAR and \textcolor{gray}{-} those of the standard BVAR. Lines denote the posterior median and shaded areas the 90\%, 95\% and 99\% credible intervals (from dark to light). Blue shaded areas indicate the NBER recession dates.}
    \label{fig:pred_dist_3}
\end{figure}

\begin{figure}[!t]
    \centering
    \begin{subfigure}[b]{0.3\textwidth}
        \centering
        \textbf{Small-sized model}
    \end{subfigure}
    \hfill
    \begin{subfigure}[b]{0.3\textwidth}
        \centering
        \textbf{Medium-sized model}
    \end{subfigure}
    \hfill
    \begin{subfigure}[b]{0.3\textwidth}
        \centering
        \textbf{Large-sized model}
    \end{subfigure}
     % UNRATE
     \begin{subfigure}[b]{0.3\textwidth}
         \centering
         \includegraphics[width=\textwidth]{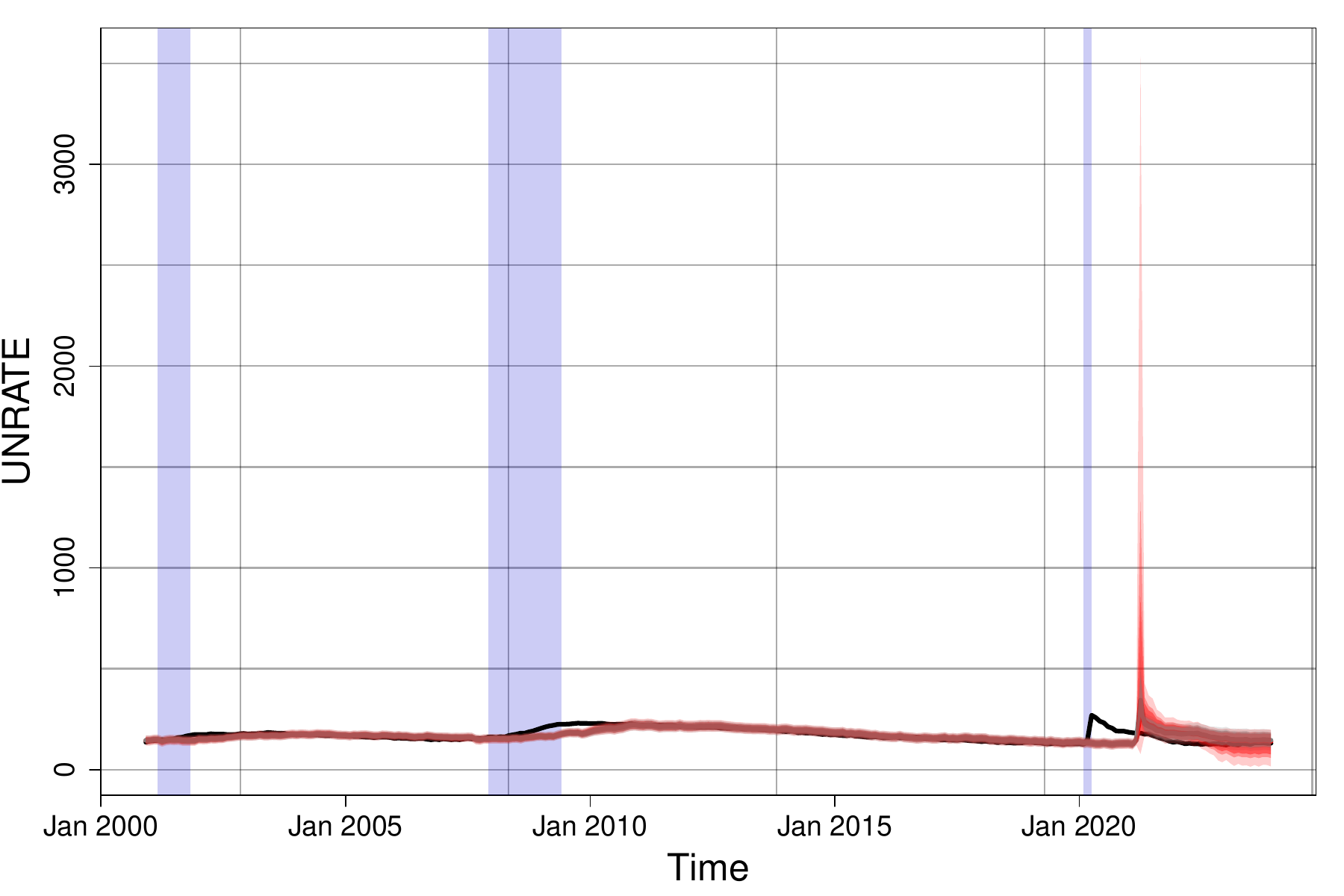}
         \caption{UNRATE}
         \label{fig:pred_dist_UNRATE_small_12}
     \end{subfigure}
     \hfill
     \begin{subfigure}[b]{0.3\textwidth}
         \centering
         \includegraphics[width=\textwidth]{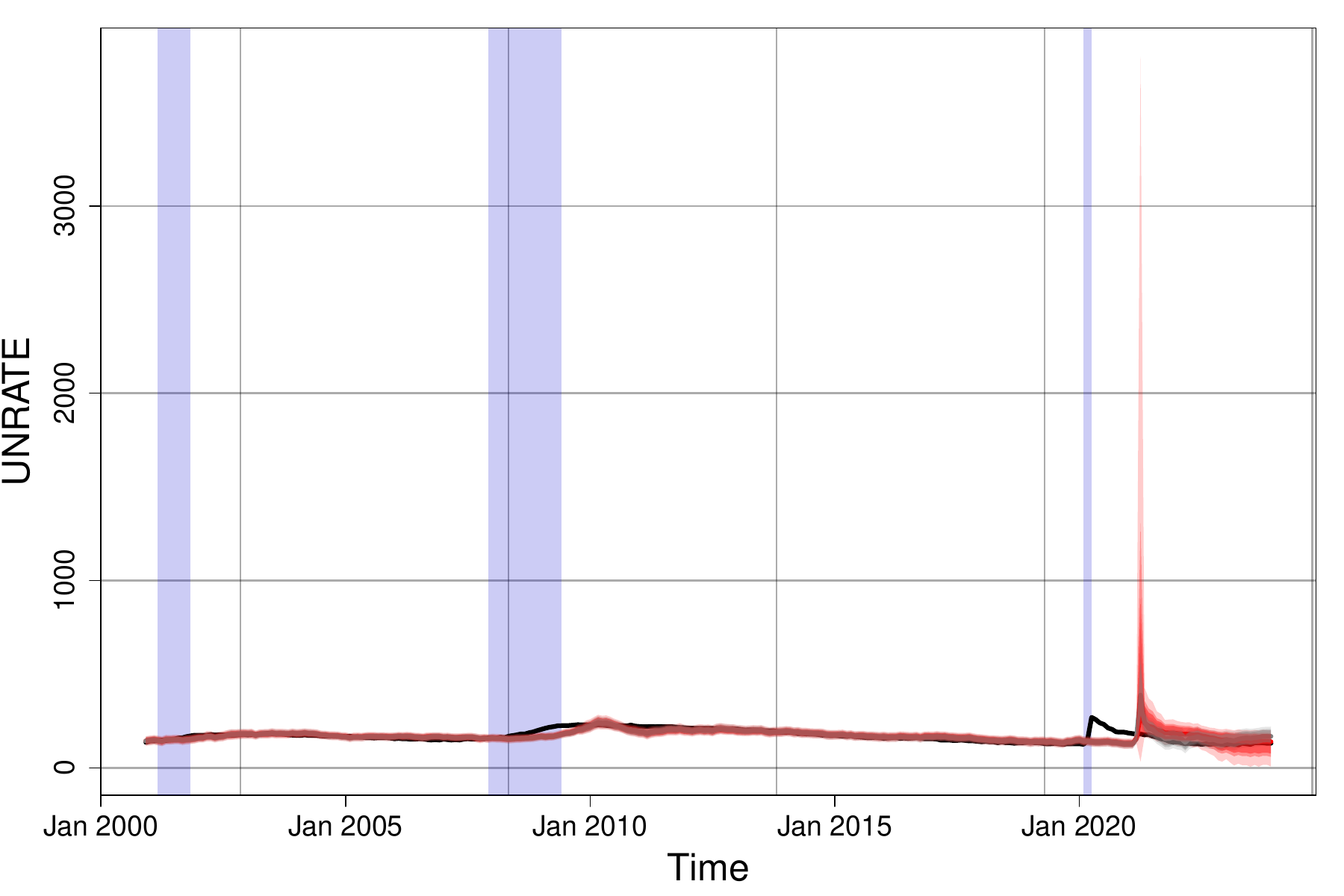}
         \caption{UNRATE}
         \label{fig:pred_dist_UNRATE_medium_12}
     \end{subfigure}
     \hfill
     \begin{subfigure}[b]{0.3\textwidth}
         \centering
         \includegraphics[width=\textwidth]{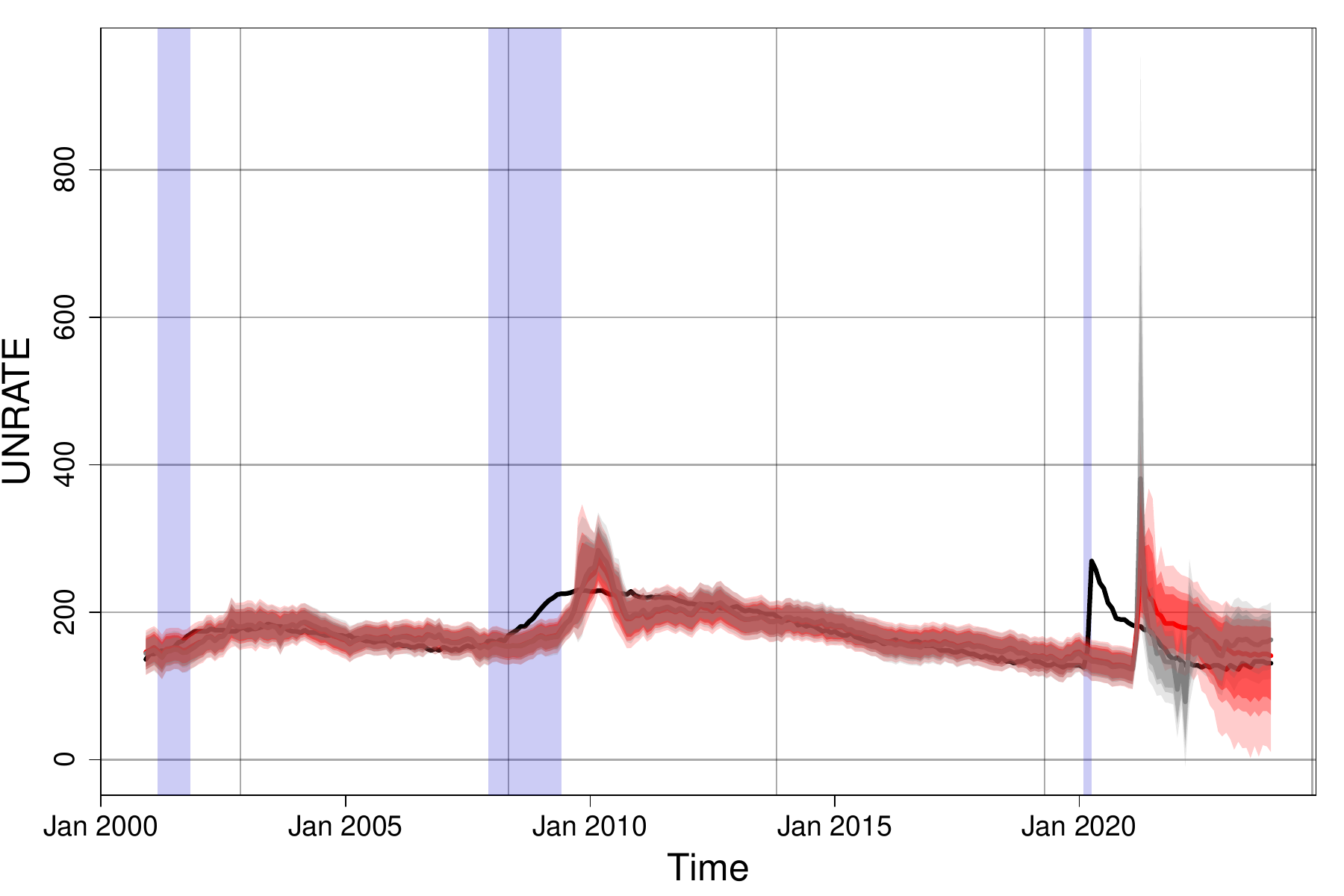}
         \caption{UNRATE}
         \label{fig:pred_dist_UNRATE_large_12}
     \end{subfigure}
     % CPIAUCSL
     \begin{subfigure}[b]{0.3\textwidth}
         \centering
         \includegraphics[width=\textwidth]{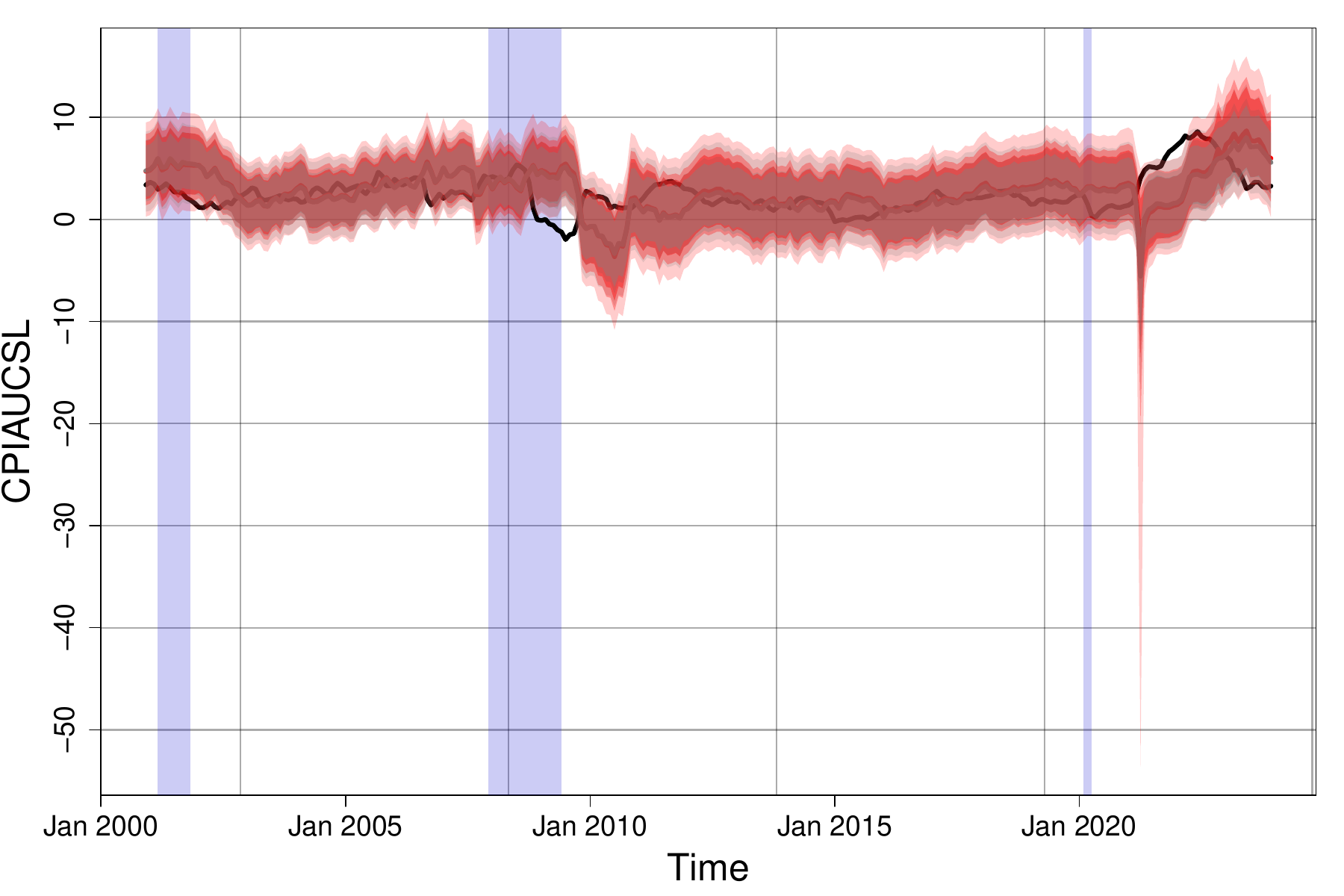}
         \caption{CPIAUCSL}
         \label{fig:pred_dist_CPIAUCSL_small_12}
     \end{subfigure}
     \hfill
     \begin{subfigure}[b]{0.3\textwidth}
         \centering
         \includegraphics[width=\textwidth]{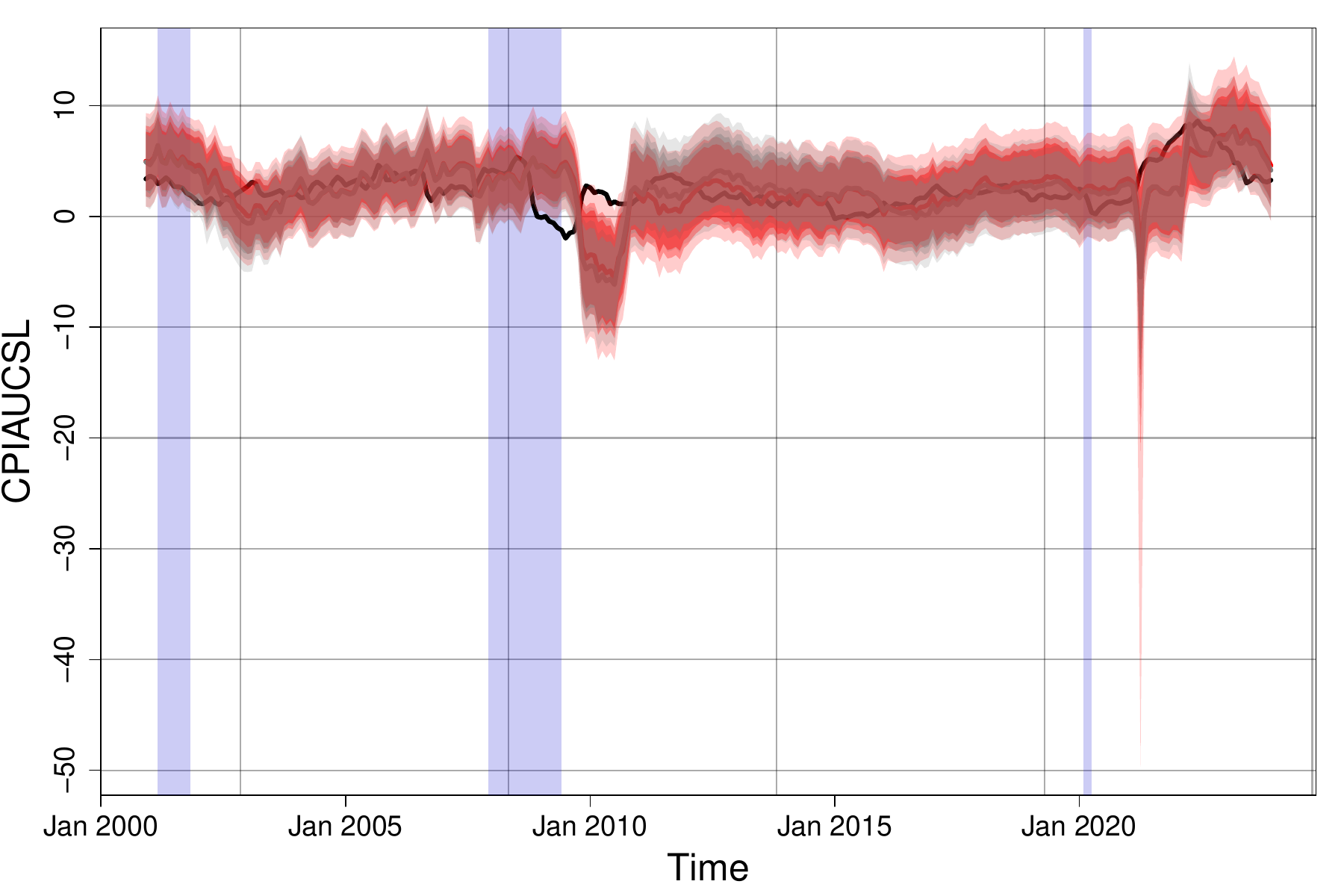}
         \caption{CPIAUCSL}
         \label{fig:pred_dist_CPIAUCSL_medium_12}
     \end{subfigure}
     \hfill
     \begin{subfigure}[b]{0.3\textwidth}
         \centering
         \includegraphics[width=\textwidth]{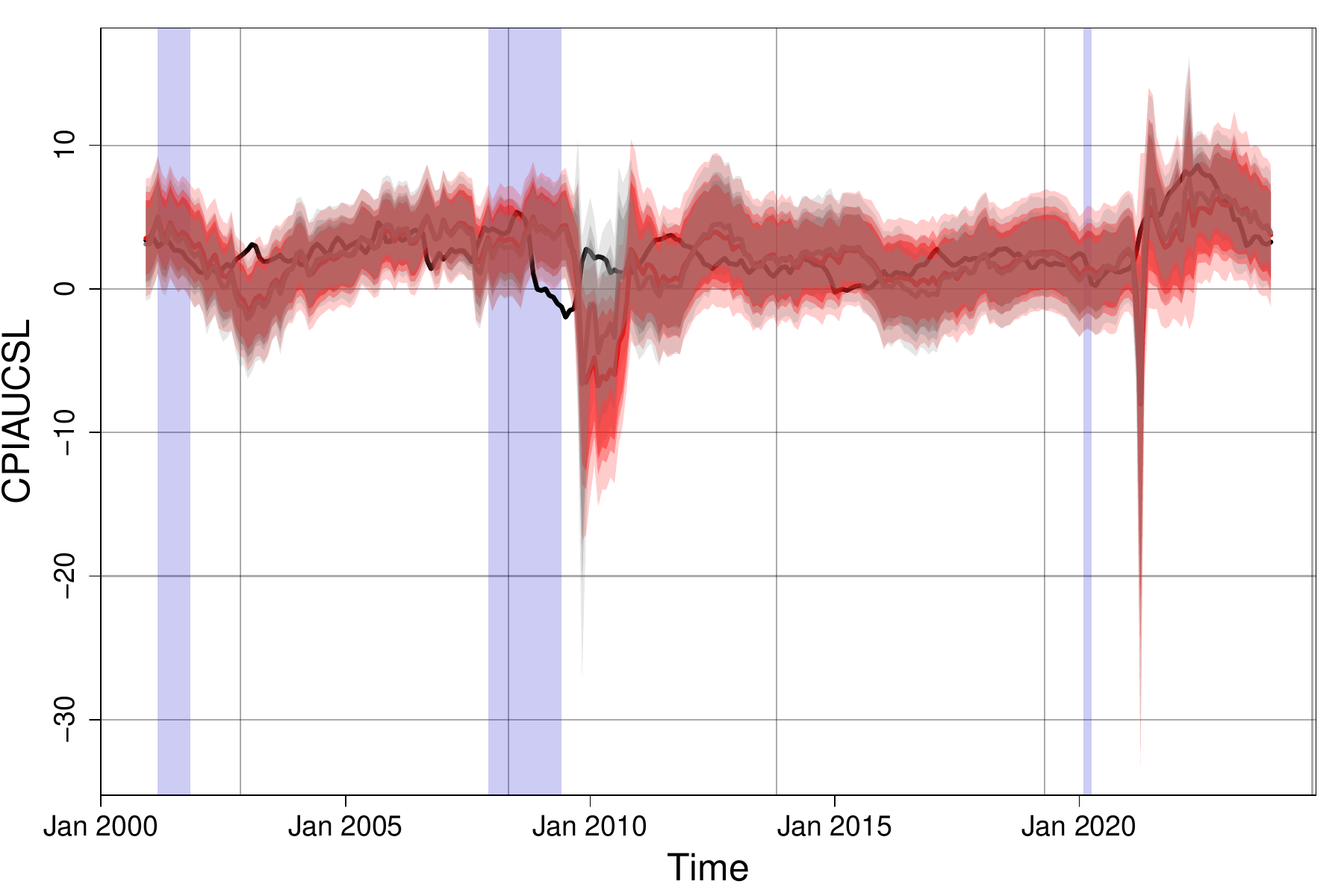}
         \caption{CPIAUCSL}
         \label{fig:pred_dist_CPIAUCSL_large_12}
     \end{subfigure}
     % FEDFUNDS
     \begin{subfigure}[b]{0.3\textwidth}
         \centering
         \includegraphics[width=\textwidth]{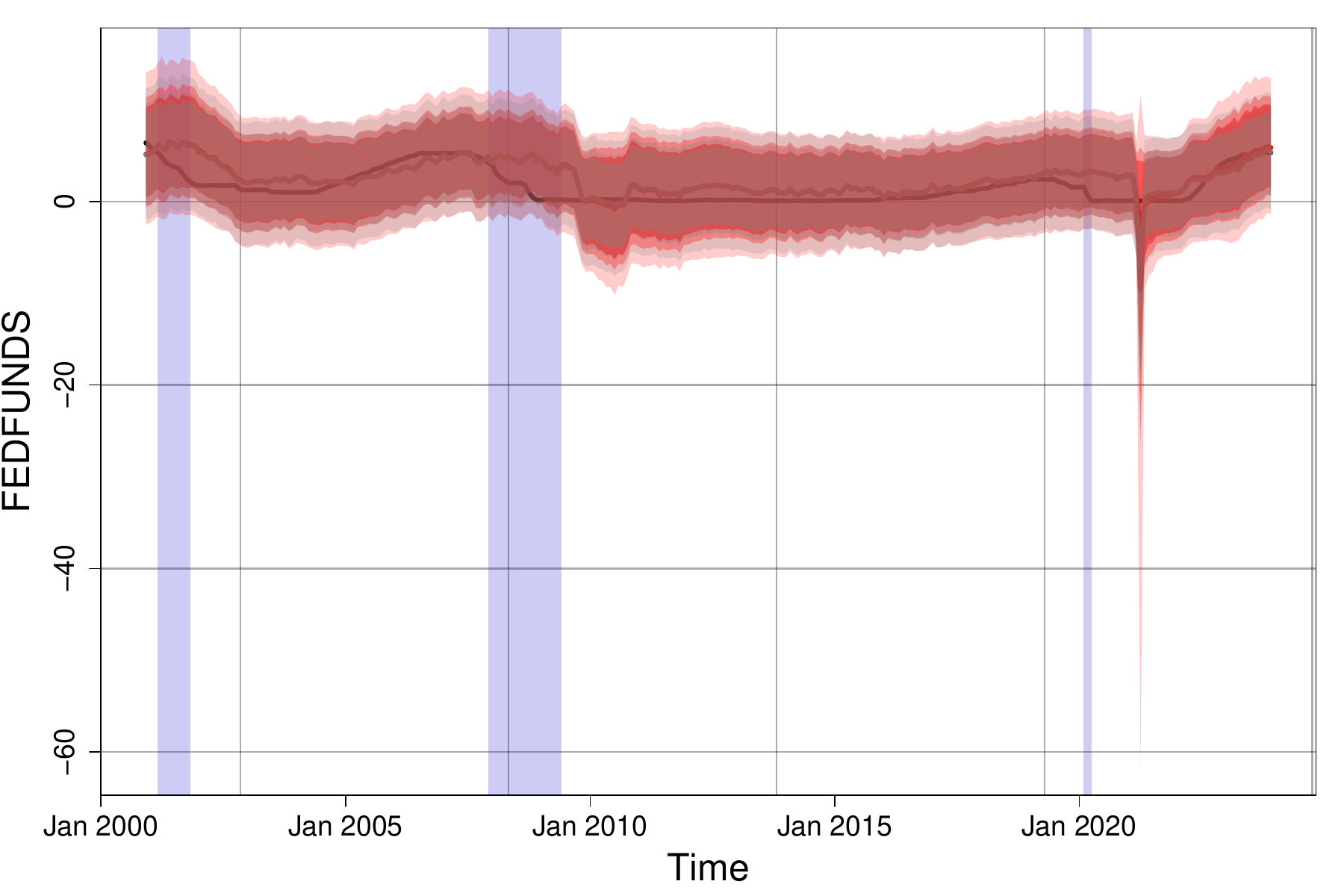}
         \caption{FEDFUNDS}
         \label{fig:pred_dist_FEDFUNDS_small_12}
     \end{subfigure}
     \hfill
     \begin{subfigure}[b]{0.3\textwidth}
         \centering
         \includegraphics[width=\textwidth]{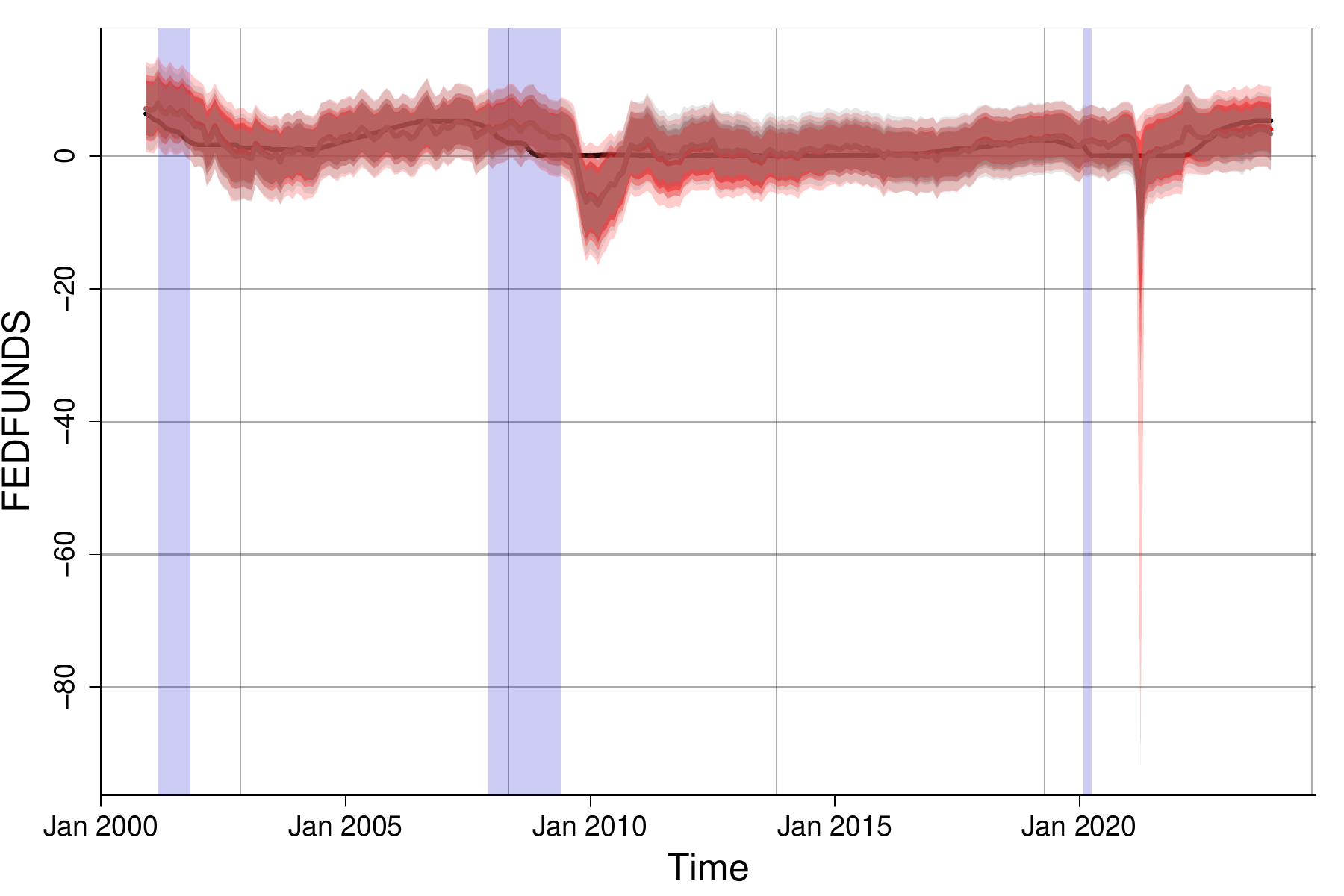}
         \caption{FEDFUNDS}
         \label{fig:pred_dist_FEDFUNDS_medium_12}
     \end{subfigure}
     \hfill
     \begin{subfigure}[b]{0.3\textwidth}
         \centering
         \includegraphics[width=\textwidth]{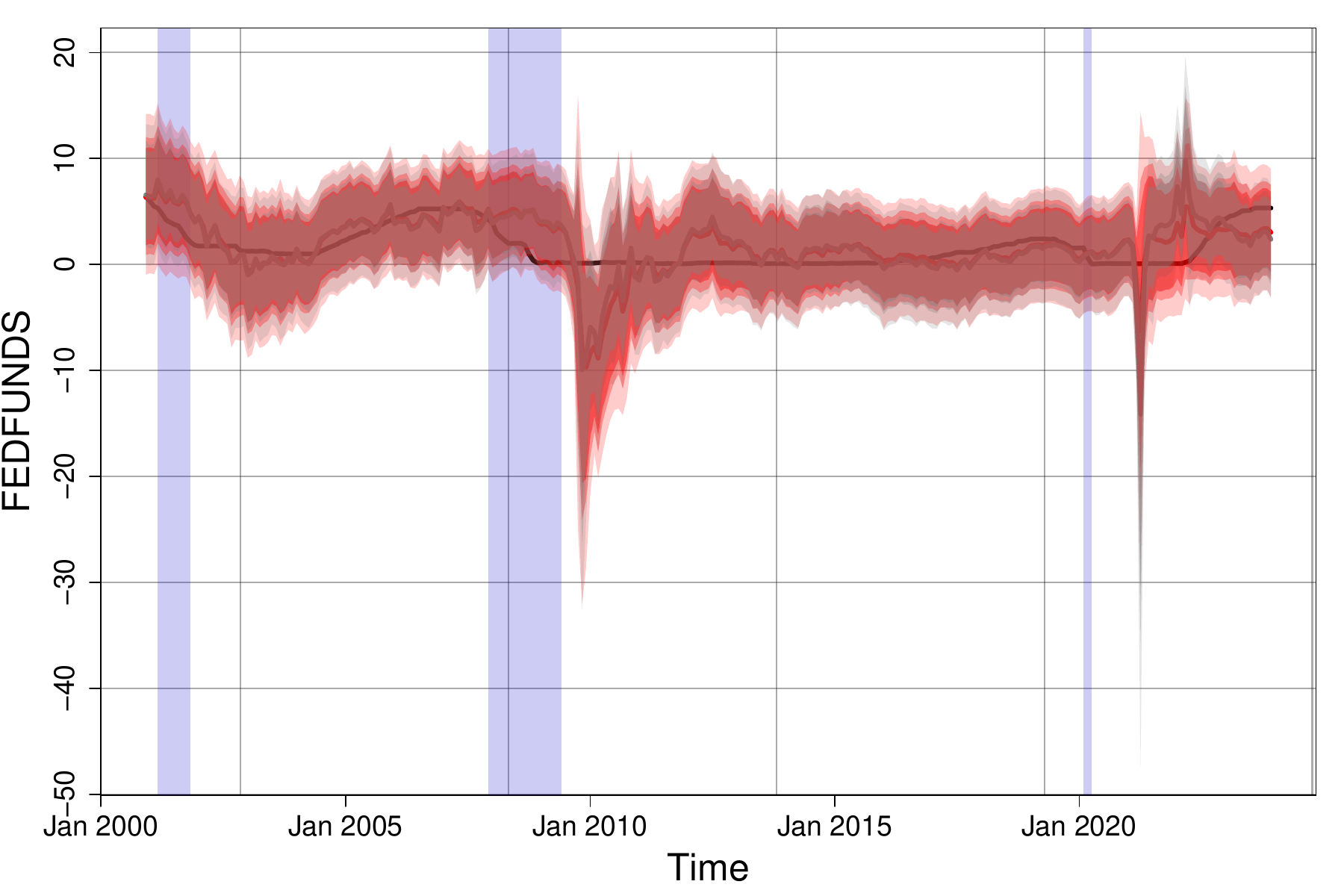}
         \caption{FEDFUNDS}
         \label{fig:pred_dist_FEDFUNDS_large_12}
     \end{subfigure}
    \caption{Observed path versus one-year-ahead predictive distributions for each of the focus variables.\\
    \textbf{Legend:} \textcolor{red}{-} indicates predictions of the cBVAR and \textcolor{gray}{-} those of the standard BVAR. Lines denote the posterior median and shaded areas the 90\%, 95\% and 99\% credible intervals (from dark to light). Blue shaded areas indicate the NBER recession dates.}
    \label{fig:pred_dist_12}
\end{figure}

Additionally, we plot the predictive distributions for each of the focus variables and model sizes at the one-month-ahead (\autoref{fig:pred_dist_1}), one-quarter-ahead (\autoref{fig:pred_dist_3}) and one-year-ahead (\autoref{fig:pred_dist_12})

\end{appendices}

\end{document}